\documentclass[%
reprint,
aps,rmp,
]{revtex4-1}

\usepackage{graphicx}
\usepackage{dcolumn}
\usepackage{bm}
\usepackage{color}

\usepackage[linktocpage]{hyperref} 

\def\bea{\begin{eqnarray}}
\def\eea{\end{eqnarray}}
\def\beq{\begin{equation}}
\def\eeq{\end{equation}}

\def\nablav{\mbox{\boldmath$\nabla$}}

\def\rot{\nablav\times}

\def\ddt{\partial_t}
\def\ddx{\partial_x}

\def\xv{\hat{\bf x}}

\def\bra{\left\langle}
\def\ket{\right\rangle}
\newcommand{\braket}[1]{\bra#1\ket}

\def\Wcm2{\mbox{W cm}^{-2}}
\def\um{\mu\mbox{m}}
\def\cm3{\mbox{cm}^{-3}}
\def\MeV{\mbox{MeV}}

\begin{document}

\title{Ion acceleration by superintense laser-plasma interaction}

\author{Andrea Macchi}
\affiliation{Istituto Nazionale di Ottica, Consiglio Nazionale delle Ricerche (CNR/INO), U.O.S. ``Adriano Gozzini'', Via G. Moruzzi 1, Pisa, Italy}
 \email{andrea.macchi@ino.it}\homepage{http://www.df.unipi.it/~macchi}
\affiliation{Department of Physics ``Enrico Fermi'', University of Pisa, Largo Bruno Pontecorvo 3, I-56127 Pisa, Italy}
\author{Marco Borghesi}
\affiliation{Centre for Plasma Physics, The Queen's University of Belfast, BT71NN Belfast, UK}
 \email{m.borghesi@qub.ac.uk}
\affiliation{Institute of Physics of the ASCR, ELI-Beamlines Project, Na Slovance 2, 18221 Prague, Czech Republic}
\author{Matteo Passoni}
\affiliation{Dipartimento di Energia, Politecnico di Milano, Via Ponzio 34/3, I-20133 Milan, Italy}
\email{matteo.passoni@polimi.it}

\date{\today}

\begin{abstract}
Ion acceleration driven by superintense laser pulses is attracting an 
impressive and steadily increasing effort. Motivations can be found in the 
potential for a number of foreseen applications and in the perspective to 
investigate novel regimes as far as available laser intensities will be 
increasing. Experiments have demonstrated in a wide range of laser and target 
parameters the generation of multi-MeV proton and ion beams with unique 
properties such as ultrashort duration, high brilliance and low emittance. 
In this paper we give an overview of the state-of-the art of ion acceleration by laser pulses  as well as an outlook on its future development and perspectives. We describe the main features observed in the experiments, the observed scaling with laser and plasma parameters and the main models used both to interpret experimental data and to suggest new research directions.

\end{abstract}

       
       
\maketitle

\tableofcontents

\section{Introduction}
\label{sec:intro}
More than half a century ago,
\textcite{vekslerSJAE57} introduced the concept of
``coherent acceleration'' of particles
as a mechanism in which the accelerating field
on each particle is proportional to the number of particles being
accelerated, in contrast to traditional techniques.
Additional elements in Veksler's vision of future accelerators
included the
automatic synchrony between the particles and the accelerating field, the
localization of the latter inside the region where the particles are, and
the production of quasi-neutral groups with
large numbers of energetic particles.

These features are realized in the acceleration of ions from
plasmas produced by intense laser pulses. There, as a very general description,
strong electric fields are generated by a collective displacement of a large
number of electrons, and such electric
fields accelerate ions until charge neutrality is restored
and ions move  together with electrons in a ballistic way.

Before the year 2000, ions having energies up to several MeVs had been observed
in several high-intensity laser-matter interaction experiments and for
different targets, including thick solid targets
\cite[and references therein]{gitomerPF86,fewsPRL94,begPoP97},
gas jets \cite[and references therein]{sarkisovPRE99,krushelnickPRL99}
and sub-micrometric clusters
\cite[and references therein]{ditmireN97,ditmireN01}.
Common to these experiments was the rather isotropic
ion emission and the resulting low brilliance, making these configurations not
attractive as ion accelerators for applications.

\begin{figure}[b!]
\includegraphics[width=0.48\textwidth]{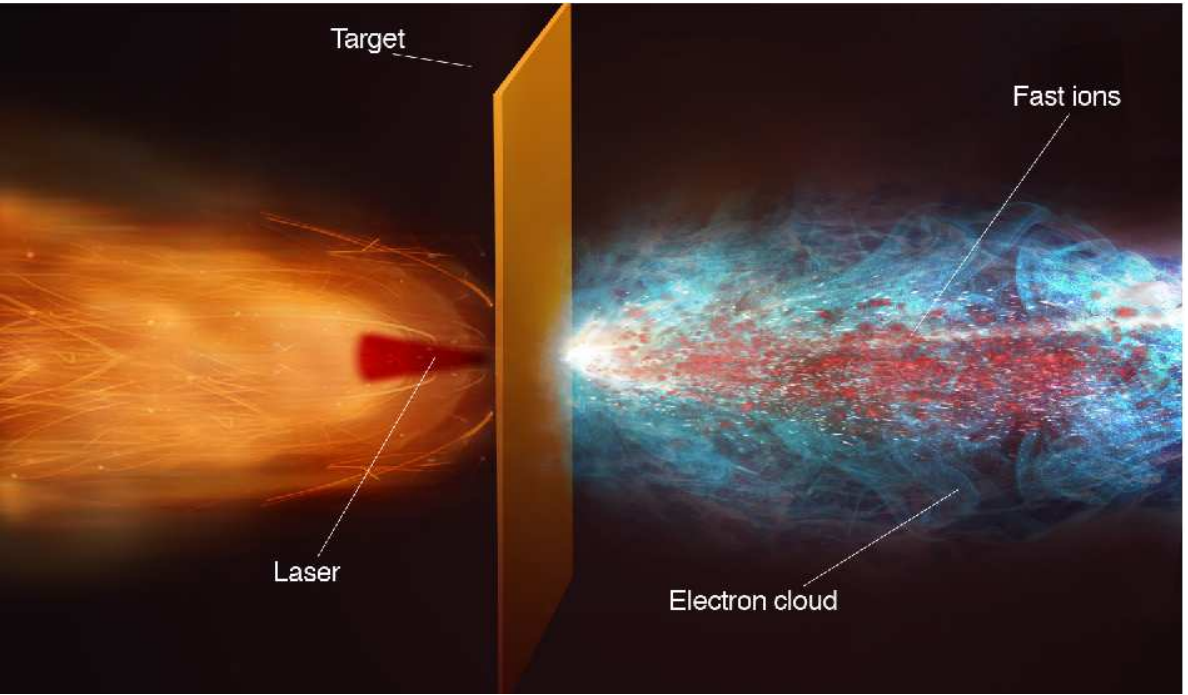}
\caption{(Color online)
Artist's view of a typical experiment on proton emission from
laser-irradiated solid targets.}
\label{fig:expscheme}
\end{figure}

\begin{figure}[t!]
\includegraphics[width=0.48\textwidth]{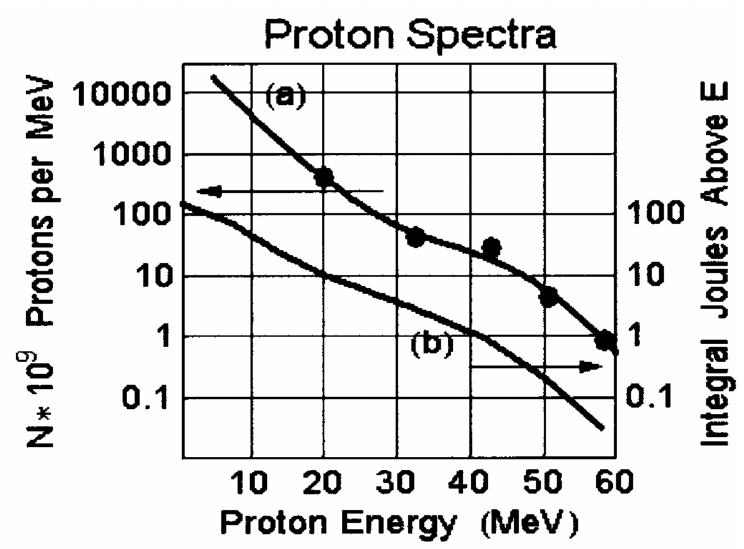}
\caption{Proton energy spectrum from the rear side of a $100~\mu\mbox{m}$ solid
target irradiated by a $423~\mbox{J}$, $0.5~\mbox{ps}$ pulse
at normal incidence, corresponding to an intensity of
$3 \times 10^{20}~\mbox{W cm}^{-2}$. The integrated energy of protons indicates
a conversion efficiency of $\simeq 10\%$ for protons above $10~\mbox{MeV}$.
Reprinted figure with permission from 
\textcite{snavelyPRL00}, Phys. Rev. Lett. \textbf{85}, 2945.
Copyright 2013 by the American Physical Society.}
\label{Fig3a_PRL_snavely_00}
\end{figure}

In 2000 three experiments
\citep{clarkPRL00,maksimchukPRL00,snavelyPRL00}
independently reported the
observation of an intense emission of multi-MeV protons from solid targets,
either metallic or plastic (CH),
of several microns thickness irradiated by high intensity laser pulses.
The basic set-up of these experiments in shown in Fig.\ref{fig:expscheme}.
The laser intensities, number of protons and
maximum ion energy observed
for the three experiments were, respectively,
$3 \times 10^{18}~\Wcm2$, $\gtrsim 10^9$ and $1.5~\MeV$ \citep{maksimchukPRL00},
$5 \times 10^{19}~\Wcm2$, $\sim 10^{12}$  and $18~\MeV$  \citep{clarkPRL00},
and
$3 \times 10^{20}~\Wcm2$, $\sim 2 \times 10^{13}$ and $58~\MeV$
\citep{snavelyPRL00}.
Fig.\ref{Fig3a_PRL_snavely_00} shows the spectrum of ions observed by
\textcite{snavelyPRL00}.
The protons were emitted as a quite collimated beam
in the forward direction with respect to the
laser pulse propagation and were detected at the
rear side of the target, opposite to the laser-irradiated surface.

The emission of protons from metallic targets whose chemical composition does
not include hydrogen may sound surprising, but it was already clear from
previous experiments that protons originated from impurities, i.e. thin layers
of water or hydrocarbons which are ordinarily present on solid surfaces.
In experiments
performed both with ``long'', nanosecond pulses
\cite[and references therein]{gitomerPF86} and ``short''
(sub--)picosecond, high-intensity pulses
\cite{fewsPRL94,begPoP97,clarkPRL00b},
protons and heavier ions were commonly detected in the backward direction
(i.e. towards the laser) with a broad angular distribution and
their origin was interpreted in terms of acceleration in the expansion of
the hot laser-produced plasma {at the front (laser-irradiated) side
of the target.}
The characteristics of the forward proton emission in the new experiments,
such as
the high degree of collimation and laminarity of the beam,
were much more impressive.

\begin{figure}[b!]
\includegraphics[width=0.48\textwidth]{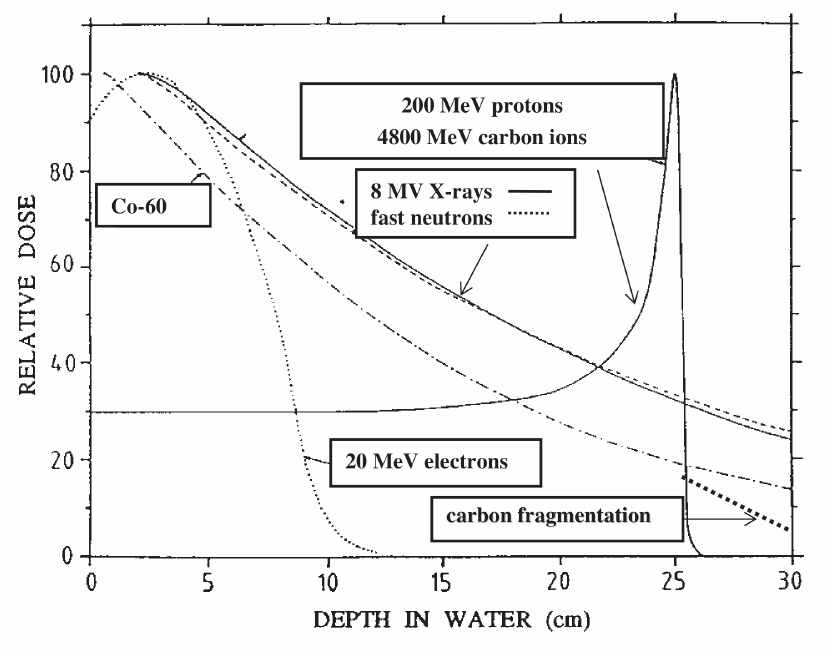}
\caption{Example of the profile of energy deposition of protons and C ions
in water, compared to those of electrons, X- and $\gamma$-rays, and neutrons.
Protons and C ions profiles are characterized by the Bragg peak at the end of
the path. The quantity plotted is the relative dose, i.e. the energy absorbed
per unit mass.
Reprinted figure from 
\textcite{amaldiRPP05}, Rep. Prog. Phys. \textbf{68}, 1861.
By permission from Institute of Physics Publishing (2013).
}
\label{fig:braggpeak}
\end{figure}

These findings generated an enormous interest
both in fundamental research and
in the possible applications.
For these latter, the most relevant and peculiar feature of
multi-MeV ions is the profile of energy deposition in dense matter. Differently
from electrons and X-rays, protons and light ions deliver most of their energy
at the end of their path (Fig.\ref{fig:braggpeak}),
at the so-called Bragg peak
\citep{ziegler-book08,knoll-book}.
The physical reason is that the energy loss is
dominated by Coulomb collisions for which the cross section strongly grows
with decreasing energy, so that the stopping process becomes progressively
more and more efficient. This property makes protons and ions very suitable
for highly localized energy deposition. The applications that were proposed
immediately after the discovery of multi-MeV proton acceleration included
ion beam cancer therapy, laser triggering and control of nuclear reactions,
production and probing of warm dense matter,
``fast ignition'' of Inertial Confinement Fusion targets and
injectors for ion accelerators. These foreseen applications are reviewed
in Sec.\ref{sec:applications}. As a particularly innovative and
successful application, ultrafast probing of plasmas by laser-driven
proton beams will be described in Sec.\ref{sec:applications_radiography}.

While the potential for applications was apparent, the details of the physics
behind proton acceleration were not clear.
A debate arose on the actual location of the region
where the protons were accelerated and, consistently, on the mechanism driving
acceleration. \textcite{clarkPRL00} and
\textcite{maksimchukPRL00} suggested that protons were accelerated at the
front side of the target, crossing the latter and being detected on
the opposite side.
In contrast, \textcite{snavelyPRL00} provided evidence that
protons were accelerated at the rear side
(see also \textcite{hatchettPoP00}).
To support the interpretation of these latter experiments (performed at
the Petawatt facility of Lawrence Livermore National Laboratory, USA)
the so-called Target Normal Sheath Acceleration (TNSA) model was introduced
by \textcite{wilksPoP01}. Briefly, TNSA is driven by the space-charge field
generated at the rear surface of the target by highly energetic 
electrons accelerated at the front surface, crossing the target bulk, and
attempting to escape in vacuum from the rear side. The basic theory of
TNSA and related models are described in detail in
Sec.~\ref{sec:TNSA}.
Most of the experiments later performed on proton acceleration by
laser interaction with solid targets have been
interpreted in terms of the TNSA framework
(Secs.~\ref{sec:TNSA_scenario} and \ref{sec:TNSA_characterization})
that has also guided
developments towards source optimization by target engineering
(Sec.~\ref{sec:TNSA_target}).

\begin{figure}
\begin{center}
\includegraphics[width=0.48\textwidth]{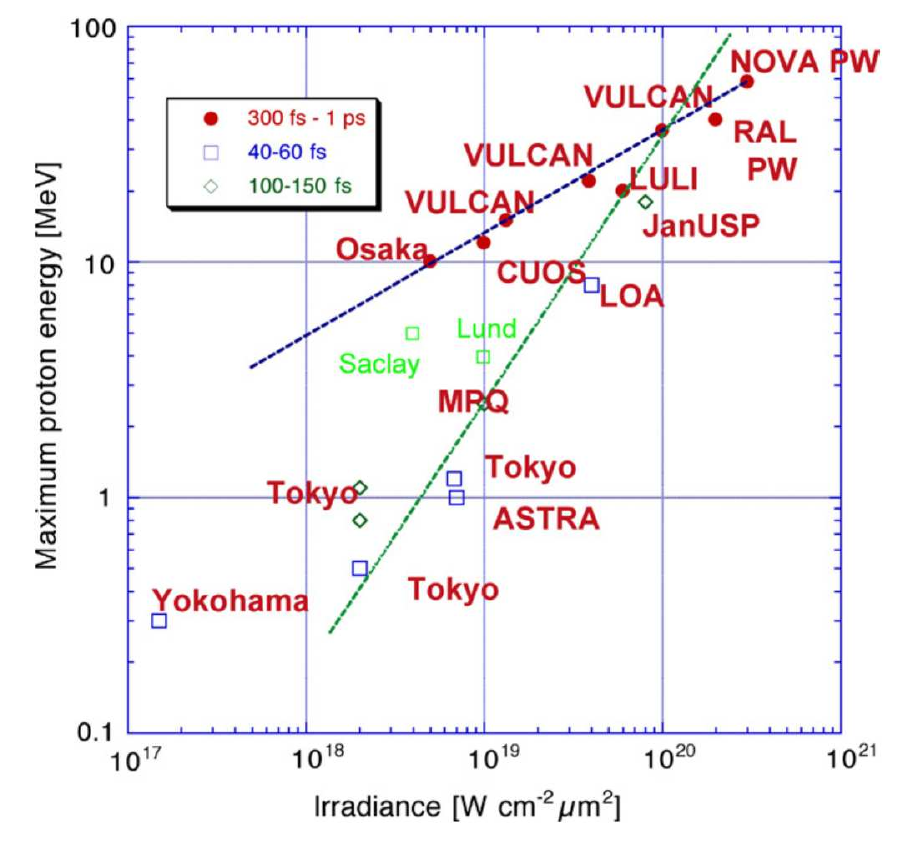}
\end{center}
\caption{(Color online)
Maximum proton energy from laser-irradiated solid targets
as a function of the laser irradiance and for three ranges
of pulse durations, and with additional data (labels ``Lund'' and ``Saclay'')
reporting later experiments up to 2008.
Two trendlines are overlayed, the shallower one corresponding to a $\sim I^{1/2}$ dependence, and the steeper one to a scaling proportional to $I$. 
Reprinted figure from 
\textcite{borghesiPPCF08}, Plasma Phys. Contr. Fusion \textbf{50}, 124040.
By permission from Institute of Physics Publishing (2013).
}
\label{Fig1_PPCF_borghesi_08}
\end{figure}

A major request for several of the foreseen applications is an increase of the
energy per nucleon up to the hundred of MeVs and beyond. The next generation
of laser facilities should allow intensities higher than the current record
of $\sim 10^{22}~\Wcm2$ \citep{yanovskyOE08}, but at present it is neither
guaranteed that
the ion energy scaling observed so far will be maintained at such
extreme intensities nor that TNSA will be still effective. An analysis of
proton acceleration experiments performed up to 2006 suggests a
$\sim (I\lambda^2)^{1/2}$ scaling of proton energy with laser intensity
$I$ and wavelength $\lambda$ up to values of
$I\lambda^2=3 \times 10^{20}~\Wcm2\mu\mbox{m}^2$
\citep{fuchsNP06,borghesiFST06}. Fig.\ref{Fig1_PPCF_borghesi_08} summarizes
such data, together with more recent results obtained with Ti:Sa-based,
ultrashort (tens fs) pulses, exhibiting a $\sim I\lambda^2$ scaling.
Measurements by \textcite{robsonNP07} at energies up to 400~J,
pulse durations between 1 and 8 ps and
intensities up to $6 \times 10^{20}~\Wcm2$ suggested a lower scaling.
It is also of crucial importance to establish the most
relevant scaling parameters
as well as improve or optimize beam emittance, brilliance and monoenergeticity
for specific applications.
For instance TNSA-generated proton beams are highly laminar and have
very low emittance (Sec.~\ref{sec:TNSA_characterization})
but the energy spectrum is ordinarily broad and thus not optimal for most
applications.

These issues motivate the search for {other ion acceleration mechanisms.}
{The latter include concepts which have been explored previously in
different contexts (e.g. astrophysics), such as Radiation Pressure Acceleration
(RPA) and Collisionless Shock Acceleration (CSA). Other proposed schemes
exploit the potential both of advanced target engineering and of nonlinear
``relativistic'' optical effects in plasmas, such as ion acceleration in
ultrathin solid targets which become transparent to intense laser pulses
(Break-Out Afterburner, BOA) {or involving low-density targets}.
The basic physics of these mechanisms and the
related experimental work, still in a preliminary stage with respect to TNSA,
will be described in Sec.\ref{sec:othermech}.}
{The development of advanced acceleration schemes is sustained by the
continuous trend towards laser pulses of higher intensity and larger energy.
A detailed account of the many running projects and developing facilities
related to optical and infrared lasers ($\lambda \sim 1~\um$)
is given in Sec.II of \textcite{dipiazzaRMP12}.
Progress towards CO$_2$ lasers ($\lambda \sim 1~\um$) having Terawatt power
\citep{haberbergerOE10} is also of growing interest for ion acceleration,
as discussed in Secs.\ref{sec:RPA_holeboring} and \ref{sec:shock}.
}

On the theoretical side, the interpretation of experiments has
revitalized classic and often controversial problems of plasma physics such as
plasma expansion into vacuum and the formation of collisionless sheaths,
at the basis of most of TNSA models,
{as well as other general physics models such as the motion of
relativistic moving mirrors, a concept already discussed in
the original work on special relativity by \textcite{einsteinAP05}, which
serves as a model for RPA.}
{Simple, analytically affordable models are extremely useful to
understand the basic acceleration mechanisms and, in particular, to provide
scaling laws which may give directions for further developments. }
{Reference models also highlight the several} connections  with
other fields, such as the physics of discharges, of ultracold plasmas, and
of particle acceleration in astrophysics.
{The theoretical discussions in Secs.\ref{sec:TNSA} and
\ref{sec:othermech} are, to a significant
extent, based on this approach.}

{Beyond simple modeling a rich and complex dynamics of laser-plasma
interaction and ion acceleration, involving collective and self-organization
effects, is apparent. Unfolding such dynamics requires the use of
self-consistent electromagnetic, kinetic simulations. To this aim,
the particle-in-cell (PIC) method
(Sec.\ref{sec:nutshell_simulations})
is by far the mostly used approach. Large-scale, multi-dimensional PIC
simulations running on parallel supercomputers are an effective support for
the design and interpretation of laser-plasma acceleration experiments,
although fully ``realistic'' simulations in three spatial dimensions and for
actual laser and target parameters are most of the time still
beyond computational possibilities.
These limitations further motivate the development of complementary,
reduced simulation models. These issues are further discussed in
Secs.\ref{sec:nutshell_simulations} and \ref{sec:TNSA_PIC}.
}

A comprehensive description of laser-plasma dynamics if far beyond the scope
of the present work and may be found in recent books and reviews
\citep{gibbon-book,mourouRMP06,mulser-book}.
In Secs.\ref{sec:nutshell_interaction} and \ref{sec:nutshell_fastelectrons}
we only describe a few basic issues of relevance for the understanding of ion
acceleration mechanisms. These latter are first introduced in a compact way
in Sec.\ref{sec:nutshell_acceleration}, leaving a detailed discussion in the
following Secs.\ref{sec:TNSA} and \ref{sec:othermech}.

\section{Laser ion acceleration in a nutshell}
\label{sec:nutshell}

\subsection{Laser interaction with overdense matter}
\label{sec:nutshell_overview_interaction}
\label{sec:nutshell_interaction}

In the present work we mostly refer to ion acceleration occurring in the
interaction with solid targets, for which the electron density $n_e$ greatly
exceeds the so-called critical or cut-off density,
\bea
n_c=\frac{m_e\omega^2}{4\pi e^2}
   =1.1 \times 10^{21}~\cm3~\left(\frac{\lambda}{1~\um}\right)^{-2} .
\eea
The condition $n_e=n_c$ is equivalent to $\omega_p=\omega$ where
$\omega_p=(4\pi n_e e^2/m_e)^{1/2}$
and $\omega=2\pi c/\lambda$ are the plasma and laser frequencies, respectively.
Since the linear refractive index of the plasma is
${\sf n}=(1-\omega^2_p/\omega^2)^{1/2}=(1-n_e/n_c)^{1/2}$,
in the $n_e>n_c$ ``overdense'' region ${\sf n}$ has
imaginary values and the laser pulse cannot propagate. All the laser-plasma
interaction occurs either in the ``underdense'' region $n_e<n_c$ or near
the ``critical'' surface at which $n_e \simeq n_c$.

Relativistic effects make the refractive index nonlinear.
Qualitatively speaking, the ``relativistic'' refractive index
describing the propagation of a plane wave
whose vector potential is ${\bf A}={\bf A}(x,t)$
is obtained from the linear expression by replacing the
electron mass with the quantity $m_e\gamma$, where the relativistic
factor $\gamma$ is given by
\bea
\gamma
=\sqrt{1+\braket{{\bf a}^2}}
=\sqrt{1+a_0^2/2},
\label{eq:gamma}
\eea
${\bf a}=e{\bf A}/m_ec^2$, and the angular brackets denote an average
over the oscillation period.
The parameter $a_0$ is the commonly used ``dimensionless''
amplitude related to the laser intensity $I$
by\footnote{Consistently with our definitions, given the value for $I$
the peak value of the dimensionless vector potential of the
plane wave will be given by $a_0$ for linear polarization and by
$a_0/\sqrt{2}$ for circular polarization.}
\bea
a_0=0.85\left({\frac{I\lambda^2_{\tiny\um}}{10^{18}~\Wcm2}}\right)^{1/2},
\label{eq:a0}
\eea
where we used $I=c\braket{E^2}/4\pi$ to relate the electric field ${\bf E}=-(1/c) \partial{\bf A}/\partial t$ to the laser
intensity $I$.

The nonlinear, ``relativistic'' index
${\sf n}_{\mbox{\tiny NL}}=({1-n_e/(\gamma n_c)})^{1/2}$ becomes imaginary
when $n_e>\gamma n_c$, showing an increase of the cut-off density for a
plane wave: this effect is known as relativistic self-induced transparency
or, briefly, relativistic transparency.
However, the problem of laser penetration inside a plasma is not
such trivial \citep{cattaniPRE00,golovizninPoP00,shenPRE01}
because of both the nonlinearity in the wave equation
and the self-consistent modification of the plasma density profile
due to radiation pressure effects.
These latter may be described via the ponderomotive force
(PF).\footnote{Throughout the
present review we refer to the ponderomotive force as the slowly-varying,
effective force describing the cycle-averaged
motion of the ``oscillation center'' of a
charged particle in an oscillating non-uniform
field, over a time scale longer than the oscillation period. ``Fast''
oscillating components are not included in the definition of ponderomotive force here adopted.}
In an oscillating, quasi-monochromatic
electromagnetic field described by a dimensionless vector potential
${\bf a}({\bf r},t)$ whose envelope is sufficiently smooth in space
and time, the relativistic PF is
[see e.g. \textcite{bauerPRL05} and \textcite{mulser-book}]
\bea
{\bf f}_p=-m_ec^2\nablav(1+\braket{\bf a}^2)^{1/2}
.
\eea
For a plane wave impinging on an overdense plasma, the resulting PF, more effective on the lightest particles,
is in the inward direction and tends to push and pile up electrons inside
the plasma. 
Ponderomotive effects will be further discussed below (see
Secs.\ref{sec:nutshell_fastelectrons}-\ref{sec:nutshell_acceleration}).

In a multi-dimensional geometry, a laser pulse of finite width may produce
a density depression around the propagation axis also because 
of ponderomotive pushing of the electrons in the radial direction. 
{Jointly with the relativistic effect and target expansion driven by 
electron heating, this mechanism} 
may lead to a transition to transparency as soon as
the electron density drops the cut-off value \citep{fuchsPoP98b}.
Investigations of ion acceleration in the transparency regime are
described in Sec.\ref{sec:OTHER_transparency}.

The penetration of the laser pulse
depends not only on the electron density but also on the target size
when the latter becomes close to, or smaller than one wavelength.
As a simple but useful example,
the nonlinear transmission and reflection coefficients
can be calculated analytically for a sub-wavelength foil
modeled as a Dirac delta-like density profile \citep{vshivkovPP98,macchiPRL09},
obtaining a transparency threshold
\bea
a_0>\pi \frac{n_e}{n_c}\frac{\ell}{\lambda}\equiv \zeta
\label{eq:SITthinfoil}
\eea
where $\ell$ is the thickness of the foil. This formula has some interest for
the interaction with ultrathin foil targets
(see Secs.\ref{sec:RPA_thin}-\ref{sec:OTHER_transparency}).

\subsection{Hot electrons}
\label{sec:nutshell_fastelectrons}

Since the laser pulse cannot penetrate into solid density regions, the
absorbed energy is there transported mostly by energetic 
(commonly named either ``hot'' or ``fast'') electrons
which may be generated during the interaction by several mechanisms.
By ``hot'' electrons in the present context one typically refers to
relativistic electrons whose
energy is of the order of the cycle-averaged oscillation
energy in the electric field of the laser in vacuum,
\bea
{\cal E}_p&=&
m_ec^2(\gamma-1)
=m_ec^2\left(\sqrt{1+a_0^2/2}-1\right),
\label{eq:ponderomotive}
\eea
where Eq.~(\ref{eq:gamma}) has been used.
Expression (\ref{eq:ponderomotive}) is also called the ``ponderomotive''
energy \citep{wilksPRL92}.
Hot electrons penetrating into solid targets
have been observed and characterized in several experiments
at very high intensities
and for different interaction conditions\footnote{See e.g.
\textcite{keyPoP98,whartonPRL98,tanimotoPoP09,chenPoP09,nilsonPRL11} and references therein.} and play a fundamental
role in applications such as laser-driven photonuclear physics and
fast ignition of fusion targets. Moreover,
as it will be discussed in Sec.~\ref{sec:TNSA},
in most of the experiments reported so far, acceleration of protons
and heavier ions is driven by hot electrons.

The process of hot electron generation turns
out to be complex and, possibly, not completely understood yet. A complete
account of past and ongoing research on the topic 
may be found in recent books
\citep{gibbon-book,mulser-book} and in a vast experimental and
theoretical literature.
Here we give a very basic discussion at a
qualitative level, focusing on those aspects which are most essential
and relevant to ion acceleration.

At the surface of an overdense plasma, electrons are driven by the 
Lorentz force ${\bf f}_L=-e({\bf E}+{\bf v}\times{\bf B}/c)$ which 
include both the incident and reflected laser pulses and the self-generated
fields. 
As a necessary condition for the efficient generation of hot electrons 
near the critical surface ${\bf f}_L$ must have an \emph{oscillating} component
directed along the density gradient $\nablav n_e$. 
This is the case for the well-known resonance absorption where the condition
 ${\bf E}\cdot\nablav n_e \neq 0$ is necessary to drive resonant 
plasma oscillations which in turn accelerate electrons. In a plane geometry 
such condition requires oblique incidence 
and $P$-polarization of the laser pulse, and that absorption is sensitive to 
the density scalelength $L_n=n_e/|\nablav n_e|$ because the driving force 
is evanescent in the resonance region.

In a sharp boundary plasma where $L_n\ll \lambda$, absorption and heating 
may arise because electron motion is not adiabatic, as electrons are driven 
from the region of strong fields to the evanescence region in a time shorter 
than $2\pi/\omega$, so that the cycle average 
$-e\braket{{\bf E}\cdot{\bf v}}$ may not cancel out.
Thus, short duration and high intensity laser pulses favor electron heating 
because the hydrodynamic expansion has both not sufficient time to wash 
out sharp density gradients ot it is dominated by 
the strong ponderomotive force that steepens the density profile.

At this point it is worth to remind that in most high-intensity
experiments the main interaction pulse is preceded by 
prepulses\footnote{{In general the main pulse is preceded by both short pulses of similar duration as the main pulse, and by a much longer pedestal due to amplified spontaneous emission.}}
which cause early plasma formation and expansion, so that the short pulse
interaction does not occur with a sharp-boundary, solid-density plasma.
However, profile steepening at the critical surface
will be still effective, thus one may expect the interactions still occurs with
a sharp density profile, having with a lower density jump with respect to 
a solid target. Occasionally, ``preplasma'' formation may also allow additional
electron acceleration mechanisms to take place in the underdense plasma
region {\citep{esareyRMP09}}, 
possibly leading to electron energies much higher than given by Eq.
(\ref{eq:ponderomotive}). In more recent experiments, advanced
pulse cleaning techniques may allow to minimize prepulse effects
(see Sec.\ref{sec:TNSA_target}).

\subsubsection{Heating models}
\label{subsec:electrostaticmodeling}
\label{subsec:JXB}

\begin{figure}
\includegraphics[width=0.25\textwidth]{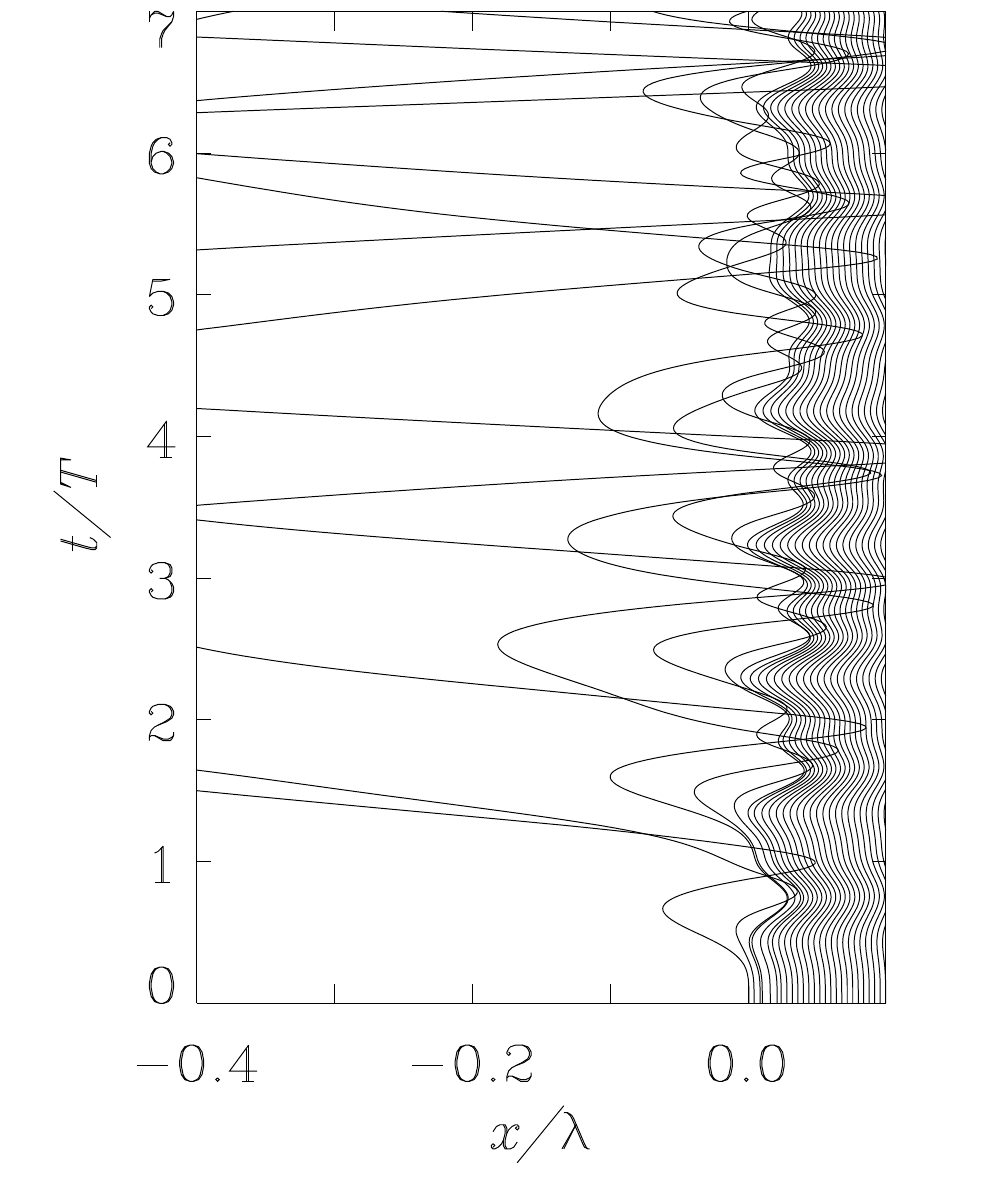}
\caption{Numerical solution of the electrostatic ``plasma sheet'' model
based on Eqs.(\ref{eq:capacitor_lagrange_osc}) plus the exchange of 
initial position for crossing plasma sheets (see text for details). 
The trajectories of a limited number of sheets (one over twenty) in the ($x,t$) 
plane are shown.
The driver field has the profile of an evanescent wave with peak 
amplitude $0.5m_e\omega c/e$ in vacuum and a $\sin^2(\pi t/2\tau)$ 
rising front with $\tau=5T$ where $T=2\pi/\omega$. A density 
$n_e/n_c=(\omega_p/\omega)^2=5$ is assumed.}
\label{fig:simpleVH}
\end{figure}

A popular electrostatic model of electron heating at a step-boundary plasma
has been proposed by \textcite{brunelPRL87}. In this model, 
electrons in this system are dragged out of the surface of
a perfect conductor by an oscillating ``capacitor field'', extending on the
vacuum side, and representing the $P$-component of the incident plus reflected 
laser electric field.   
Electrons are considered to be ``absorbed'' when, after having performed
about half of an oscillation on the vacuum side, they re-enter the target
there delivering their energy, which is of the order of the oscillation energy
in the external field.\footnote{This effect is also commonly refereed to as 
``vacuum  heating''. See \textcite{gibbon-book} for a discussion also on the 
origin of the name.} The model thus accounts in a simplified way for the
pulsed generation (once per cycle) of hot electrons
directed into the target and having an energy,
roughly speaking, close to the ``vacuum'' value (\ref{eq:ponderomotive}).
This simple model is not self-consistent because, for instance, the
capacitor field is assumed to vanish inside the target, implying the presence 
of a surface charge density. Nevertheless,
following \textcite{mulserLPB01} it is possible
to provide a ``minimal'' 1D model, still in the capacitor approximation,
where the electrostatic field is calculated self-consistently and
an acceleration of electron bunches similar to that inferred by
Brunel is apparent.
We take the electric field as the sum of the
electrostatic and a driver fields, e.g. $E_x=E_e+E_d$ where
$E_d=\tilde{E}_d(t)\sin\omega_0 t$ with
$\tilde{E}_d(t)$ a suitable temporal envelope,
a step-boundary density profile $n_i=n_0\Theta(x)$
($Z=1$ for simplicity) and a ``cold'' plasma, i.e. we neglect
thermal pressure. 
Following these assumptions
we write Maxwell's equations for the electrostatic field and Euler's equation
for the electron fluid having velocity $v_x$
\begin{eqnarray}
\partial_x E_e &=& 4\pi\rho=4\pi e [n_0\Theta(x)-n_e],
\label{eq:capacitor_E1D}\\
\partial_t E_e  &=& -{4\pi}J_x={4\pi}en_ev_x, \label{eq:capacitor_j1D}\\
\frac{dv_x}{dt} &=& (\ddt+v_x\ddx)v_x=-\frac{e}{m_e}(E_e+E_d).
\label{eq:capacitor_Euler1D}
\end{eqnarray}
Switching to Lagrangian variables $x_0$ and
$\xi=\xi(x_0,t)$ defined by $x=x_0+\xi$, $d\xi/dt=v_x$,
a straightforward calculation along with the constraint of
$E_e$ being continuous at $x=x_0+\xi=0$ yields the following equations of motion
describing electrostatic, forced oscillations of
electrons across a step-like interface:
\bea
\frac{d^2\xi}{dt^2}=\left\{ \begin{array}{lr}
             -\omega_p^2\xi  -eE_d/m_e & (x_0+\xi>0) \\
             +\omega_p^2 x_0 -eE_d/m_e & (x_0+\xi<0)
            \end{array}
\right. .
\label{eq:capacitor_lagrange_osc}
\eea
From Eqs.(\ref{eq:capacitor_lagrange_osc}) we see that electrons
crossing the boundary ($x=x_0+\xi<0$) feel a secular force $\omega_p^2 x_0$
leading to dephasing from $E_d$ and acceleration \citep{mulserLPB01}.
Eqs.(\ref{eq:capacitor_lagrange_osc}) can be solved
numerically for a discrete but
large ensemble of electron ``sheets''
(corresponding to a set of values of $x_0>0$),
with the prescription to exchange the values of $x_0$ for two crossing sheets
to avoid the onset of singularity in the equations.\footnote{This numerical implementation basically corresponds to the pioneering, elementary model of plasma simulation formulated by \textcite{dawsonPF62}.}
Representative trajectories of electrons moving across the
interface are found as in Fig.\ref{fig:simpleVH}.
Electrons whose trajectory extends in vacuum for half or one period of the
driving field and then re-enter at high velocity inside the plasma are
observed. Similar trajectories are found in
electromagnetic and self-consistent simulations
(Sec.\ref{subsec:numerics}).

Notice that the ``cold'' plasma assumption is consistent with the requirement
that the external field should be strong enough to overcome the
potential barrier which, in an equilibrium state, confines warm electrons 
inside a bounded plasma (such barrier corresponds to a Debye sheath, see also
Secs.\ref{subsubsec:rearsurfaceacceleration} and \ref{sec:TNSA_QS}).
{For $\omega_p\gg\omega$ and nearly total reflection, the laser field 
component normal to the surface has an amplitude 
$E_{\perp}\simeq 2E_0\sin\theta$, with $E_0=(4\pi I/c)^{1/2}$ the amplitude in 
vacuum.  
The sheath field $E_s \simeq T_e/(e\lambda_D)=(4\pi n_0T_e)^{1/2}$, so that 
the condition $E_{\perp}>E_s$ may be rearranged as 
$4(I/c)\sin^2\theta>n_0T_e$. This implies (at non-grazing incidence) the
radiation pressure to exceed the thermal pressure and thus} 
to counteract the thermal expansion and to steepen the density profile, 
making the assumption of  a step-like plasma more self-consistent. 

For $S$-polarization or normal incidence there is no component of the electric
field perpendicular to the surface. However, for high intensities the 
magnetic force term becomes important and may drive electron oscillations 
along the density gradient also for normal incidence. This effect 
is commonly named as ``${\bf J}\times{\bf B}$'' heating \citep{kruerPF85}.
By considering the driver capacitor field as a model for the magnetic force 
component, the related electron dynamics may still be described using  
the above outlined models, but with two significant differences. First,
to lowest order the magnetic force oscillates at $2\omega$, thus leading 
to the generation of hot electron bunches twice per laser period.
Second, the oscillating component perpendicular to the surface vanishes for 
circular polarization (and normal incidence), so that hot electron generation
might be strongly \emph{suppressed} in such conditions. In fact, 
the vector potential representing a plane, elliptically 
polarized field may be written as 
\bea
{\bf A}(x,t)=\frac{A(x)}{\sqrt{1+\epsilon^2}}
       (\hat{\bf y}\cos\omega t+\epsilon\hat{\bf z}\sin\omega t),
\label{eq:ellipticpol}
\eea
Using ${\bf B}=\rot{\bf A}$ and ${\bf p}_{\perp}=e{\bf A}/c$ for the transverse
momentum of electrons, the 
$-e({\bf v}\times{\bf B}/c)$ force can be written as
\begin{eqnarray}
-e\frac{\bf v}{c}\times{\bf B}
=-\hat{\bf x}\frac{e^2\partial_x A^2(x)}{4m_e\gamma c^2}
    \left(1+\frac{1-\epsilon^2}{1+\epsilon^2}\cos 2\omega t\right),
    \label{eq:f02w}
\end{eqnarray}
showing that the oscillating component vanishes 
for circular polarization ($\epsilon=1$).\footnote{A more detailed analysis shows that electron heating is quenched when the parameter $\epsilon$ exceeds some threshold value, see \textcite{rykovanovNJP08,macchiCRP09}.} 

The integral over $x$ of $n_ef_{px}$, where
${f}_{px}=\braket{f_x}=\braket{-e({\bf v}\times{\bf B}/c)_x}$
is the steady ponderomotive force density on electrons, equals the total 
radiation pressure on the target surface.
For circular polarization and normal incidence we thus expect radiation 
pressure to push the target while electron heating is quenched. 
These conditions have been investigated in order to optimize radiation 
pressure acceleration of ions versus other mechanisms driven by hot electrons, 
see Sec.\ref{sec:RPA}.

\subsubsection{Simulations, multi-dimensional effects and simple estimates}
\label{subsec:numerics}

A more quantitative description of laser absorption and hot electron
generation requires numerical simulations.
To address electromagnetic effects in his model
\textcite{brunelPF88} performed two-dimensional (2D) PIC simulations in a 
plane wave, oblique incidence geometry.
Several later studies using 1D simulations with the
``boosted frame'' technique \citep{bourdierPF83} 
are summarized and reviewed by \textcite{gibbonPP99}.
The absorption degree of a $P$-polarized laser
pulse is quite sensitive to the incidence angle and the density
scalelength, with the latter varying on the timescale
of ion motion \citep{gibbonPRL94} yielding a time-dependent absorption.
Experimental attempts \citep{flaccoJAP08,mckennaLPB08,bataniNJP10}
have been made to vary the density scalelength
in order to increase absorption in hot electrons 
and consequently to enhance ion acceleration 
(see Sec.\ref{sec:TNSA_optimization2}).
Hot electron generation tends to become more efficient for lower plasma
densities, and particularly close to the critical density $n_c$, as it is 
observed that stronger coupling and volumetric heating occurs near the 
transmission threshold. A ``near-critical'' plasma may be either
produced by the laser prepulse or by using a special target material, e.g. 
a low-density foam (see Sec.\ref{sec:OTHER_underdense}).

2D simulations reveal additional effects, as for instance
the deformation of the plasma surface due to radiation pressure-driven
``hole boring'' (see also Sec.\ref{sec:RPA_holeboring})
that changes the local incidence angle \citep{wilksPRL92}, 
leading to increased absorption and
providing a dynamic ``funnel'' effect collimating the electron
flow inside the target \citep{ruhlPRL99}.
{A similar dynamics occurs in microcone targets which have proved to be 
effective in enhancing hot electron generation
\citep[and references therein]{sentokuPoP04,nakamuraPRL09,gaillardPoP11}.}

Absorption is also sensitive to small-scale surface deformations, either
self-generated or pre-imposed, so that the use of microstructures on the front 
target surface of has been also suggested as a way to enhance hot electron
generation: see e.g. \textcite{klimoNJP11} and references therein. 
Another possible approach is the use of grating surfaces where the resonant
excitation of surface plasma waves may also lead to very high absorption
\citep{raynaudPoP07,bigongiariPoP11}.

\begin{figure}
\begin{center}
\includegraphics[width=0.48\textwidth]{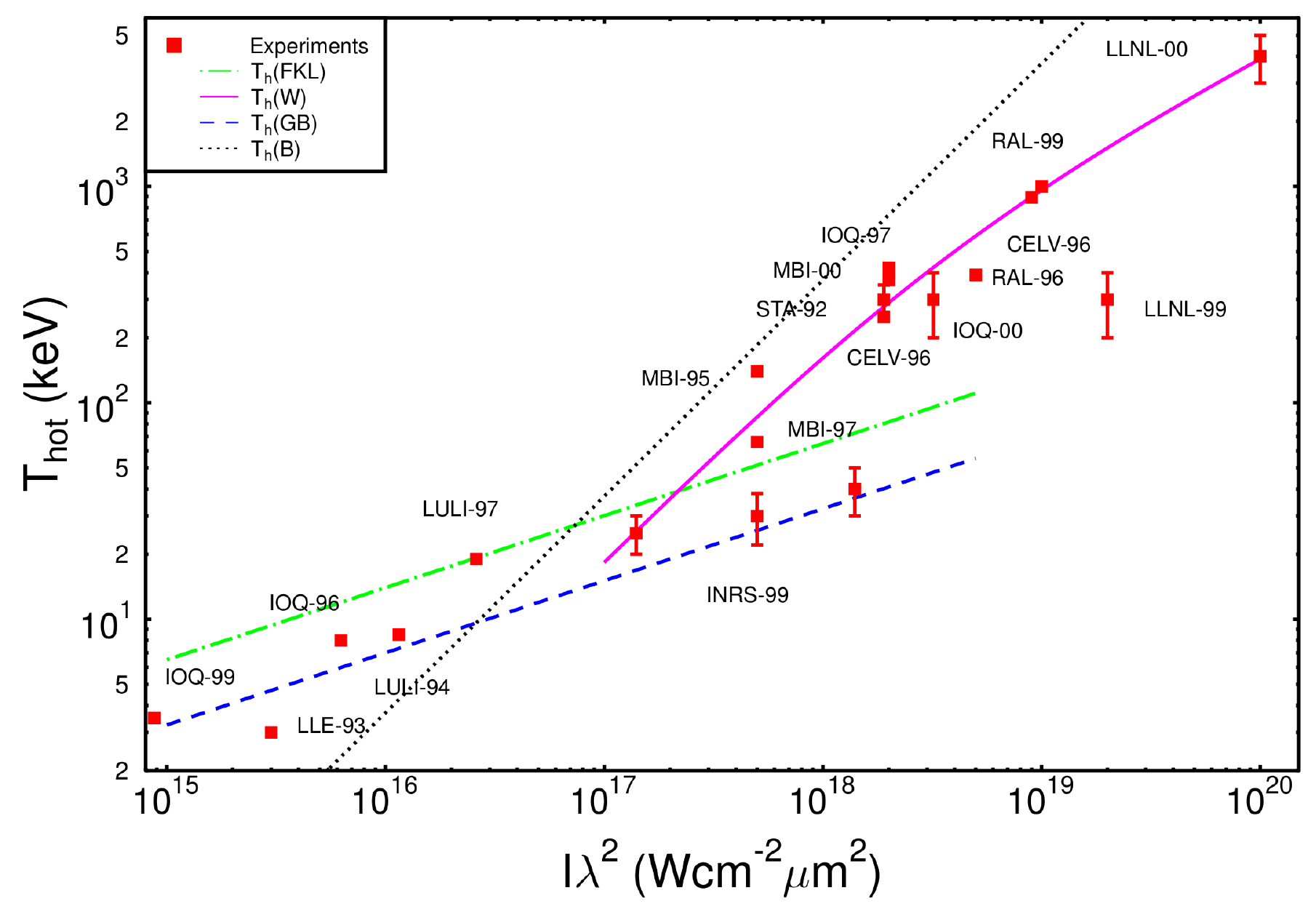}
\caption{(Color online)
Hot electron temperature as a function of irradiance from experiments
of sub-ps laser-solid interaction. See table~5.2 in \textcite{gibbon-book} for
details on experimental parameters, diagnostic methods and references.
The lines give scaling laws derived from different models
[FKL: \textcite{forslundPRL77}, W: \textcite{wilksPRL92}, 
GB: \textcite{gibbonPRL92}, B: \textcite{brunelPRL87}].
Reprinted figure from \textcite{gibbon-book}, \emph{Short Pulse Laser Interaction with Matter} (Imperial College Press).
By permission from Imperial College Press and World Scientific (2013).}
\label{Fig5-19_gibbon-book}
\end{center}
\end{figure}

The high sensitivity of hot electron generation to laser and plasma parameters
partly accounts for data scatter
and differences observed in the many experimental investigations reported
in the literature, with the above mentioned prepulse effects bringing
additional complexity. For these reasons, absorption values and
characteristics of the hot electron distribution are often taken into
account in a phenomenological way.
It has been often considered acceptable to assume the hot
electron distribution to be Maxwellian with a temperature $T_h$ given by
Eq.(\ref{eq:ponderomotive}) as a function of the laser irradiance.
Fig.\ref{Fig5-19_gibbon-book} presents a collection
of temperature measurements obtained for sub-picosecond pulses up to the year
2000 \citep{gibbon-book};
these data broadly support a scaling of $T_h$ as $(I\lambda^2)^{1/2}$.
The total
fractional absorption in hot electrons $\eta_h$ is usually estimated to be in
the $10\%-30\%$ range, with experimental indications of possibly quite higher
values at ultra-relativistic intensities \citep{pingPRL08}.
An energy flux balance condition such as $\eta_h I \simeq n_h v_h T_h$
(with $v_h \simeq c$ at ultra-high intensities)
may then be used to estimate the ``initial'' density of hot electrons $n_h$,
which usually is not larger than $n_c$, consistently with the argument that
$n_h$ cannot exceed the density of the region where hot electrons are
generated.

Inside the target, the effective density may become
different from the above estimate for $n_h$
in particular conditions due to, e.g., the angular divergence
of the electron flow  or to electron refluxing effects
depending on the electron time of flight and recirculation and thus on the
target thickness \citep{mackinnonPRL02}.
Still one might roughly estimate the total number of hot electrons $N_h$ by an
energy balance relation $N_h\sim\eta_h U_L/T_h$ where $U_L$ is the energy
of the laser pulse. The angular divergence $\theta_{\mathrm{div}}$
is also estimated from experiments to range between $20$ and $60$ degrees
and to increase with irradiance \cite[and references therein]{greenPRL08}, 
although such estimates might depend on the
accuracy of the sheath field modeling  
\cite{ridgersPRE11}.

\subsubsection{Hot electron transport in solid matter}
\label{sec:nutshell_fastelectrons_transport}

Transport of hot electrons in solid matter has been investigated extensively
also because of its relevance for the electron-driven Fast Ignition scheme in
ICF [see \textcite{freemanFST06} for a survey].
Key issues characterizing this regime are the very high values of the
currents and the effect of self-generated fields.
From the above estimates it can be inferred that near the front surface
of the target the current density ${\bf j}_h=-en_hv_h$ associated to hot
electrons may reach values up to
$j_h \sim en_cc \simeq 4.8 \times 10^{12}~\mbox{A cm}^{-2}$, corresponding to
a total current of $\sim 15~\mbox{MA}$ over a circular spot of
$10~\um$ radius. This huge current must be locally neutralized by a return
current ${\bf j}_r$ such ${\bf j}_h+{\bf j}_r\simeq 0$, otherwise either the
electric field generated by the charge unbalance or the magnetic field
generated by the free flowing current ${\bf j}_h$ would be strong enough to
stop the hot electrons \citep{daviesPRE97,passoniPRE04}.
{The free, ``cold'' electrons contributing to the return current are 
either present as conduction electrons in metals or produced by field and
collisional ionization in insulators \citep{tikhonchukPoP02}.
Additional complexity is introduced by effects such as target heating
and hot electron refluxing, which have been inferred in several experiments
\citep[and references therein]{belleiNJP10,nilsonPoP11,quinnPPCF11}.}
Filamentation instabilities and dependence on the target material
have also been extensively studied 
\citep[and references therein]{fuchsPRL03,manclossiPRL06,mckennaPRL11}. 
Simulation models accounting
for both collisional effects and self-consistent generation of quasi-static
fields are needed for quantitative investigations.\footnote{See e.g.
\textcite{gremilletPoP02,bellPPCF06,evansPPCF07,klimoPRE07,solodovPoP09,kempPoP10} and references therein.}
Finally, it is noticeable that at least a fraction of hot
electrons propagate coherently through the target conserving the temporal
periodicity of the driving force, i.e. as bunches with $\omega$ or $2\omega$
rate depending on incidence angle and polarization, as inferred by
optical transition radiation measurements \citep{popescuPP05}.

\subsection{Ion acceleration mechanisms}
\label{sec:nutshell_acceleration}

\begin{figure}[t!]
\includegraphics[width=0.48\textwidth]{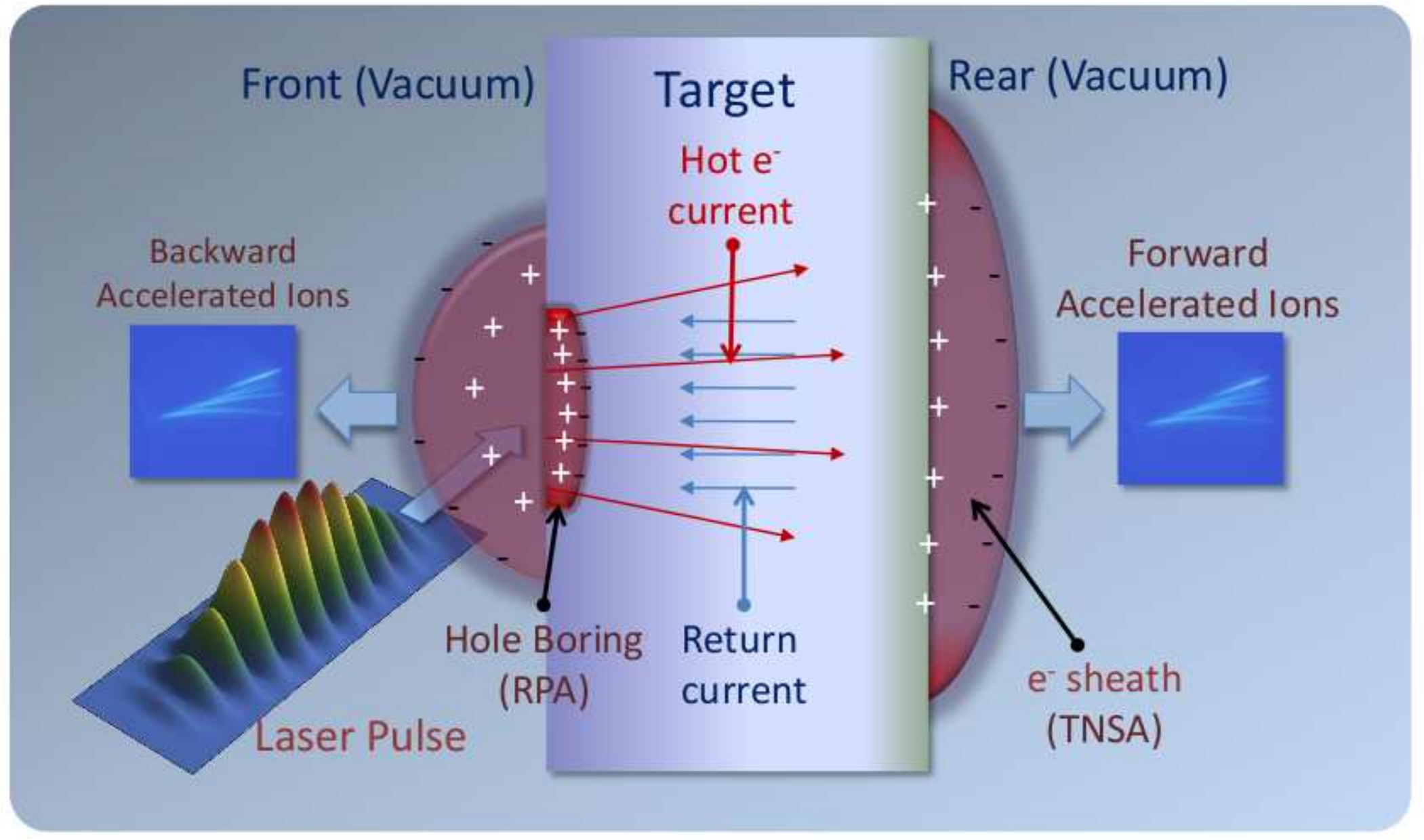}
\caption{(Color online)
Cartoon showing {some of} the possible acceleration mechanisms in the interaction with a thick solid target, including TNSA at the rear side (Sec.\ref{subsubsec:rearsurfaceacceleration}), hole boring RPA at the front side (Sec.\ref{subsubsec:frontsurfaceacceleration}), and backward acceleration in the plasma blow-off (see e.g. \textcite{clarkPRL00b}). Also shown are the hot electron flow leading to sheath formation and expansion at the rear side, and the associated return current.}
\label{fig:scheme_acceleration_mechanisms}
\end{figure}

In this Section we give an overview of ion acceleration mechanisms
including both those proposed to explain early experimental results in solid
targets and those investigated later, either following inspiration from
theoretical work or testing novel target designs.
Some of the mechanisms described below and the target regions where they
are active are indicated in the cartoon of
Fig.\ref{fig:scheme_acceleration_mechanisms}.
Ion acceleration models will be described more in detail in
Secs.\ref{sec:TNSA} and \ref{sec:othermech} along
with the most relevant experiments.

\subsubsection{Rear surface acceleration}
\label{subsubsec:rearsurfaceacceleration}

As outlined in Sec.\ref{sec:nutshell_fastelectrons},
a very intense current of high-energy hot electrons may be generated at
the front side of the target and eventually reach the rear side. There,
as the hot electrons cross the rear side boundary and attempt to escape in
vacuum at the rear side, the charge unbalance generates
a sheath field $E_s$, normal to the rear surface.
Since $E_s$ must backhold electrons with a typical ``temperature'' $T_h$,
the typical spatial extension of the sheath $L_s$ will be related to $E_s$ by
\bea
eE_s \sim \frac{T_h}{L_s}.
\label{eq:roughsheath}\label{eq:Eself}
\eea
From dimensional arguments, assuming a steep interface and
$n_h$ and $T_h$ as the only parameters,
$L_s$ may be roughly estimated as the Debye length of hot electrons,
$L_s \sim \lambda_{Dh}=({T_h}/{4\pi e^2 n_h})^{1/2}$.
Assuming the simple scalings of Sec.\ref{sec:nutshell_fastelectrons} for $T_h$,
taking a laser irradiance $I\lambda^2=10^{20}~\mbox{W cm}^{-2}$ 
and a fractional absorption $\eta_h=0.1$ we find
$T_h \simeq 5.1m_ec^2=2.6~\mbox{MeV}$,
$n_h \sim 8 \times 10^{20}~\cm3$, $\lambda_{Dh}=4.2\times 10^{-5}~\mbox{cm}$ and
$E_s \sim 6 \times 10^{10}~\mbox{V cm}^{-1}$.
This huge field will backhold
most of the escaping electrons, ionize atoms at the rear surface and start
to accelerate ions. 
{As a rough estimate, a test ion crossing the sheath would acquire the 
energy ${\cal E}_i \sim ZeE_sL_s=ZT_h$, resulting in MeV energies and a scaling 
as $I^{1/2}$ if $T_h \simeq {\cal E}_p$ given by Eq.(\ref{eq:ponderomotive})
holds.}
Protons from a thin layer of hydrogen-containing impurities
on the surface will be in a very favorable condition for acceleration because
of both their initial position, located at the maximum of the field, and their
{highest} charge-to-mass ratio so that they will be more 
rapid than heavier ions in following electrons and screening the sheath field. 
This is the qualitative
scenario for TNSA of protons as introduced by
\textcite{wilksPoP01} to explain their experimental results on proton
acceleration \citep{snavelyPRL00,hatchettPoP00}. 

The essential features of the TNSA mechanism have been supported
by several experiments
and TNSA has become the reference framework to interpret observations of
multi--MeV protons from the target rear side. Various schemes for
beam optimization and control have been designed on the basis of TNSA. A detailed discussion of main experimental findings is reported in
Secs.\ref{sec:TNSA_scenario}, \ref{sec:TNSA_characterization} and \ref{sec:TNSA_target}. 

From a theoretical viewpoint, there are two main categories of models
which describe TNSA, namely ``static'' and ``dynamic'' models which, depending on the starting assuptions, allow to provide simplified analytical descriptions to be used to interpret experimental data. 
{These models and related numerical investigations are presented in Sec.\ref{sec:TNSA_modeling}.}

\subsubsection{Front surface acceleration}
\label{subsubsec:frontsurfaceacceleration}

Already in the first measurements of proton acceleration in the forward
direction, the possibility of a contribution originating at
the front surface of the target was also conceived
\citep{clarkPRL00,maksimchukPRL00}. 
As a consequence, mechanisms leading to
ion acceleration in such region have also been extensively investigated.

At the front surface, the intense radiation pressure of the laser pulse
pushes an overdense target inwards, steepening the density profile and bending
the surface; this process is {commonly named} 
as ``hole boring''.
The recession velocity {$v_{hb}$} of the plasma surface
may be estimated
by balancing the electromagnetic and mass momentum flows, 
${I}/{c} \sim n_i(m_i v_{hb})v_{hb}$.
This corresponds to an energy per nucleon 
${\cal E}_{i}={m_p}v^2_{hb}/2 \sim {I}/{(A n_i c)}$.
The scaling with the laser
intensity $I$ is more favorable than the $I^{1/2}$ scaling for TNSA,
and suggests Radiation Pressure Acceleration (RPA) effects to become more 
important for higher intensities.
More accurate, relativistic and dynamic modeling is
presented in Sec.\ref{sec:RPA_holeboring} along with related 
experimental indications. 

Radiation pressure action and hot electron temperature may also lead to 
the generation of collisionless shock waves \cite{tidman-krall} with high Mach number $M$. Such waves are associated to the reflection of ions from the shock front, resulting in a velocity $v_i=2Mc_s$ and an energy per nucleon ${\cal E}_i=2m_pM^2c_s^2=2(Z/A)M^2T_h$, being $c_s=\sqrt{ZT_h/Am_p}$ the ion sound velocity. Such Collisionless Shock Acceleration (CSA) scenario and related experiments are discussed in Sec.\ref{sec:shock}.

{Finally, the possibility of front side (or bulk) acceleration being favored by resistivity effects is discussed in Sec.\ref{sec:OTHER_resistive}.}

\subsubsection{Acceleration in ultrathin, mass-limited and low-density targets}
\label{subsubsec:acceleration_target}

Both TNSA and RPA may have different features in targets having peculiar geometrical and physical properties, if compared to the 
solid targets used in the 2000 experiments 
\cite{clarkPRL00,maksimchukPRL00,snavelyPRL00}
which were several micron thick and much wider than the laser spot diameter.
Experimental investigations of ``ultrathin'', sub-micrometric targets 
requires extremely ``clean'', prepulse-free
pulses to avoid early target evaporation and thus became possibile only
recently thanks to the development of
advanced techniques (see Sec.\ref{sec:TNSA_target}).
The use of ``mass-limited'' targets which also have limited lateral
dimensions (in the sub mm-range) allows 
the refluxing and concentration of hot electrons in
a small volume, and may lead 
to higher ion energies via TNSA. 
These {studies} will be presented in Sec.\ref{sec:TNSA_target}.

For RPA, a sufficiently thin foil target is expected to be accelerated 
as a whole.
Assuming the foil to be a perfect mirror of thickness $\ell$, 
its non-relativistic motion may be simply described by the equation 
$m_in_i\ell dV/dt=2I/c$ from which we obtain an energy 
${\cal E}_i=m_pV^2/2=(2/m_i)(F/\ell c)^2$ where $F=\int I dt$ 
is the laser pulse fluence. 
This is the basics of the ``Light Sail'' (LS) regime of RPA 
(Sec.\ref{sec:RPA_lightsail}) that seems very promising for the foreseen 
fast scaling and the intrinsic monoenergeticity. 

For extremely thin (a few nm) targets,
the breakthrough of the laser pulse through the foil due to relativistic 
transparency may stop LS-RPA, 
but at the same time lead to strong heating of electrons. This effect opens
up a regime of enhanced acceleration possible, which has been also named 
Break-Out Afterburner (BOA) and will be discussed in 
Sec.\ref{sec:OTHER_transparency}.

In general, reducing the effective size of the target allows for
laser pulse penetration, volumetric heating, and energy confinement,
possibly allowing for efficient ion acceleration even at low laser pulse
energies. As a famous example, the interaction of ultrashort,
moderate intensity ($\simeq 10^{16}~\mbox{W cm}^{-2}$) pulses
with sub-wavelength clusters allowed acceleration of ions up to energies
sufficient to produce nuclear fusion reactions \citep{ditmireN97,ditmireN01}.
A limitation on the use of such clusters as ion sources is the isotropic
ion emission and the resulting low brilliance. 
``Droplet'' targets with size of the order of one wavelength have been
investigated as a trade-off approach, as discussed in 
Sec.\ref{sec:TNSA_optimization_source}.

As mentioned above, special target materials may be used to produce plasmas
with density close to $n_c$ (for laser wavelengths $\lambda \sim 1~\um$)
in order to enhance the generation of hot electrons
which drive TNSA (see Sec.\ref{sec:OTHER_underdense}).
Gas jet targets {have been} also used both with $\lambda \sim 1~\um$
lasers to investigate ion acceleration in {underdense} plasmas 
(Sec.\ref{sec:OTHER_underdense})
and with CO$_2$ lasers ($\lambda \sim 10~\um$) 
for studies of RPA and CSA in 
moderately overdense plasmas 
(see Secs.\ref{sec:RPA} and \ref{sec:shock}). 
Apart from the possibility to vary the background density,   
using flowing gas jets as targets is of interest because they enable the 
interaction with a pure proton plasma and are suitable 
for high repetition rate operation as needed for most 
foreseen applications (Sec.\ref{sec:applications}).

\subsection{Particle-in-cell simulations}
\label{sec:nutshell_simulations}

The particle-in-cell (PIC) method \citep{dawsonRMP83,birdsall-langdon},
already mentioned in the Introduction, is the most widely used approach to
the kinetic simulation of plasmas. The PIC method provides a 
solution to the Maxwell-Vlasov system using a Lagrangian approach, with 
fields and currents allocated on a fixed grid and the phase space represented
by an ensemble of computational particles. Thus, the PIC method is mostly 
appropriate to describe collisionless laser-plasma interaction dynamics, 
although models are available either to implement
either collisions [see e.g. \textcite{fiuzaPPCF11} and references therein]
or ionization [see e.g. \textcite{petrovPPCF09} and references therein].

\begin{figure}[b]
\begin{center}
\includegraphics[width=0.48\textwidth]{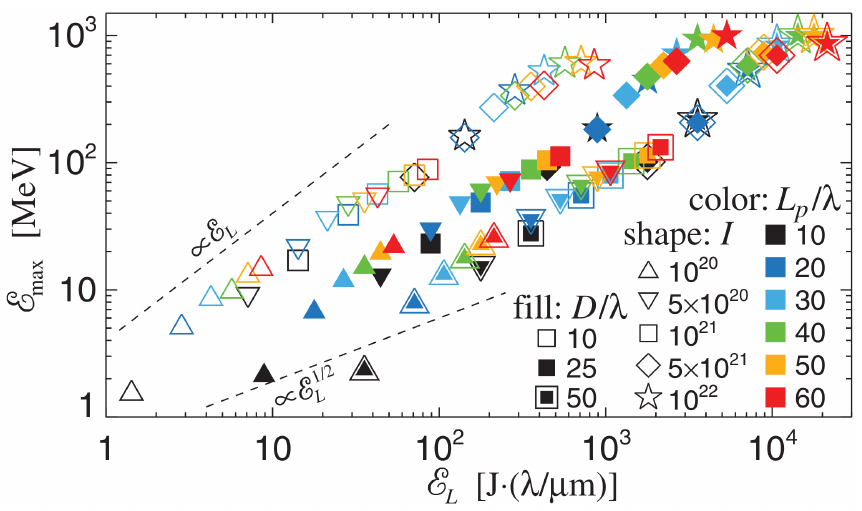}
\caption{(Color online)
Results from multiparametric 2D simulations for a double layer target \citep{esirkepovPRL02,esirkepovPRL06}. The maximum energy of protons accelerated
from the rear layer is shown as a function of laser pulse energy ${\cal E}_L$ and for different values of the intensity $I$, pulse length $L_p$ and focal spot diameter $D$. The target density and thickness are $n_e=100n_c$ and $\ell=\lambda$, respectively.
Reprinted figure with permission from 
\textcite{esirkepovPRL06}, Phys.Rev. Lett. \textbf{96}, 105001.
Copyright 2013 by the American Physical Society.}
\label{Fig3_PRL_esirkepov_06}
\end{center}
\end{figure}

PIC simulations of laser interaction with solid density plasmas at peak 
densities typically exceeding $10^2n_c$ are a very demanding task even when
most powerful supercomputers are used. At least, one has to resolve 
temporal scales $\sim \omega_p^{-1}$ and spatial scales $\sim c/\omega_p$ 
where $\omega_p \sim n_e^{-1/2}$ is the plasma frequency. Thus, when approaching
parameters of a real experiment, relevant lenghts such as the laser beam 
waist may correspond to thousand of gridpoints in each spatial direction, and 
typical dynamic times to thousands of timesteps. In addition,
kinetic effects such as generation of hot electron tails in the distribution
function and large density variations need very large numbers of particles 
to be properly resolved. For these reasons, ``realistic'' 3D simulations 
with proper resolution are typically beyond computational possibilities.
This issue forces most of the times either to use a reduced dimensionality
or to relax the actual parameters to some extent, e.g. by assuming relatively 
low densities or short scales. For some peculiar problems, development of 
hybrid modeling may be appropriate, as discussed in Sec.\ref{sec:TNSA_PIC}.

Despite the above mentioned limitations, several groups have been able to 
perform large scale 3D simulations relevant to ion acceleration regimes such 
as, e.g., TNSA \citep{pukhovPRL01}, RPA \citep{esirkepovPRL04,tamburiniPRE12}
and BOA \citep{yinPRL11}. Use of parallel supercomputers has also allowed 
extended multi-parametric studies (Fig.\ref{Fig3_PRL_esirkepov_06}) in order to 
infer scaling laws and to evidentiate a transition from TNSA to RPA dominance 
at high intensities \citep{esirkepovPRL06}.
These and other numerical results will be discussed in Secs.\ref{sec:TNSA} and
\ref{sec:othermech}.

\subsection{Ion diagnostics}
\label{sec:nutshell_diagnostics}

The specific properties of laser-driven ion beams  (e.g. broad spectrum, 
high flux, significant divergence) have required either modifications of 
established {diagnostics} techniques or the development of new ones.

\begin{figure}
\begin{center}
\includegraphics[width=0.48\textwidth]{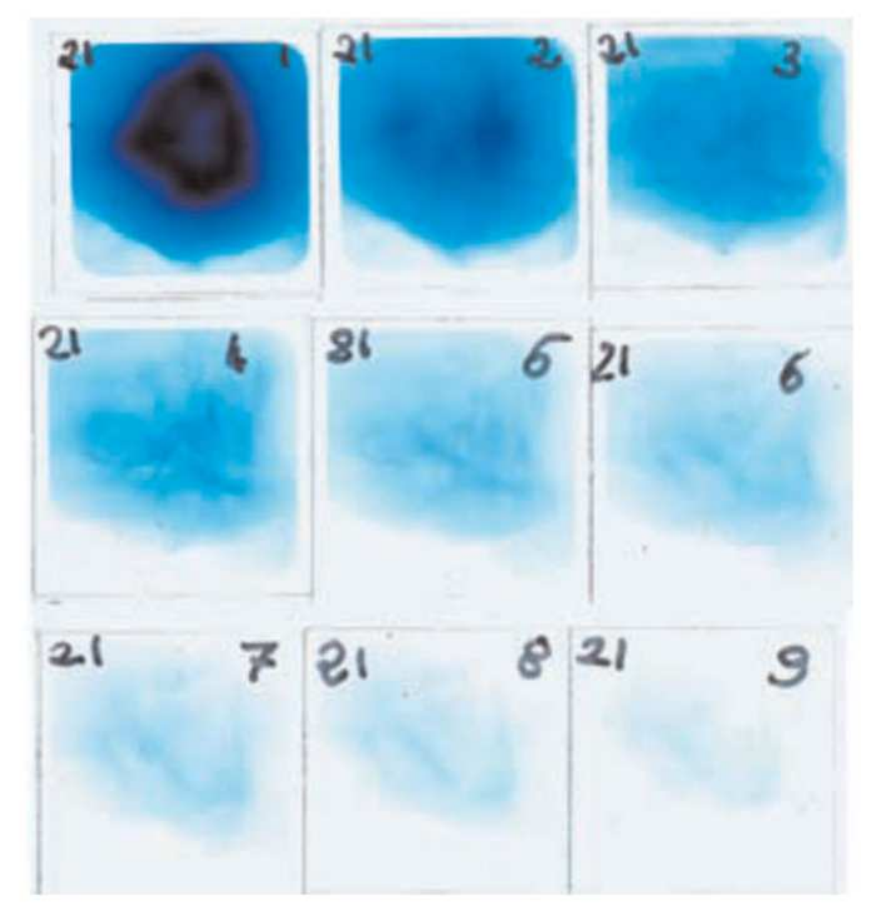}
\end{center}
\caption{(Color online)
A typical RCF stack obtained in an experiment with the TARANIS laser
at Queen's University, Belfast. Higher numbers correspond to higher energy 
protons with layer 1 corresponding to $\sim$ 1~MeV and layer 9 to $\sim$ 12~MeV.
Below each film the corresponding Bragg peak energy for protons is reported.
Reprinted figure from 
\textcite{dzelzainisLPB10}, Las. Part. Beams \textbf{28}, 451.
By permission from Cambridge University Press (2013).
}
\label{fig:dzelzainisLPB10}
\end{figure}

Radiochromic film (RCF) \citep{mclaughlinPRD96} is a detector which is favoured
by many experimenters, since the early work by \textcite{snavelyPRL00},
mainly due to simplicity of use. This is a high-dose, high-dynamic range film, 
widely used in medical context for X-ray dosimetry
\citep{niroomand-radMP98}.
The films consist of one or more active layers containing a microcrystalline monomeric dispersion buried in a clear plastic substrate. 
Typical examples are the HD810, MD55 and EBT2 Gafchromic varieties. 
After interaction with ionizing radiation, the active material undergoes 
polymerization and the film changes its colour from nearly transparent to blue. 
The consequent change in Optical Density 
can be calibrated against the dose released in the film, and therefore provide information on the flux of particles directed at the layer.
Tipically, RCFs are used in a stack arrangement, so that each layer acts as a filter for the following ones in the stack. 
The signal in a given layer will be due only to ions having energy $E\geq E_B$, 
where $E_B$ is the energy of the ions which reach their Bragg peak within the 
layer (see Fig.\ref{fig:dzelzainisLPB10}).
In first approximation,
for an exponential--like spectrum as those typically produced by TNSA,
the dose deposited in a layer can be taken as proportional to the number of protons with $E\sim E_B$,
allowing a rough spectral characterization of the beam.
Various, more refined procedures 
have been developed for deconvolving the spectral information 
(either integrated across the beam or angularly resolved)
in multilayer RCF data 
\citep{breschiLPB04,heyRSI08,nuernbergRSI09,kirbyLPB11}.

Plastic track detectors such as CR39 \citep{engeRM95},
{which have the advantage of being insensitive to X-rays and electrons,}
have been widely used 
also in multi-stack arrangements, e.g. by \textcite{clarkPRL00}.
CR39  layers (typically $0.25-1~\mbox{mm}$ in thickness) require etching in a 
NaOH solution after exposure to ions, so that the damage tracks created by the 
particles can be revealed thanks to the different etching rate in the track 
compared to the undamaged bulk
\citep{seguinRSI03}.
After etching, the single tracks can be counted, which provides a direct 
measurements of the number of protons hitting the detector. 
CR39 works better for low particle fluxes, as at high flux (tipically above
$\sim 10^8~\mbox{particles~cm}^{-2}$) or for long etching times
the tracks start to overlap, leading to saturation
\citep{gaillardRSI07}.

The interaction of laser-driven high-energy ions with secondary targets can 
initiate a number of nuclear reactions (Sec.\ref{sec:APP_nuclear}), 
which can been used to diagnose the beam properties
{with the ability to provide absolute particle numbers with a linear response 
and virtually no saturation at high flux.}
{The}
$^{63}\mbox{Cu}(p,n)^{63}\mbox{Zn}$
reaction in copper stacks has been used to quantify the proton numbers through 
measurement of $\beta^{+}$ decay of $^{63}\mbox{Zn}$ nuclei,
using a NaI detector-based coincidence counting system
\citep{spencerNIMB01,santalaAPL01}.
Techniques employing a single Cu layer, in which a range of isotopes resulting from proton-induced nuclear reactions is analysed in order to reconstruct the proton spectrum, have also been used \citep{yangAPL04}.
Spectral resolution is provided by a combination of filtering and known thresholds for the considered reaction. The above described approaches 
provide particle flux integrated over the whole beam cross section.
Contact radiography of $(p,n)$-generated isotopes in activation samples (where the activated foil is placed in contact with RCF) has been developed
\citep{clarkeNIMA08,offermannJPCS10}
as a way to achieve 2D images with high spatial resolution and 
extremely high dynamic range.
Neutron spectra produced through fusion reactions of the type
$\mbox{D}(d,n)^3\mbox{He}$ have been used as a diagnostic of laser-driven deuterium ions inside a laser-irradiated target
\citep{habaraPRE04,habaraPoP03,habaraPRE04b}.

Obtaining {spectra with high energy resolution} requires the use of 
magnetic dispersion techniques. 
In simple magnetic spectrometers [see e.g. \textcite{chenRSI08}]
the ions, spatially selected by an entrance slit or pinhole, are dispersed 
along one spatial direction according to their energy by a 
$\sim 1~\mbox{T}$ magnetic field ${\bf B}$
This arrangement, which discriminates particles according to their energy but not to their charge/mass ratio, is adequate for diagnosing
the high-energy proton spectrum in ``standard'' TNSA experiments 
in which protons are the dominant accelerated species 
\citep{hegelichPRL02}.

\begin{figure}[b!]
\begin{center}
\includegraphics[width=0.48\textwidth]{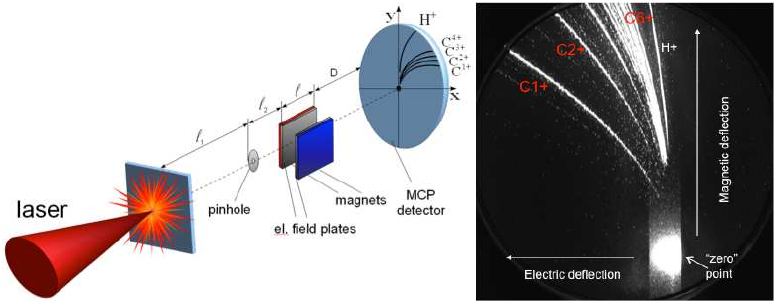}
\caption{(Color online)
Left frame: schematic of a Thomson parabola (courtesy of S.~ter-Avetisyan).
Right frame: a typical example of ion traces obtained with the Thomson parabola.
}
\label{fig:thomson}
\end{center}
\end{figure}

A more complete spectral characterization of multi-component ion beams can be 
obtained with Thomson Parabola spectrometers, 
{based on the principle for mass spectrometry introduced by \textcite{thomsonPM11}. 
A schematic of the device is shown in Fig.\ref{fig:thomson}~a).
Ions are deflected by parallel ${\bf E}$  and ${\bf B}$ fields
(with $E \sim 10^4$~V/m \citep{sakabeRSI80}) resulting in}
a characteristic deflection pattern in which species with different 
charge/mass ratio form separated parabolic traces in the detector plane,
as shown in the typical image of Fig.\ref{fig:thomson}~b).
{Modified magnetic and Thomson spectrometers, having imaging and angular resolution capability, have been also developed \citep{jungRSI11,chenRSI10,ter-avetisyanPoP09}}.

The detectors used in conjunction with these spectrometers are typically 
either CR39, Image Plates (IP), scintillating plates or MicroChannel Plates 
(MCPs). Photostimulable IPs are film-like radiation image sensors, developed 
for X-ray medical imaging, which are composed of specially designed phosphors that trap and store radiation energy in metastable excited states, and
can be absolutely calibrated in terms of particle flux
\citep{mancicRSI08,freemanRSI11}.
Scintillating plates \citep{greenNJP10} or MCPs \citep{ter-avetisyanJPD05}
are favoured in situations where online detection is required 
(e.g. high-repetition laser systems), as the scintillator screen or the MCPs 
phosphor are imaged on a CCD and the detector does not require replacing after 
exposure.
Scintillators can also be used for beam profiling
\citep{sakakiAPE10},
with potential for energy range selection \citep{greenSPIE11}.

A different approach also allowing online beam monitoring is the use of Time of Flight (TOF) techniques, where the broadband ions are left to propagate over a given distance and then detected employing scintillating plates coupled to a Photo Multiplier
\citep{nakamuraJJAP06},
Faraday Cups or semiconductor detectors
\citep{margaroneJAP11}.
The time-varying signal produced by the detectors maps the ion energy spectrum, although the finite response time of the detector and realistic propagation distances limit the use of these techniques for measurements up to a few
$\mbox{MeV}/\mbox{nucleon}$ energy.
{State-of-the art TOF-MCP detectors allow for measurements of protons with a kinetic energy up to $\sim 20$~MeV/nucleon \citep{fukudaPRL09}.}

\section{Target Normal Sheath Acceleration}
\label{sec:TNSA}

\subsection{TNSA Scenario. Main experimental observations}
\label{sec:TNSA_scenario}

As anticipated in Sec.\ref{sec:nutshell_acceleration}
the TNSA process \citep{wilksPoP01}
is a consequence of the huge charge
separation generated by hot electrons reaching the rear side of the target.
There, a cloud of relativistic electrons is formed, extending out of the
target for several Debye lengths, and giving rise to an extremely
intense electric field, mostly directed along the normal to the
surface.
A consequent distinctive feature is that ions are accelerated perpendicularly
to the surface, with high beam collimation.
The electric field generated at the rear surface will depend on
parameters of the electron distribution (temperature, number, divergence)
as well as parameters of the surface itself
(mostly its density profile, as detailed below).

\begin{figure}[b!]
\includegraphics[width=0.48\textwidth]{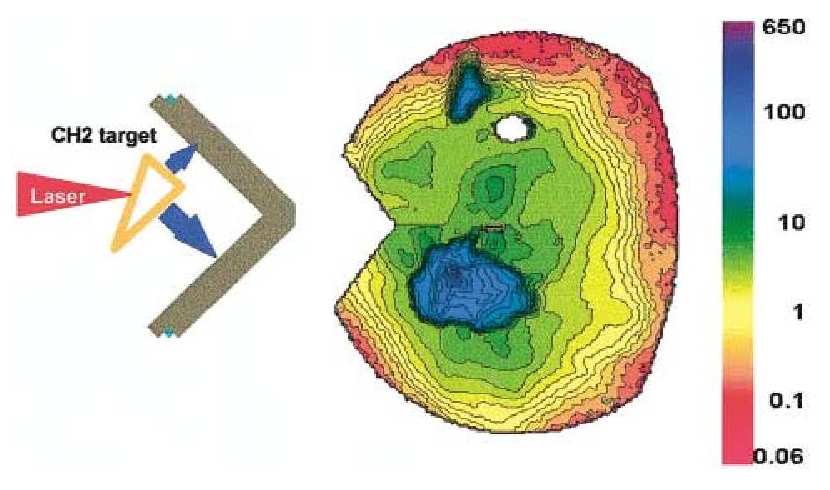}
\caption{(Color online) 
Proton emission from a wedge target effectively having two
rear surfaces. Two separate spots are produced on the detector, showing that
most of the protons originate from the rear side of the target.
Reprinted figure with permission from 
\textcite{snavelyPRL00}, Phys. Rev. Lett. \textbf{85}, 2945.
Copyright 2013 by the American Physical Society.}
\label{Fig4_PRL_snavely_00}
\end{figure}

The acceleration is most
effective on protons, which can be present either in the form of surface 
contaminants or among the constituents of the solid target, 
as in plastic targets. The
heaviest ion populations 
provide a positive charge with much
more inertia, thus creating the charge separation which generates the
accelerating field. Part of the heavy population can be also
effectively accelerated, on a longer time scale, if the proton number is not
high enough to balance the charge of the escaping hot electrons,
and especially if impurity protons are removed before the 
interaction, for example by pre-heating the target
\citep{hegelichPRL02}. 
In this way, ions of several different species may be accelerated 
\cite{hegelichPoP05}.

{Several} observations strongly
supported the TNSA scenario taking place at the rear side.
Already \textcite{snavelyPRL00} gave clear evidence that the emission was 
normal to the rear surface using wedge targets which effectively have more 
than one rear surface. Two separate proton beams were observed
in the directions normal to the two rear surfaces 
of the wedge (Fig.\ref{Fig4_PRL_snavely_00}).

\begin{figure}[b!]
\includegraphics[width=0.48\textwidth]{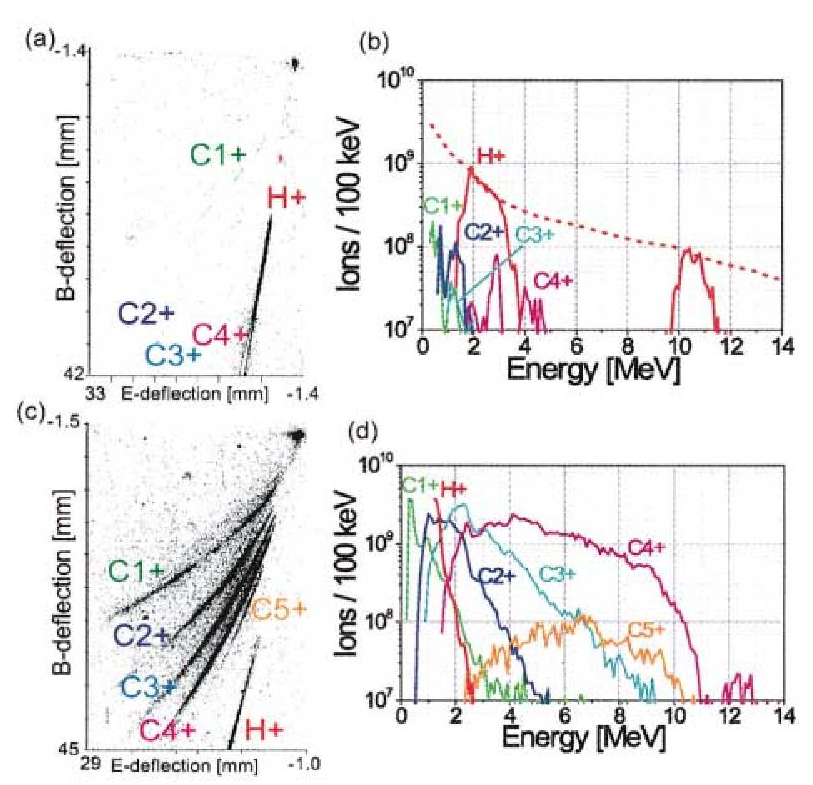}
\caption{(Color online) Effect of impurity removal on Carbon ion spectra. Frames
a) and b) show C ions traces (from CR-39 tracke detectors) and spectra from
Al foils coated with a C layer on the rear side, in the presence of
hydrocarbon contaminants on the surface. In frames c) and d), the contaminants
had been previously removed by resistive heating.
Reprinted figure with permission from 
\textcite{hegelichPRL02}, Phys. Rev. Lett. \textbf{89}, 085002.
Copyright 2013 by the American Physical Society.}
\label{Fig1_PRL_hegelich_02}
\end{figure}

\begin{figure*}[t!]
\includegraphics[width=0.98\textwidth]{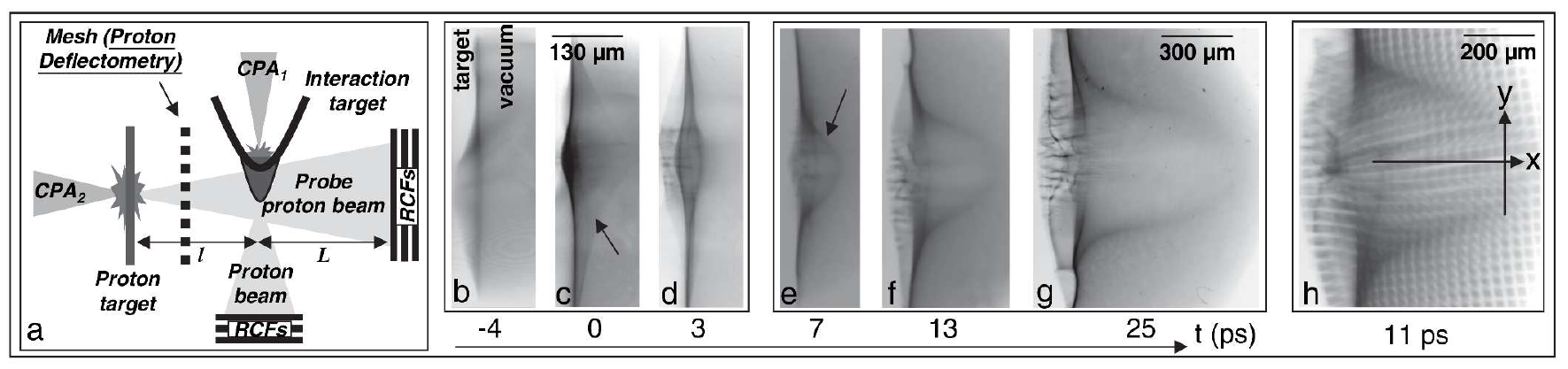}
\caption{Proton probing of the expanding sheath at the rear surface of a
laser-irradiated target. Frame~a): set-up fo the experiment. A proton beam
is used as a transverse probe of the sheath.
Frames~b)-g): Temporal series of images produced by to the
deflection of probe protons in the fields, in a time-of-flight arrangement.
The probing times are relative to the peak of the interaction.
Frame~h): a deflectometry image where a mesh is posed between the probe and
the sheath plasma for a quantitative measure of proton deflections.
Reprinted figure with permission from 
\textcite{romagnaniPRL05}, Phys. Rev. Lett. \textbf{95}, 195001.
Copyright 2013 by the American Physical Society.
}
\label{Fig1_PRL_romagnani_05}
\end{figure*}

\textcite{mackinnonPRL01} reported experimental observations of the interaction
of ultraintense laser pulses using targets with and without preformed plasmas on
the rear surface of the foil. The peak and mean energies of the proton beam 
were found to depend strongly on the plasma scalelength at the rear of the 
target. While an energetic proton beam was obtained with an
unperturbed rear surface, no evidence of high energy protons was recorded
when a large local scalelength in the
ion density at the rear surface was induced, consistently with the dependence
of the accelerating field on the scalelength in
Eq.(\ref{eq:Eself}).
             
\textcite{hegelichPRL02} used as targets Al and W foils, 
resistively heated to remove Hydrogen contaminants and coated on the rear
side with thin C and CaF$_2$ layers, respectively. 
The observation of high-energy C, Ca and F ions
out of these prepared source layers prove the existence of an effective
rear-surface acceleration mechanism (Fig.\ref{Fig1_PRL_hegelich_02}).
Further evidence was given by the work
of \textcite{allenPRL04} who showed that, removing contamination from the back
surface of Au foils strongly reduced the total yield of accelerated protons, 
while removing
contamination from the front surface of the target had no observable effect on
the proton beam.

A further proof that ions are accelerated at the rear side was given by the
observation that a structuring (i.e. grooving) of the rear surface produced
modulations in the proton beam \citep{cowanPRL04}.
This effect also evidences the high laminarity
of the beam and allows to measure its emittance, as discussed in
Sec.\ref{sec:TNSA_characterization}. 

Direct experimental evidence of the generation
of an initial intense sheath field at the rear surface and a late time field
peaking at the beam front was provided by  \textcite{romagnaniPRL05} using
the proton imaging technique (see Sec.\ref{sec:applications_radiography}).
In other words, TNSA provided itself an unique diagnostic which allowed a 
direct experimental confirmation of the essential nature of the acceleration 
process.
Fig.\ref{Fig1_PRL_romagnani_05} shows a temporal series of ``proton images'' in which the propagation of the bell-shaped front of ion expansion can be observed.

More recent developments have been obtained playing with the detailed properties of both the laser pulse and the solid target, showing the possibility to significantly control and optimize the TNSA process, and the capability of achieving interesting and promising variations of the main scheme, also in the light of possible specific applications.
These topics will be presented in Sec.\ref{sec:TNSA_target}.

\subsection{Beam properties characterization}
\label{sec:TNSA_characterization}

Several experiments have
investigated in detail the properties of the TNSA ion beams.
The energy spectrum of the beams is typically broadband, 
up to a cut-off energy {(see Fig.\ref{Fig3a_PRL_snavely_00})}.
The particle number per MeV can be roughly approximated by a
quasi-thermal distribution with a sharp cutoff at a maximum energy
\citep{kaluzaPRL04,fuchsPRL05}
which scales with the laser parameters as will be discussed in
Sec.~\ref{sec:TNSA_comparison}.
Many experiments have reported spectral observations for a wide range of
laser and target parameters, and partial surveys have been provided in a
number of publications
\citep{borghesiFST06,fuchsNP06,zeilNJP10}.

A number of experimental studies have been devoted to the investigation 
of the spatial and angular characteristics of the emitted beams. 
These latter are closely dependent on the electron sheath spatial distribution, 
and consequently on the target properties (resistivity, surface roughness, etc) 
affecting the electron propagation.

It was observed relatively early on that the use of conducting targets leads to 
smooth proton beam profiles with a sharp boundary, as detectable, for example, 
in RCF images \citep{snavelyPRL00,fuchsPRL03},
while using dielectric targets creates non-homogeneities in the proton density 
across the beam section.
In the latter case, the transport of the electron current is prone to electromagnetic instabilities, which break the hot electron flow into filaments.
This leads to an uneven electron sheath at the target rear
\citep{manclossiPRL06}
and consequently to a modulated proton beam cross section
\citep{rothPRSTAB02,fuchsPRL03}.
The close correlation between proton beam properties and electron beam 
transport characteristics has indeed been exploited in a number of experiments, 
which have used the proton beam as a diagnostic for the electron beam behaviour 
inside the target, revealing, beside the aforementioned differences related to 
the target conductivity
\citep{fuchsPRL03},
effects of magnetic collimation on the beam transport
\citep{yuanNJP10,gizziPRSTAB11}
or the role of  lattice structure in dielectric targets
\citep{mckennaPRL11}.
Other factors that can lead to structured beam profiles even in conducting 
targets are surface roughness at the target rear, resulting in a randomized 
local orientation of the protons
\citep{rothPRSTAB02},
and intensity modulations in the focal spot which can be coupled to the protons 
via structured electron beams in medium-$Z$ thin targets
\citep{fuchsPRL03}.

The existence of a sharp angular boundary in the proton angular distribution (clearer in higher-$Z$ and thicker targets) is consistent with a bell-shaped transverse distribution of hot electrons in the rear surface sheath due to the fact that the density will naturally be higher along the laser axis and decrease with transverse radius.
Protons are accelerated normal to the local iso-density contour, and the presence of an inflexion point in the sheath therefore results in a maximum angle of acceleration
\citep{fuchsPRL03}.
Comparison of experimental data with simple electrostatic models indicate that the shape of the accelerating sheath is generally Gaussian
\citep{fuchsPRL03,carrollPRE07}
as also observed directly in sheath imaging data
\citep{romagnaniPRL05}, see Fig.\ref{Fig1_PRL_romagnani_05}.

\begin{figure}[b]
\begin{center}
\includegraphics[width=0.48\textwidth]{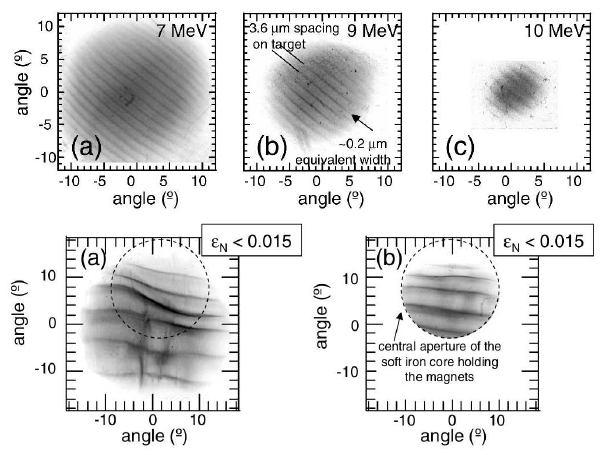}
\end{center}
\caption{Top frames: modulations in the proton distribution, for different energies, on a RCF detector from a target with micro-grooves imprinted on the rear side. The target is a $18~\um$ thick Al foil irradiated at $10^{19}~\mbox{W cm}^{-2}$.
Bottom frames: effect of electron removal by magnetic fields, showing that the proton beam emittance is not significantly affected. The images are for
$6.5~\mbox{MeV}$ protons and the target thickness is $40~\um$.
Reprinted figure with permission from 
\textcite{cowanPRL04}, Phys. Rev. Lett. \textbf{92}, 204801.
Copyright 2013 by the American Physical Society.}
\label{Fig_PRL_cowan_04}
\end{figure}

A modulation of the proton beam angular distribution can be introduced 
purposefully by nanostructuring the target surface.
A technique based on micro-machining shallow grooves on the rear surface of 
the target, introduced by \textcite{cowanPRL04},
has successfully been used 
in several experiments for diagnosing the emission properties of the beam
\citep{nuernbergRSI09}.
From these patterned targets, a periodic modulation of the beam angular 
envelope arises during TNSA due to the local perturbation of the target normal 
direction, which causes an initial beam microfocusing at the groove locations.
As the sheath expands, the local modulations are added over the global 
divergence of the beam \citep{ruhlPoP04} and are observable as a modulation of 
the proton dose on the detector 
(Fig.\ref{Fig_PRL_cowan_04}).
The modulations can be used as a spatial fiducial from which one can infer the 
dimensions of the area from where ions are accelerated, i.e. the proton or ion 
source size \citep{brambrinkPRL06}.
Similar information has been obtained by considerations based on the projection 
by the ion beam of patterned objects e.g. metal meshes \citep{borghesiPRL04}
or knife-edges \citep{schreiberAPB04}.

A crucial property of laser-driven ion beams is their laminarity.
In an ideal laminar source, there is a linear correlation between the radius 
within the source from where a particle is emitted and its angle of emission.
The degree of laminarity of charged-particle beams is typically expressed in 
terms of their transverse emittance, a quantity which is proportional to the 
area of the bounding ellipsoid of the distribution of particles in phase space
\citep{humphries-book}.
The highest quality ion beams have the lowest values of transverse and 
longitudinal emittance, indicating a low effective transverse ion temperature 
and a high degree of angle-space and time-energy correlation, respectively.
Transverse emittance has been measured in a number of experiments.
Methods based on mesh projection (which is broadly equivalent to the 
established ``pepper-pot'' method) indicate that
$\epsilon < 0.1 \pi~\mbox{mm mrad}$
\citep{borghesiPRL04,nishiuchiPoP08,ceccottiPPCF08}.
The above discussed groove imaging technique 
allows a full reconstruction of the transverse phase space, and possibly a more 
precise estimation of the {transverse} emittance
\citep{cowanPRL04,brambrinkPRL06,nuernbergRSI09}
{which,} for protons of up to $10~\mbox{MeV}$, 
has been estimated  as $0.004~\mbox{mm mrad}$, i.e. 100-fold better than 
typical RF accelerators and at a substantially higher ion current (kA range).

It has also been found that the removal of the co-moving electrons after 1~cm 
of the quasi-neutral beam expansion did not significantly increase 
the measured proton transverse emittance, as shown in Fig.\ref{Fig_PRL_cowan_04}
\citep{cowanPRL04}.
This last observation is important since, in order to take advantage of the 
exceptionally small proton beam emittance in future applications, e.g. to 
capture them into a post-accelerator, removal of the co-moving electrons 
without significantly perturbing the protons is crucial.

The ultra-low emittance stems from the extremely strong, transient acceleration that takes place from a cold, initially unperturbed surface and from the fact that during much of the acceleration the proton space charge is neutralized by the co-moving hot electrons.
Using the ion beam as a projection source, having a low-emittance beam is equivalent to projecting from a \emph{virtual} point-like source
located in front of the target, 
with much smaller transverse extent than the ion-emitting region on the target surface
\citep{borghesiPRL04,nuernbergRSI09}.
As will be discussed in Sec.\ref{sec:applications_radiography}, this property of laser-driven ion beams allows to implement point-projection radiography with high spatial resolution.

\subsection{TNSA modeling}
\label{sec:TNSA_scenario_modeling}
\label{sec:TNSA_modeling}

The experimental observations and the considerations summarized in 
Sec.\ref{sec:TNSA_scenario} suggest the following starting assumptions, 
leading to the formulation of a relatively simple system of equations which 
can be investigated analytically and numerically \cite{passoniPRE04}.
First of all, we assume an electrostatic approximation, so that the electric 
field ${\bf E}=-\nablav\phi$ where the potential $\phi$ satisfies Poisson's 
equation
\bea
\nabla^2\phi=4\pi e(n_e-\sum_jZ_jn_j) , \,
\label{eq:poisson}
\eea
with the sum running over each species of ions, having density $n_j$ and 
charge $Z_j$.
As a consequence of the laser-solid interaction, the electron density $n_e$ may 
be described as composed of at least two qualitatively distinct populations, 
which will be labeled {\it cold} and {\it hot}
in the following, having densities
$n_c$ and $n_h$ such that $n_e=n_c+n_h$.
In the simplest approach, thermal effects are neglected for the cold 
population, while $n_h$ is given by a one-temperature Boltzmann distribution,
\begin{equation}
\label{nch}
n_h=n_{0h} {\rm e}^{{e\phi}/{T_{h}}} \, .
\end{equation}
This expression can be a reasonable first approximation to account for the 
presence of the self-consistent sheath field and has  actually been used in 
many works on TNSA\footnote{See e.g. 
\textcite{passoniPRE04,moraPRL03,albrightPRL06,nishiuchiPLA06,robinsonPRL06}.} 
but, as discussed below, it can lead to serious problems when the main goal is 
the estimation of the maximum energy of the accelerated ions.
Alternatively, the electron dynamics can be included via either fluid or 
kinetic  equations.
It is most of times appropriate to
consider two different ion species, a light ({\it L}) and a heavy ({\it H}) 
population: in this way it is possible, e.g. to model the acceleration of light 
species present on the surfaces of a solid target made of heavy ions.

Depending on the description of the ion populations, two main categories of TNSA models, to be discussed in detail in Secs. \ref{sec:TNSA_scenario_QS} and \ref{sec:TNSA_scenario_fluid} respectively, may be identified as follows.
The first includes {\it static} models in which it is assumed that the light 
ions, or at least the most energetic ones, are accelerated in the early stage of the formation of the sheath, so that the latter may be assumed as stationary.
In these conditions, the effects of the light ions on the electrostatic potential are usually neglected, while the heavy ion population of the target is considered immobile. The aim is thus to provide the most accurate description of the sheath depending on assumptions on the hot electron distributions.
The second category includes {\it dynamic} models where the system is described as a neutral plasma in which the ions acquire kinetic energy in the course of the sheath evolution.
In several cases a unique ion component is considered.
This approach is strongly connected to the classic problem of plasma expansion in vacuum, first considered by \textcite{gurevichJETP66}.
In a cold fluid description, neglecting relativistic effects the ions
are described by the equations
\bea
{\frac{\partial{\bf u}_j}{\partial t}}
+{\bf u}_j\cdot\nablav{\bf u}_j
&=& - \frac{Z_j e}{m_j} \nablav\phi
\, ,
\label{f1dion2} \\
{\frac{\partial n_j}{\partial t}} +
\nablav\cdot\left ({n_j{\bf u}_j} \right ) &=& 0 \qquad (j=L,H)
\, ,
\label{f1Dion1}
\eea
where ${\bf u}_j={\bf u}_j({\bf r},t)$ is the fluid velocity.
If the ions are described kinetically, their Vlasov equation for the
phase space distribution $f_j=f_j({\bf r},{\bf v},t)$ is
\begin{equation}
\label{V1Dion}
\frac{\partial f_j}{\partial t}
+{\bf v}\cdot\nablav{f_j}
-\frac{Z_j e}{m_j}\nablav\phi\cdot\frac{\partial f_j}{\partial{\bf v}} = 0.
\end{equation}
Most of general studies of plasma expansion and related ion acceleration 
developed both before and after TNSA experiments\footnote{General studies of plasma expansion in vacuum include \textcite{gurevichJETP66,allenJPP70,widnerPF71,crowJPP75,pearlmanPRL78,denavitPF79,moraPF79,dorozhkinaPRL98}. Early papers focused on modeling ion acceleration in laser-produced plasmas include \textcite{pearlmanPRL78,wickensPRL78,truePF81,kishimotoPF83}. More recent works stimulated by the TNSA experiments include \textcite{moraPRL03,kovalevPRL03,moraPRE05,bettiPPCF05,ceccheriniLP06,peanoPRE07}.}
as well as more specific models of TNSA\footnote{A list of papers describing TNSA models mostly based on a static modeling include \textcite{passoniPRE04,passoniLPB04,schreiberPRL06,albrightPRL06,lontanoPoP06,robinsonPRL06,passoniPRL08}.}
so far proposed in the literature
can be considered as suitable simplifications of the previous equations, falling into one of the two above mentioned categories (or suitable combinations of them) and obtained adding further, physically motivated assumptions.
Most of these models assume a 1D geometry, consistently with the electrostatic approximation, and planar in most cases.
This latter assumption, when applied to TNSA modeling, requires the rear surface to be sufficiently flat and the electron cloud to be spatially uniform in the plane normal to the ion motion.

Notice that all the models proposed to describe TNSA are, to a large extent, phenomenological, i.e. they need as input parameters physical quantities which are not precisely known.
Since these descriptions give
a simplified picture of the acceleration process, the ``best'' model in this context may be considered the one which provides the best fit of experimental data with the lowest set of laser and target parameters.
This issue will be discussed in Sec.\ref{sec:TNSA_scenario_comparison}.
In principle, these difficulties could be overcome performing ``realistic'' numerical simulations, but actually also the latter always consider ``model'' problems because of intrinsic difficulties in the numerical study of these phenomena, such as for example the large variations of density from the solid target to the strongly rarefied expansion front.
At present, a complementary use of simple models, presented in Secs.\ref{sec:TNSA_QS}--\ref{sec:TNSA_multispecies}, and advanced simulations, discussed in Sec.\ref{sec:TNSA_PIC}, seems the most suitable option to theoretically approach TNSA.

\subsubsection{Quasi-static models}
\label{sec:TNSA_scenario_QS}
\label{sec:TNSA_QS}

Starting approximations of static models consist in assuming, on the time scale of interest (i.e. in the sub-ps regime), immobile heavy ions, an isothermal laser-produced hot electron population, and the light ions to be sufficiently few to neglect their effect on the evolution of the potential so that they can be treated as test particles.
In this limit, if Eq.(\ref{nch}) is used to describe hot electrons and neglecting thermal effects for cold electrons, the potential in planar geometry is determined by
\bea 
\label{PBQS}\label{PB1}
{\frac{\partial^2 \phi}{\partial x^2}}
&=& 4\pi e[n_{0h}{\rm e}^{{e\phi}/{T_{h}}}-(Z_Hn_{0H}-n_{0c})]
\nonumber \\
&=& 4\pi en_{0h}\left[{\rm e}^{{e\phi}/{T_{h}}}-\Theta(-x)\right] \, ,
\eea
where we assumed the background charge to fill the $x<0$ region with uniform
density.
The corresponding electron density and electric field
can be calculated, as well as the energies of test ions moving
in such potential.
This can be considered the simplest self-consistent approach to theoretically describe the TNSA accelerating field.
The solution of Eq.(\ref{PBQS}), in the semi-infinite region $x>0$ 
is \citep{crowJPP75}
\begin{equation}
\label{SolCrow}
\phi(x)=-\frac{2 T_h}{e} \left[ \ln\left(1+\frac{x}{\sqrt{2
\mathrm{e}}\lambda_{Dh}}\right) -1 \right] \, ,
\end{equation}
where $\lambda_{Dh}=\sqrt{T_h/(4\pi e^2n_{0h})}$.
The field reaches its maximum at the surface and is given by
\bea
E(0)=\sqrt{\frac{2}{\mbox{e}}}E_0 \, , \qquad E_0=\frac{T_h}{e \lambda_{Dh}},
\label{eq:E0}
\eea
which justifies the simple estimates used in Sec.\ref{sec:nutshell_acceleration}.
However, the electrostatic potential (\ref{SolCrow})
leads to an infinite acceleration of a test proton which is initially at zero
energy in $x = 0$.
The reason is that the apparently reasonable choice of the Boltzmann relation 
poses sever difficulties to the analysis \citep{passoniLPB04} because, 
in order to have an electron density equal to zero at infinity, 
the self-consistent electrostatic potential must diverge at large distance 
from the target [mathematically, $\phi \rightarrow -\infty$ as 
$x \rightarrow +\infty$, see Eq.(\ref{nch})].
This is not a pathological consequence of the one dimensional approximation
but it is related instead to the fact that the Boltzmann relation implies
the existence of particles with infinite kinetic energy, which is not physically meaningful [see also $\S$ 38 of \textcite{landauSP1}].
This unphysical behavior can be avoided by assuming an upper energy cut-off
${\cal E}_c$ in the electron distribution function, so that
$e\phi \rightarrow -{\cal E}_c$ as $x \rightarrow +\infty$ and the electric
field turns to zero at a finite distance. The cut-off assumption can be
justified as a consequence of the laser-solid interaction
producing electrons with a maximum kinetic energy
and of the escape from the system of the most energetic ones
\citep{lontanoPoP06,passoniPRL08}.
Experimental indications of target charging due to electron escape
have been found by, e.g. \textcite{karPRL08a,quinnPRL09}.
The finite range of the electric field driving TNSA
is also apparent in direct measurements
\citep{romagnaniPRL05}.

As a first development, still using the Boltzmann relation, 
it can be assumed the 1D solution given by Eq.(\ref{PBQS}) 
to hold only up to a longitudinal distance roughly equal to the transverse 
size of the sheath, because at larger distances 3D effects should be taken 
into account, contributing to remove the divergence 
\citep{nishiuchiPLA06}. 
Alternatively, by assuming that the hot electron population occupies 
only a finite region of width $h$, the solution of 
Eq.(\ref{V1Dion}) 
in the vacuum region $0<x<h$, together with the corresponding electric 
field and electron density can be determined 
\cite{passoniLPB04}.

Another possibility, explored by \textcite{schreiberPRL06} has been to heuristically assume that the hot electrons expansion in vacuum creates a cylindrical quasi-static cloud in the vacuum, behind the target, and a circular positive surface charge on its rear face.
The generated electrostatic potential is evaluated on the symmetry axis,
along which the most energetic ions are accelerated. 
The total surface charge and the radius of the distribution are model 
parameters estimated from experiments 
(see also Sec.~\ref{sec:TNSA_comparison}).

In order to consistently overcome the previously discussed limits,
\textcite{lontanoPoP06}
proposed to solve the Poisson equation by assuming that a quasi-stationary 
state is established where only those electrons ({\it trapped electrons}) 
with negative total energy $W=m c^2 (\gamma-1) - e \phi$ are retained, while 
those with positive total energy are lost from the system. 
The corresponding trapped electron density, given by 
$n_h=\int_{W<0} f_e(x,p) dp$, is included in the Poisson equation
and the corresponding analytical solutions determined 
\citep{lontanoPoP06,passoniPRL08,passoniNJP10}.
As a general feature, the potential, the electrostatic field and the 
electron density 
distributions go to zero at a finite position $x_f$ of the order of several 
hot Debye lengths. 

\begin{figure}[t]
\centering
\includegraphics[width=0.40\textwidth]{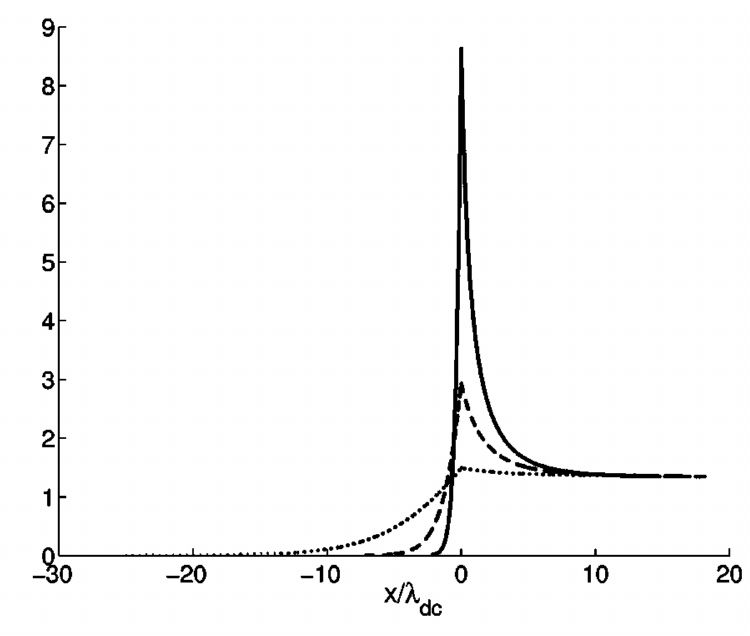}
\caption{Electric field profile in a sheath with two electron temperatures.
The field is normalized to ${T_h}/{e\lambda_{Dh}}$ and is shown
for cold-to-hot electron temperature
ratio $b=T_{c}/T_{h}=0.01$ and for different values of the pressure
ratio $ab=p_{0c}/p_{0h}=1$ (dotted line), $ab=10$ (dashed line)
and $ab=100$ (solid line). The $x$ coordinate is normalized to the
\textit{cold} electron Debye length $\lambda_{Dc}$ corresponding
to $ab=10$.
Reprinted figure with permission from \textcite{passoniPRE04}, Phys. Rev. E \textbf{69}, 026411.
Copyright 2013 by the American Physical Society.}
\label{fig2T}
\end{figure}

If both electron populations, hot and cold, are considered, it is possible to find an implicit analytical solution of Eq.(\ref{PB1}) both inside the target and in the vacuum region.
Using a two temperature Boltzmann relation to describe the electron density, that is $n_e=n_{0h}\exp \left( {e \phi}/{T_h}\right)+n_{0c}\exp \left({e \phi}/{T_c}\right)$, the electric field profile turns out to be governed by the parameters $a\equiv n_{0c}/n_{0h}$ and $b\equiv T_{c}/T_{h}$, as shown in Fig.\ref{fig2T} \citep{passoniPRE04}.
The presence of the cold electron population strongly affects the spatial profiles of the field, which drops almost exponentially inside the target over a few cold electron Debye lengths. An estimate of $T_c$, as determined by the ohmic heating produced by the return current (Sec.\ref{sec:nutshell_fastelectrons_transport}), is required. A simple analytical model of the process has been proposed \citep{daviesPRE03,passoniPRE04}, to which we refer for further details and results.

The quasi-static approach allows to draw several general properties of the 
accelerating TNSA field.
The spatial profile is characterized by very steep gradients, with the field 
peaking at the target surface and decaying typically over a few $\um$ distance.
The most energetic ions, accelerated in the region of maximum field, cross the 
sheath in a time shorter than the typical timescale for plasma expansion, 
electron cooling and sheath evolution.
As a consequence the static approximation will be more accurate for the faster 
ions.
Assuming a time-independent field also requires
the electron cloud not to be affected by the ions flowing through it, 
which implies the number of the accelerated ions to be much smaller than that 
of the hot electrons, $N_i \ll N_e$.
A quasi-static model not requiring this assumption was proposed by 
\textcite{albrightPRL06} who included effects of the accelerated ion charge 
on the electric field by modeling the layer of light ions 
(having areal charge density $Q_L$) as a surface layer of density 
$n_L=(Q_L/Z_Le)\delta(x-x_L)$.
Eq.(\ref{PB1}) is then solved as a function of the instantaneous position $x_L$.
An extension of this model, using an adiabatic descritpion of the hot electrons, was proposed by \textcite{andreevPRL08}, to investigate the variation of the maximum light ion energy as a function of the heavy ion target thickness.

On the basis of the above discussions we expect static models to be most 
reliable for the estimate the cut-off in the ion energy spectrum. 
This estimate requires as an input a few parameters, depending on the model. 
This issue will be discussed in Sec.\ref{sec:TNSA_comparison}.

\subsubsection{Plasma expansion into vacuum}
\label{sec:TNSA_scenario_fluid}
\label{sec:TNSA_fluid}
\label{sec:TNSA_expansion}

A description of ion acceleration over long times and/or in conditions such 
that the quasi-static modeling of Sec.\ref{sec:TNSA_QS} is not valid anymore 
demands for the inclusion of the ion dynamics. The description may be based 
either on a fluid modeling,
using Eqs.(\ref{f1dion2}--\ref{f1Dion1}), or on a kinetic one using 
Eq.(\ref{V1Dion}).

\begin{figure}
\includegraphics[width=0.48\textwidth]{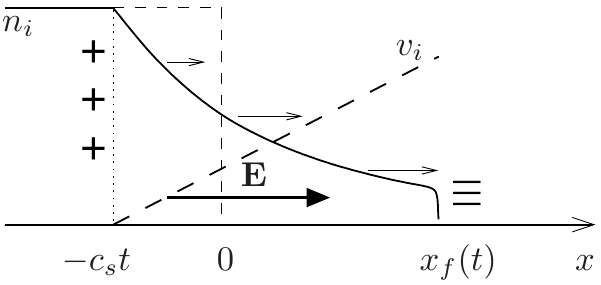}
\caption{Sketch of the density and velocity profiles from the self-similar
solution for isothermal plasma expansion, Eqs.(\ref{eq:nself}--\ref{eq:Eself}).
The front of charge separation at $x=x_f(t)$
and the rarefaction front at  $x=-c_st$ are also indicated.
The electric field is uniform in the $-c_st<x<x_f(t)$ region.}
\label{fig_plasmaexpansion_labels}
\end{figure}

The simplest approach is obtained using a 1D fluid approach, invoking 
quasi-neutrality, using Eq.(\ref{nch}) and assuming a single ion and 
electron population expanding in the  semi-infinite space $x>0$.
Eq.(\ref{PB1}) is substituted by the simpler
condition $n_e=Z_i n_i$, the index $i$ denoting the single ion component.
The boundary conditions are that the electron density should remain equal 
to the background
value well inside the plasma, so $n_e(-\infty) = n_0$, and should vanish in 
vacuum far from the surface, $n_e(+\infty) = 0$. 
Together with Eqs. (\ref{f1dion2},\ref{f1Dion1}), 
the resulting system admits the classical self-similar solution first found by
\textcite{gurevichJETP66}, 
\bea
n_i = n_0 \exp \left(-\frac{x}{c_st}-1\right) \, , \qquad
\label{eq:nself}
u_i = c_s+\frac{x}{t} \, ,
\label{eq:uself}
\eea
where $x/t$ is the self-similar variable, 
$L=n_i/|\partial_x n_i|=c_st$ is the local density scalelength, 
and the expressions are valid for $x>-c_st$.
The profiles corresponding to Eqs.(\ref{eq:nself}) 
are sketched in Fig.\ref{fig_plasmaexpansion_labels}.

As a consequence of the quasi-neutral approximation, the physical quantities 
describing the plasma dynamics present several diverging behaviors, like the 
unlimited increase of $u_i$ with $x$.
This implies that the neutral solution must become invalid at some point, which 
can be estimated by equating the local density scalelength $L$ to the local 
Debye length $\lambda_D$.
This provides $x_f(t)=c_st[2\ln(\omega_{pi}t)-1]$, the corresponding 
velocity $u_f=d x_f/dt=2c_s\ln(\omega_{pi}t)$ and the electric field at the ion 
front, $E_f=E(x_f)=2E_0/\omega_{pi}t$, where $E_0=({4\pi n_0 T_h})^{1/2}$.
This estimate gives twice the self-similar field $E={T_h}/{(ec_st)}$.
The argument also defines the front of the fastest ions moving at velocity 
$u_f$ and thus it gives also the high-energy cut-off in the energy spectrum 
of the ions in this description.

Eqs.(\ref{eq:nself})
are also singular for $t\rightarrow 0$, i.e. at the earliest instants of the 
expansion in which quasi-neutrality also breaks down. In general, in the sub-ps 
regime the inertia of ions is important and the assumption of quasi-neutrality 
must be consistently abandoned and ultimately a self-consistent analysis can be 
developed through numerical simulations (Sec.\ref{sec:TNSA_PIC}).
Still assuming, for the sake of simplicity, that only a single ion population 
and a single-temperature Boltzmann electron population are present, and 
$n_i(t=0)=n_0\Theta(-x)$,  Eq.(\ref{SolCrow}) can be used to define the 
initial conditions for the
electric field at the time $t=0$ at which the ion acceleration process begins. 
The following interpolation formulas for the electric field and ion velocity 
at the ion front
\bea
E(t) &\simeq& \sqrt{\frac{2}{\mbox{e}}}\frac{E_0}{\sqrt{\tau^2+1}}
\label{eq:moraint1} \, , \\
u_f(t) &\simeq& 2c_s\ln\left(\tau+\sqrt{\tau^2+1}\right) \, ,
\label{eq:moraint2}
\eea
where $\tau=\omega_{pi}t/\sqrt{2{\mbox e}}$,
give the correct behavior at $t = 0$ for both
the electric field [see Eq.(\ref{eq:E0})] and the front velocity,
and reduce to previous expressions for  $\omega_{pi} t\gg1$.
These formulas fit well numerical calculations by \textcite{moraPRL03}
using a Lagrangian fluid code. Related results of these studies using the fluid and the kinetic descriptions can be found in the literature.\footnote{See e.g.
\textcite{crowJPP75,widnerPF71,pearlmanPRL78,denavitPF79,moraPRL03} for the case of a single electron population and
\textcite{bychenkovPoP04,moraPRE05,tikhonchukPPCF05} for the case of
two electron components.}

The major drawback of Eq.(\ref{eq:moraint2}) is that the maximum velocity of ions, and hence the cut-off energy, diverges logarithmically with time.
This is not surprising, being an unavoidable consequence of the isothermal assumption and the chosen boundary conditions: the system has an infinite energy reservoir in the electron fluid and thus it is able to accelerate ions indefinitely.
Nevertheless, the simplicity of Eq. (\ref{eq:moraint2}) has proven to be attractive, thus it has been suggested to insert a phenomenological ``maximum acceleration time'' $t_{acc}$ at which the acceleration should stop.
Such a formula has been used in attempts to fit experimental data \citep{fuchsNP06}. We will come back to this point in Sec. \ref{sec:TNSA_scenario_comparison}.
There is no easier way to remove this unphysical behavior from the 1D planar 
model but to give a constraint of finite energy (per unit surface). In this way,
the electron temperature decays in time due to the plasma expansion and to 
collisional and radiative losses. The electron cooling cooperates with the 
effects of finite acceleration length and maximum electron energy in the 
determination of a finite value for the maximum energy gain.

The expansion of plasma slabs (foil) of finite thickness, and hence of finite 
energy, has been considered analytically and numerically. In these models 
The electron temperature is taken as a function of time, $T_h=T_h(t)$, 
determined either by the energy conservation equations 
\citep{moraPRE05,bettiPPCF05}, or with ad-hoc modeling of $T_h(t)$ 
\citep{bychenkovPoP04}.

Analytical solutions for the plasma expansion can be found, in the quasi-neutral approximation, also for the kinetic Vlasov equation (\ref{V1Dion}), using
either the self-similar theory \citep{dorozhkinaPRL98}
or a re-normalized group theoretical approach
\citep{kovalevJETPL01,kovalevJETP02}.
Two-temperature electron distributions have also been considered.\footnote{See e.g. \textcite{bezzeridesPF78,wickensPRL78,gurevichPRL79,truePF81,kovalevJETP02,diasPRE11}.}

\subsubsection{Multispecies expansion}
\label{sec:TNSA_multispecies}

We now describe the expansion of a two species plasma, in which 
the dynamics of a heavy ion component is considered in addition to 
light ions. The most peculiar effect of the presence of two (or more) ion 
species, for appropriate parameters, is the appearance of spectral peaks, 
which are of interest both as a strong experimental signature and for 
application purposes.

\begin{figure}[t]
\begin{center}
\includegraphics[width=0.48\textwidth]{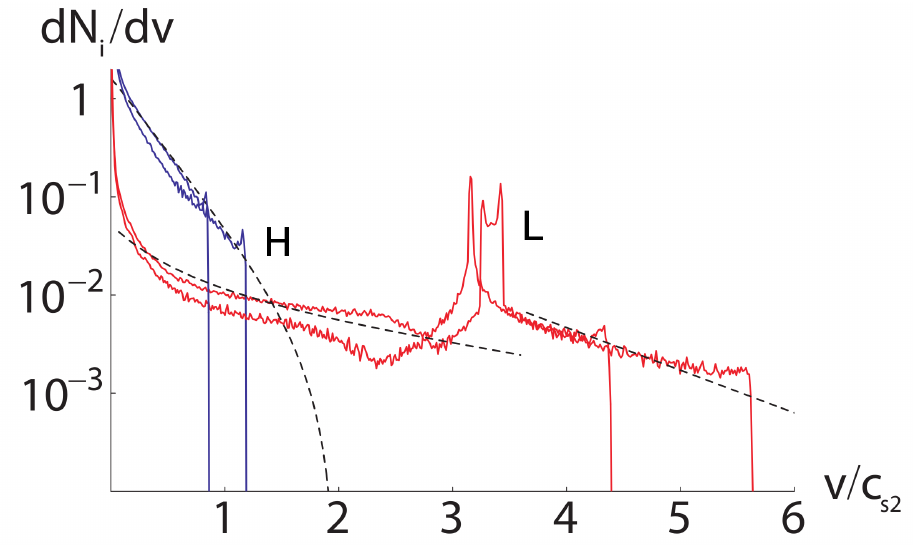}
\end{center}
\caption{(Color online) Velocity spectrum of heavy ions (H, blue thick lines)
and light ions (L, red thick lines) at two different times
from the numerical simulation of the expansion of a two-species plasma using
a Boltzmann-Vlasov-Poisson model. 
Black dashed lines are analytical profiles based on self-similar solutions.
The spectrum of light ions shows a peak typical of a multispecies expansion.
Reprinted figure from 
\textcite{tikhonchukPPCF05}, Plasma Phys. Contr. Fusion \textbf{47}, B869.
By permission from Institute of Physics Publishing (2013).
}
\label{Fig2_PPCF_tikhonchuk_05}
\end{figure}

The problem of two species expansion was studied by several authors since
longtime \citep{gurevichJETP73,bezzeridesPF78,srivastavaPF88}. Here we mostly
follow the more recent work of \textcite{tikhonchukPPCF05}, where a simplified
description is given based on the ordering assumptions
\bea
\alpha=\frac{A_H/Z_H}{A_L/Z_L} \gg 1 , \qquad 
N=\frac{Z_Hn_H}{Z_Ln_L} >\alpha .
\eea
These conditions state that the $H$ species is quite heavier than the $L$ one,
that the concentration of the latter is small, and that the $L$ ion plasma
frequency is higher so that the dynamics of L ions is faster. 
These assumptions allow to
assume that, near the rarefaction front, the effect of $L$ ions is unimportant
and that the dynamics of the $H$ ions can be described as a single species
expansion as in Sec.\ref{sec:TNSA_expansion}, where the
relevant parameter is the $H$ ion sound speed
$c_H=\sqrt{Z_HT_h/A_Hm_p}<c_L$, the $L$ ions sound speed. 
$L$ ions are treated as test particles in this region, where
they are accelerated by the electric field $E \propto c_H^{-1}$
[see Eq.(\ref{eq:Eself})], which is thus stronger than that would
be created in the expansion of the $L$ ions alone. The $L$ ion
velocity and density in this region can be obtained using the 1D fluid,
self-similar equations with the above given electric field
\citep{tikhonchukPPCF05}, obtaining for the velocity profile 
$v_L \simeq c_L\sqrt{2}\left(1+{x}/{c_Ht}\right)^{1/2}$.
Noticeably, the $L$ ions velocity and density profiles vary slowly in space
compared to the $H$ ion ones, and the $L$ ion flux is almost constant. Beyond
the $H$ ion front, only $L$ ions are present and they can be described again
by a single species expansion, $v_L \simeq c_L+x/t$ [see Eq.(\ref{eq:uself})].
However, matching of the
velocity profiles in the region behind the $H$ ion front implies the existence
of a transition region where the velocity is approximately constant. This
corresponds to a plateau region in the phase space and in a peak in the
$L$ ion energy spectrum. The heuristic reason for plateau formation is that the
$L$ ions are accelerated more efficiently behind the $H$ ion front than ahead
of it. Fig.\ref{Fig2_PPCF_tikhonchuk_05} shows the velocity spectrum from
numerical results \textcite{tikhonchukPPCF05}
using a Boltzmann-Vlasov-Poisson model \citep{bychenkovPoP04} based on
Eqs.(\ref{eq:poisson}--\ref{nch}--\ref{V1Dion}), compared
with analytical estimates from the self-similar solution.

According to the above model the peak energy of $L$ ions is
\bea
{\cal E}_L \simeq Z_LT_h\ln\left(4\sqrt{2\alpha}N/\mbox{e}\right) \, .
\eea
As an important indication from this model, the mass ratio and the relative
concentration of the two species might be engineered to optimize the $L$ ion
spectrum.
Several simulation studies\footnote{See e.g. \textcite{kempPoP05,brantovPoP06,robinsonPRL06,robinsonPRE07,psikalPoP08,robinsonPPCF09c,bradyPPCF11}.}
have been devoted to this issue and to the modeling of observations of
multispecies spectra in both planar and spherical (droplet) targets
(see Sec.\ref{sec:TNSA_optimization}).

\subsubsection{Numerical simulations}
\label{sec:TNSA_scenario_PIC}
\label{sec:TNSA_PIC}

Already in their simplest formulation TNSA models
are highly nonlinear and the set of
available analytical solutions is limited. A numerical approach can be used
to overcome these limitations and to address
additional effects.

Referring to the 1D problem of plasma expansion, an hydrodynamic two-fluid
approach may be used to take charge separation effects into account as reported 
by \textcite{moraPRL03}. The hydrodynamic model, however, cannot take into 
account kinetic effects such as non-Maxwell distribution and breakdown of 
equilibrium
conditions. To address these latter effects a numerical solution of the
Vlasov equation for the distribution function of electron and ions in phase
space is needed. 
{To this aim the PIC approach (Sec.\ref{sec:nutshell_simulations}) 
may be used.}
The drawback is the much larger computational
request with respect to hydrodynamics simulations. The reason is that to obtain full numerical convergence and accurate, low-noise
results a very large number of particles should be used to resolve the
strong density variations in the plasma expansion.

\begin{figure}[t]
\begin{center}
\includegraphics[width=0.48\textwidth]{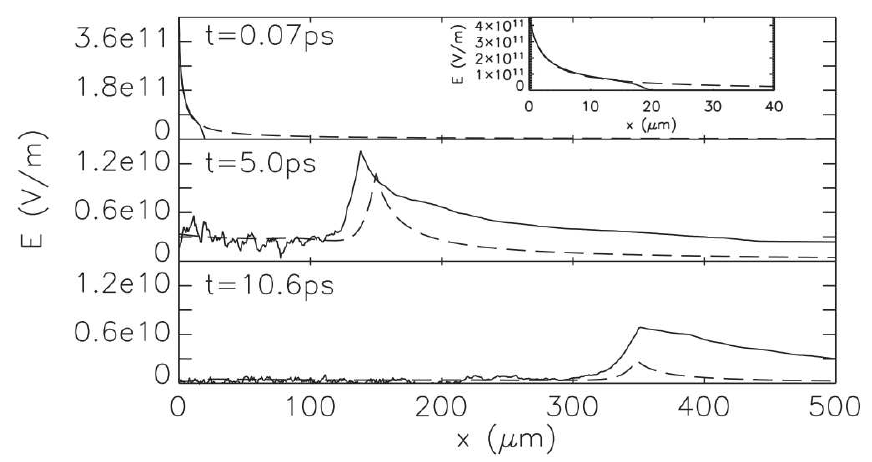}
\caption{Electric field profiles at different times from the
numerical simulation of the collisionless expansion of a slab of
warm plasma. Thick and dashed lines show results from a PIC code
\citep{bettiPPCF05} and a hydrodynamics code \citep{moraPRL03},
respectively. Both simulations assume a $40~\um$ thick proton plasma slab with
initial density $n_0=3 \times 10^{19}~\mbox{cm}^{-3}$ and electron temperature
$T_{e0}=500~\mbox{KeV}$. The inset shows the detail of the field distributions
at early times, with the field in the PIC simulation extending over a finite
distance. 
Reprinted figure with permission from 
\textcite{romagnaniPRL05}, Phys. Rev. Lett. \textbf{95}, 195001.
Copyright 2013 by the American Physical Society.
}
\label{Fig2_PRL_romagnani_05}
\end{center}
\end{figure}

In its simplest formulation the 1D simulation of collisionless plasma expansion
takes a single ion species into account and a limited set of parameters, such
as the initial electron temperature and the initial thickness of the plasma;
this corresponds to fix the total energy of the system. Such
simplified simulations already reproduce qualitative features observed in the
experiment and may match measured quantities such as the ion front velocity with
a proper choice of initial parameters.
As an example Fig.\ref{Fig2_PRL_romagnani_05} shows simulation results
performed to support experimental observations by \textcite{romagnaniPRL05},
using both an hydrodynamics and a PIC code. The two approaches use different
initial conditions, i.e. a Boltzmann equilibrium for fixed ions and a
zero charge density distribution, respectively. The latter condition enables
to resolve in the PIC calculation the propagation of the electron front,
resulting in the electric field vanishing at the front position and
showing a strong temporal maximum at the earliest instants, in agreement with
experimental observations.

The use of supercomputers allows to perform multi-dimensional PIC simulations
and to simulate the laser-plasma interaction and the generation of hot
electrons, rather than imposing a priori their number and temperature.
{The computational challenges and limitations of such large-scale 
simulations have been discussed in Sec.\ref{sec:nutshell_simulations}. In 
addition, most PIC simulations do not include collisions, which may play an 
important role in the transport of hot electrons through the target 
(Sec.\ref{sec:nutshell_fastelectrons_transport}).}
Nevertheless, PIC simulations have been vastly used as a valuable support in 
the interpretation of measurements of ion acceleration and 
were able to reproduce at least qualitatively several observed features
of the TNSA picture, see e.g. \textcite{wilksPoP01,pukhovPRL01,fuchsPRL05}.

As an alternative to the PIC method, \textcite{gibbonPoP04} used a gridless,
electrostatic ``tree'' particle code to simulate ion acceleration from wire
targets.
Such a code has the advantages of an unlimited spatial region for
particles and of ``automatic'' inclusion of collisions, at the cost of
being purely electrostatic so that the laser-plasma interaction may be
modeled only phenomenologically and magnetic field generation is not included.

\subsection{Comparison between models and experiments}
\label{sec:TNSA_modelvsexperiment}
\label{sec:TNSA_scenario_comparison}
\label{sec:TNSA_comparison}

TNSA has been deeply investigated in a very large number of experiments, performed, in the past decade, in many laser facilities all over the world.
The maximum observed value of the ion energy ${\cal E}_{\mathrm{max}}$
has been probably the 
most characterizing parameter of such experiments.
Another important feature, mostly in the light of potential applications, is 
represented by the form of the energy spectrum.

All this effort resulted in an extensive collection of experimental data, 
against which the predicting capability of the TNSA theoretical models can be 
tested. Moreover, a new generation of laser facilities will be soon available, 
and it will be possible to investigate a wider range of experimental parameters.
Therefore the challenge of satisfactorily predicting the result of a TNSA 
experiment, providing sufficiently reliable scaling laws to extrapolate 
guidelines for the future experiments, is even more important.

Experimentally, great effort has been put in properly addressing the 
correlation among the above mentioned ion properties and the main laser and 
target parameters.
Due to the importance of the laser irradiance in establishing the regime of 
interaction (see Sec.\ref{sec:nutshell}), in the literature
it has become common to report the maximum ion (mainly proton) energy 
${\cal E}_{\mathrm{max}}$ as a function of this parameter 
(see e.g. Fig.\ref{Fig1_PPCF_borghesi_08}).
Collections of experimental data for ${\cal E}_{\mathrm{max}}$ 
have been reported in several papers\footnote{See e.g. \textcite{krushelnickPPCF05,borghesiFST06,borghesiPPCF08,fuchsNP06,robsonNP07,peregoNIMA11}.}
On the other hand, it is apparent that the irradiance is not the only laser
parameter playing a role in determining ${\cal E}_{\mathrm{max}}$.
In particular, 
it has been established by many experimental data that, for fixed irradiance, 
more energetic pulses lead to higher ${\cal E}_{\mathrm{max}}$. 
Moreover, already from the early experiments the strong influence 
both of the laser prepulse level and duration and of 
the target properties on ${\cal E}_{\mathrm{max}}$
has been evident.
We will devote Sec.\ref{sec:TNSA_target} to a dedicated discussion on these 
topics, while here we will mainly point out the general difficulties which can 
be encountered in the attempt of providing predictions of 
${\cal E}_{\mathrm{max}}$
for comparison with experimental data.

For all the models introduced in Sec.\ref{sec:TNSA_modeling},
${\cal E}_{\mathrm{max}}$ can be evaluated once the required parameters of the
physical system are known. This is actually a very delicate and often
controversial issue because the input parameters are different in number, 
nature and reliability.
Some models use laser and target parameters characterizing the experimental
set-up, 
which are then known or controlled with well-defined precision, like e.g. 
the mean irradiance, spot radius, energy and duration of the laser pulse, and 
the thickness, chemical composition and impurity proton surface density of the  
target.
Other models use as parameters physical quantities determined by
interaction and transport processes,
in particular hot electron properties (see Sec.\ref{subsec:numerics})
such as conversion efficiency $\eta_h$, temperature $T_{h}$,
density $n_{h}$, cut-off energy
and beam divergence angle $\theta_{\mathrm{div}}$.
These quantities may in principle be modeled and/or measured but most times are not precisely known.
Finally, some models include purely phenomenological parameters
such as the ion acceleration time $t_{\mathrm{acc}}$
(Sec. \ref{sec:TNSA_expansion}) and numerical parameters determined by
fitting on some set of experimental data or numerical simulations.
The experimental and theoretical uncertainities and the basically different
nature between model parameters inevitably impose some limitations to
the conclusions that one could draw on the basis of a quantitative comparison.

We will briefly touch this problem following the work by 
\textcite{peregoNIMA11}.
The descriptions which have been selected there are the fluid expansion models 
proposed by \textcite{moraPRL03,moraPRE05}, the quasi-static approaches of 
\textcite{schreiberPRL06} and \textcite{passoniPRL08} and the ``hybrid'' 
descriptions published by \textcite{albrightPRL06} and \textcite{robinsonPRL06} 
(see Sec.~\ref{sec:TNSA_modeling}). 
The calculations used to implement these models and evaluate 
${\cal E}_{\mathrm{max}}$ are also summarized. To perform the comparison a 
database containing an extensive collection of published experimental 
parameters and results, referring to a wide range of laser and 
target parameters, has been considered. 
This analysis shows that, despite all the uncertainties, the predictions of 
the TNSA models can be considered quite good, and in some cases remarkable, 
for a wide range of experimental parameters.
In particular, quasi-static models, especially the one proposed by 
\textcite{passoniPRL08}, are more suitable for the prediction of 
${\cal E}_{\mathrm{max}}$.
These conclusions are strongly affected by the estimates of the required 
parameters, and a more realistic approach to evaluate these quantities could 
improve the predicting capability of both expansion and hybrid models.

\begin{figure}[b]
\begin{center}
\includegraphics[width=0.48\textwidth]{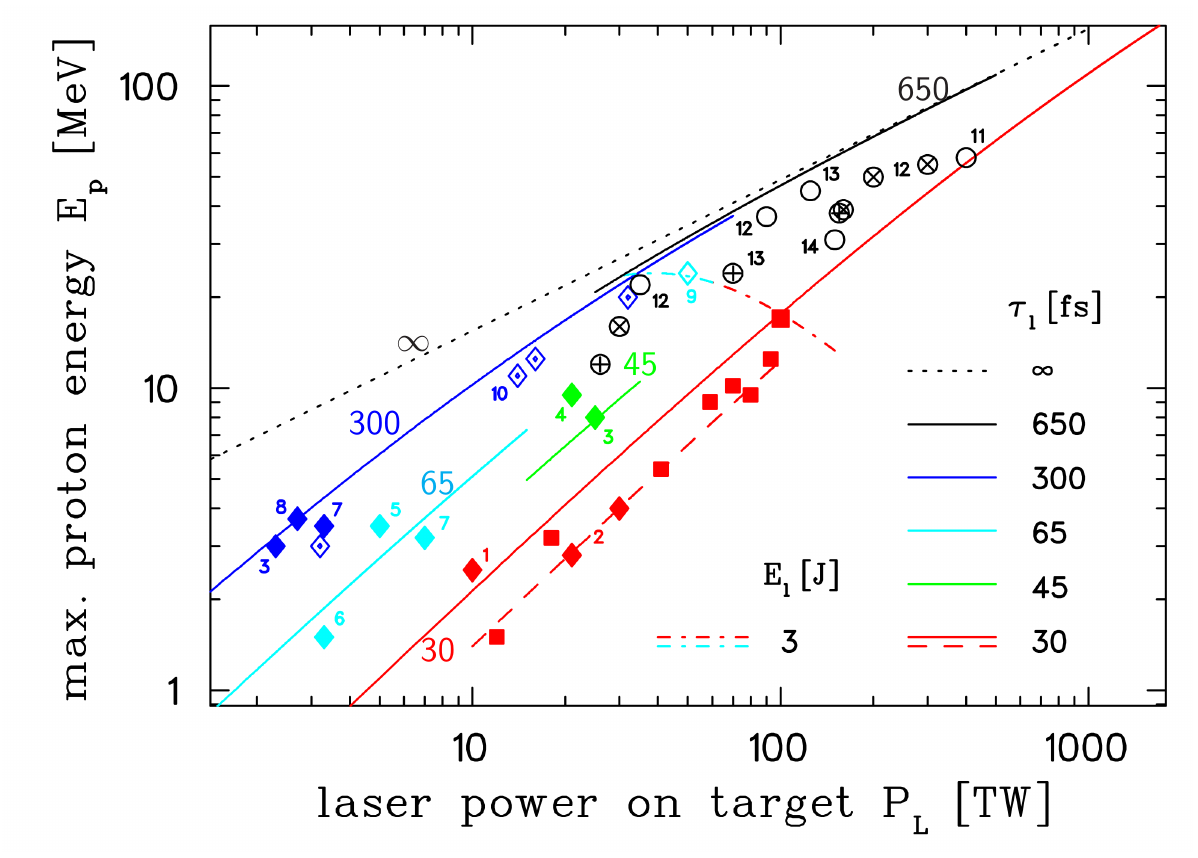}
\caption{(Color online)
Experimental scaling of proton energy cut-off with laser power and
pulse duration. Red square points 
are data from experiments performed with the DRACO
laser at FZD (Dresden),
showing a linear scaling with power in the short pulse ($30~\mbox{fs}$)
regime. Other points are data from other laboratories, see
\textcite{zeilNJP10} for references and details.
The fitting lines correspond to the
static model by \textcite{schreiberPRL06} with different colors (labels) 
corresponding to different values of the pulse duration $\tau_1$ as given in
the legenda.
Reprinted figure from \textcite{zeilNJP10}, New J. Phys. \textbf{12}, 045015.
By permission from Institute of Physics Publishing (2013).
}
\label{Fig4_NJP_zeil_10}
\end{center}
\end{figure}

The possibility to compare model predictions with experimental parametric 
studies with well defined and controlled laser conditions, aiming at 
providing reliable and clear scaling laws, can 
enhance significantly the effectiveness of the analysis.
Fig.\ref{Fig4_NJP_zeil_10} shows results
from a parametric study of the dependence of ${\cal E}_{\mathrm{max}}$
on laser power and duration \citep{zeilNJP10}. Several other 
parametric investigations of the dependence of ${\cal E}_{\mathrm{max}}$ on 
laser pulse irradiance, duration, energy and fluence have been reported 
\citep{robsonNP07,flaccoPRE10,fuchsNP06,nayukiJAP06,flippoRSI08}
as well as attempts in interpreting part of these findings
\citep{passoniAIP09,passoniNIMA10,zaniNIMA11}.

\subsection{Experimental optimization}
\label{sec:TNSA_optimization}
\label{sec:TNSA_optimization2}
\label{sec:TNSA_target}

After the first years of research, the combined vigorous development both in
laser technology and advanced target manufacturing allowed the investigation
of TNSA exploring a continuously increasing range of laser and target
parameters. Actually, in most cases the two sets are intimately related. For
example,
the use of ``extreme'' geometrical target properties, like thicknesses in the
sub-micrometric range, requires the availability of extraordinarily ``clean'',
prepulse-free pulses to avoid early target evaporation and deformation.
Such pulses can be obtained with recently developed techniques, like the
double plasma mirror
\citep[and references therein]{dromeyRSI04,fuchsJPIV06,thauryNP07},
Optical Parametric Amplification \citep{shahOL09}
 or Crossed Polarized Wave (XPW) generation
\citep[and references therein]{jullienOL05,zaouterOL11}.

\subsubsection{Energy cut-off enhancement}

\label{sec:TNSA_optimization_energy}

\textcite{mackinnonPRL02} studied the dependence of the ion acceleration on
the target thickness, with the aim of addressing the role played by the
electron temporal dynamics and its effect on the formation of the sheath
accelerating electric field.
The experimental results showed an increase in the peak proton energy
from 6.5 to 24~MeV when the thickness of the Al foil target was
decreased from 100 to 3 $\mu$m. These data clearly indicate that an increase
in the target thickness imply a lower mean density of the hot electrons at the
surface and a consequent lowering of the peak proton energy.

The influence of the laser prepulse due to amplified spontaneous emission
(ASE) on the acceleration of protons in thin-foil experiments has been
investigated in detail by \textcite{kaluzaPRL04}.
In this experiment Al foils of different thickness (from 0.75 to 86 $\mu$m)
were used in connection with the possibility of changing the duration of the
ASE prepulse. The results characterized
an optimal value for the target thickness,
strongly depending on the prepulse duration, at which the TNSA process leads to
the highest proton energies. For the thinner targets, a prepulse-induced plasma
formation at the rear side was able to effectively suppress TNSA, in agreement
with the considerations developed in Secs.\ref{sec:nutshell_acceleration} and
\ref{sec:TNSA_scenario}. Related experimental work, where a wide range of laser
parameters and different target materials have been considered, can be found
in the literature \citep{spencerPRE03,fuchsNP06},

\begin{figure}[t]
\begin{center}
\includegraphics[width=0.40\textwidth]{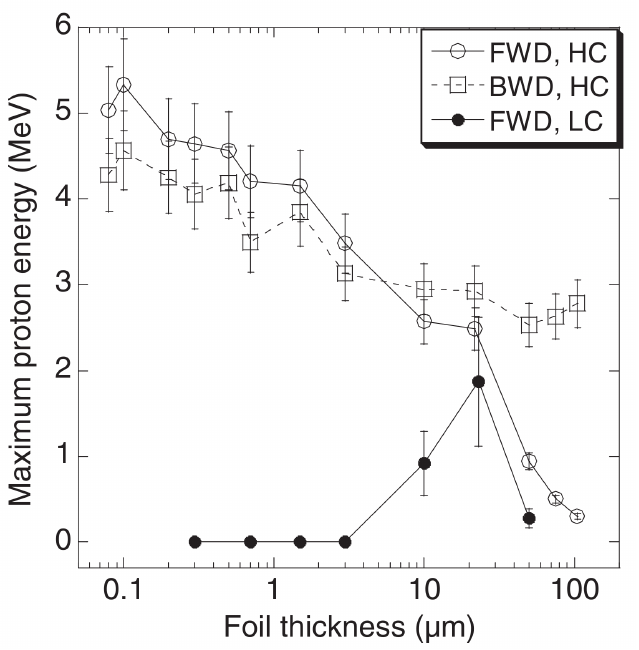}
\caption{Maximum detectable proton energy as a
function of target thickness for high-contrast (HC) and low-contrast (LC)
conditions.
Data are shown for both backward (BWD) and forward (FWD) directed ions,
respectively, showing the symmetrical behavior of TNSA for HC and ultrathin
targets.
{The LC results show the existence of an ``optimal'' thickness determined by the laser prepulse causing early target distruption, similarly to \textcite{kaluzaPRL04}.}
The laser pulse had $65~\mbox{fs}$ duration,
$(0.5\div 1)\times 10^{19}~\mbox{W cm}^{-2}$ intensity,
$45^{\circ}$ incidence and $P$-polarization.
Reprinted figure with permission from \textcite{ceccottiPRL07}, 
Phys. Rev. Lett. \textbf{99}, 185002.
Copyright 2013 by the American Physical Society.}
\label{Fig1_PRL_ceccotti_07}
\end{center}
\end{figure}

Effective suppression of the laser prepulse level, that is the adoption
of ultrahigh laser contrast can significantly alter the physical picture,
since ultrathin targets, down to the sub-$\mu$m level, can keep their
integrity until the interaction with the main pulse.
In these conditions a more effective acceleration process can be expected
because the refluxing and concentration of hot electrons in a smaller volume
may lead to the establishment of a stronger electric field and, consequently,
to higher ion energies.
These ideas have been successfully tested by \textcite{neelyAPL06}, where
Al target with thicknesses as low as $20$ nm have been used in combination
with 33~fs pulses having ASE intensity contrast reaching 10$^{10}$.
A significant increase of both maximum proton energy and laser-to-proton energy conversion efficiency was found at an optimum thickness of $100$ nm. Similar results have been obtained by \textcite{anticiPoP07} and \textcite{ceccottiPRL07}.
As a further interesting feature of this latter experiment, a symmetrical TNSA on both front and rear sides has been demonstrated, as shown in Fig.\ref{Fig1_PRL_ceccotti_07}, when a sufficiently high ($>10^{10}$) laser contrast is used.
This result confirms the universality of the TNSA process, which may occur also at the front side (accelerating ions in the backward direction) if the density profile is sharp enough.
Very recently, using a laser pulse with similar contrast,
40~fs duration, $10^{21}~\mbox{W cm}^{-2}$ and irradiating targets of 800~nm
thickness, \textcite{oguraOL12} reported proton energies up to $40~\mbox{MeV}$,
the highest value reported so far for pulse energies below 10~J.

\begin{figure}[b]
\begin{center}
\includegraphics[width=0.40\textwidth]{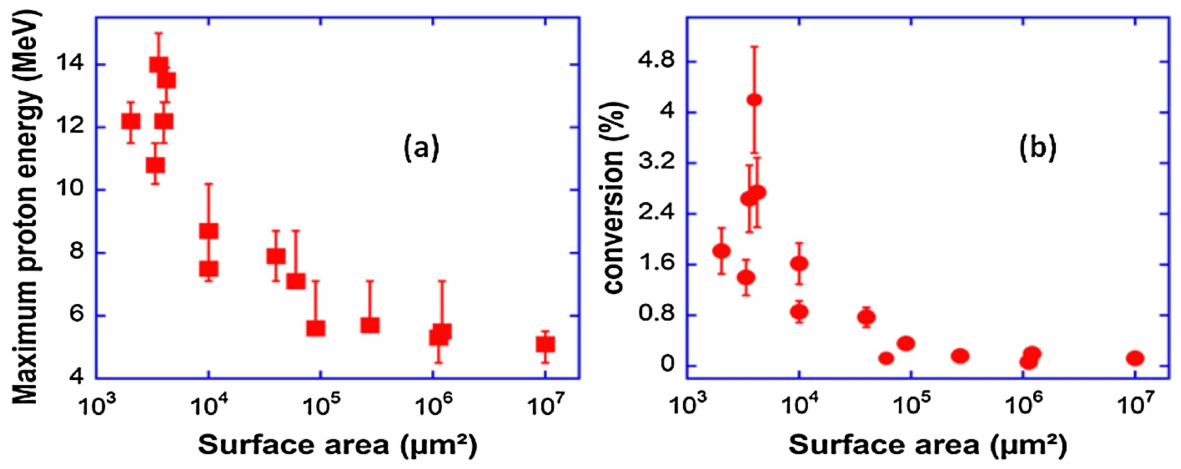}
\caption{(Color online) Experimentally observed (a) cut-off proton energies and (b) conversion efficiency (for $>1.5~\mbox{MeV}$ protons)
for $2~\um$ thick Au targets as a function of surface area,
evidencing the effect of electron refluxing.
The laser pulse had $400~\mbox{fs}$ duration,
$2\times 10^{19}~\mbox{W cm}^{-2}$ intensity, $45^{\circ}$ incidence and
$P$-polarization.
Reprinted figure with permission from 
\textcite{buffechouxPRL10}, Phys. Rev. Lett. \textbf{105}, 015005.
Copyright 2013 by the American Physical Society.}
\label{Fig3_PRL_buffechoux_10}
\end{center}
\end{figure}

Another possible strategy to exploit the effectiveness in the formation of the accelerating field in mass-limited targets is to reduce the lateral dimensions. Numerical investigations \citep{psikalPoP08} have actually shown that a reduced surface leads to higher densities of hot electrons at the rear side of target and, thus, to higher accelerating electric fields.
\textcite{buffechouxPRL10} experimentally confirmed these findings showing that in targets having limited transverse extent, down to tens of $\mu$m, the laser-generated hot electrons moving with a component of the velocity along the lateral direction can be reflected from the target edges during time scales of the same order of the acceleration of the most energetic ions.
This transverse refluxing can result in a hotter, denser and more homogeneous electron sheath at the target-vacuum interface.
A significant increase in the maximum proton energy (up to threefold), as well as increased laser-to-ion conversion efficiency (up to a factor 30), can be obtained in these conditions, as shown in Fig.\ref{Fig3_PRL_buffechoux_10}. Similar results, obtained with different laser and target parameters, have been found by \textcite{trescaPPCF11}, who also measured an increase in the maximum energy of protons accelerated from the edges of the target with decreasing target area.

Several other attempts have been made to increase either the energy density of the hot electrons in the sheath and, consequently, the maximum proton energy.
Following from the indications of \textcite{kaluzaPRL04},
\textcite{mckennaLPB08} have investigated whether there exists an optimum
density profile at the front of the target which maximizes the laser
absorption.
The proton cutoff energy was increased by 25\% with respect to a sharp
interface case at ``intermediate'' scale length (tens of $\um$s). In such
conditions, the higher conversion efficiency into fast electrons was
attributed to the self-focusing of the driver pulse.
Other studies of controlled prepulse effects on ion acceleration have been
reported by \textcite{flaccoJAP08,bataniNJP10}.

Recently, an energy cut-off
increase up to 67.5 MeV, 35\% higher than for flat foil shots,
has been demonstrated by \textcite{gaillardPoP11} using specially
devised targets, namely flat-top hollow microcone targets \citep{flippoPoP08},
which are a modification of conical targets used in Fast Ignition experiments
(Sec.\ref{sec:APP_fastignition}).
The laser pulse is focused inside the target, and starts interacting with the
walls of the cone that it grazes while focusing down towards the flat top
section. The remarkable reported result,
obtained with 80~J of laser energy on the Trident laser at LANL, is attributed to {an efficient} mechanism of electron acceleration taking place on the inner cone walls, named ``direct laser-light-pressure acceleration''.
The resulting increase in number of high energy electrons results in the increase of the maximum proton energy.

The use of targets with various structure
has also been investigated with the particular aim
to increase the ion energy already at relatively low laser intensities
(below $10^{18}~\mbox{W cm}^{-2}$), using e.g. double layer targets
\citep{badziakPRL01} and more recently nanowire-covered targets
\citep{ziglerPRL11} for which surprisingly high energies up to 5.5--7.5~MeV
for a $5 \times 10^{17}~\mbox{W cm}^{-2}$, 40~fs laser pulse has been reported.

\subsubsection{Source spectrum manipulation}
\label{sec:TNSA_optimization_source}

Various approaches have been proposed in order to manipulate the spectrum of
TNSA protons and ions, in most cases with the intent to obtain narrow band
peaks but also with the aim to enhance
proton numbers throughout the whole spectrum or in some spectral bands, as
required by specific applications.
We will firstly review a number of approaches in which the proton spectrum is
modified at the source, leaving to Sec.\ref{sec:TNSA_optimization_control}
approaches which act on the proton beam post-acceleration.

Spectral peaks can appear as a consequence of multispecies plasma expansion
{(Sec.\ref{sec:TNSA_multispecies})}.
This effect has been invoked to explain {observations} in proton beams
from thin foils, where the peaks appear as modulation of a continuum
exponential spectrum \citep{allenPoP03} and in experiments employing droplets
of heavy water, where peaks are observed in the
deuterium spectrum \citep{ter-avetisyanPRL06}.
Spectral peaks have been observed in experiments employing high-$Z$
metallic targets where a plastic layer ($0.5~\um$ PMMA) was coated as a dot on
the rear surface of a $5~\um$ Ti foil \citep{schwoererN06,pfotenhauerNJP08}.
These results, obtained on the 10~TW JETI laser in Jena, were explained on the
basis of the proton depletion approach first suggested by
\textcite{esirkepovPRL02}.
\textcite{robinsonPRE07}
suggested instead that the proton density in the multispecies plastic layer
is the important factor in determining the appearance of the spectral peak.
Experimental implementation
{required} the removal of the native contaminant layer present at the
surface {and} resulted in peaks in the proton spectra at
$\sim 2~\mbox{MeV}$, with $\sim 10\%$ spread and good reproducibility
\citep{pfotenhauerNJP08}.

\begin{figure}[t]
\begin{center}
\includegraphics[width=0.48\textwidth]{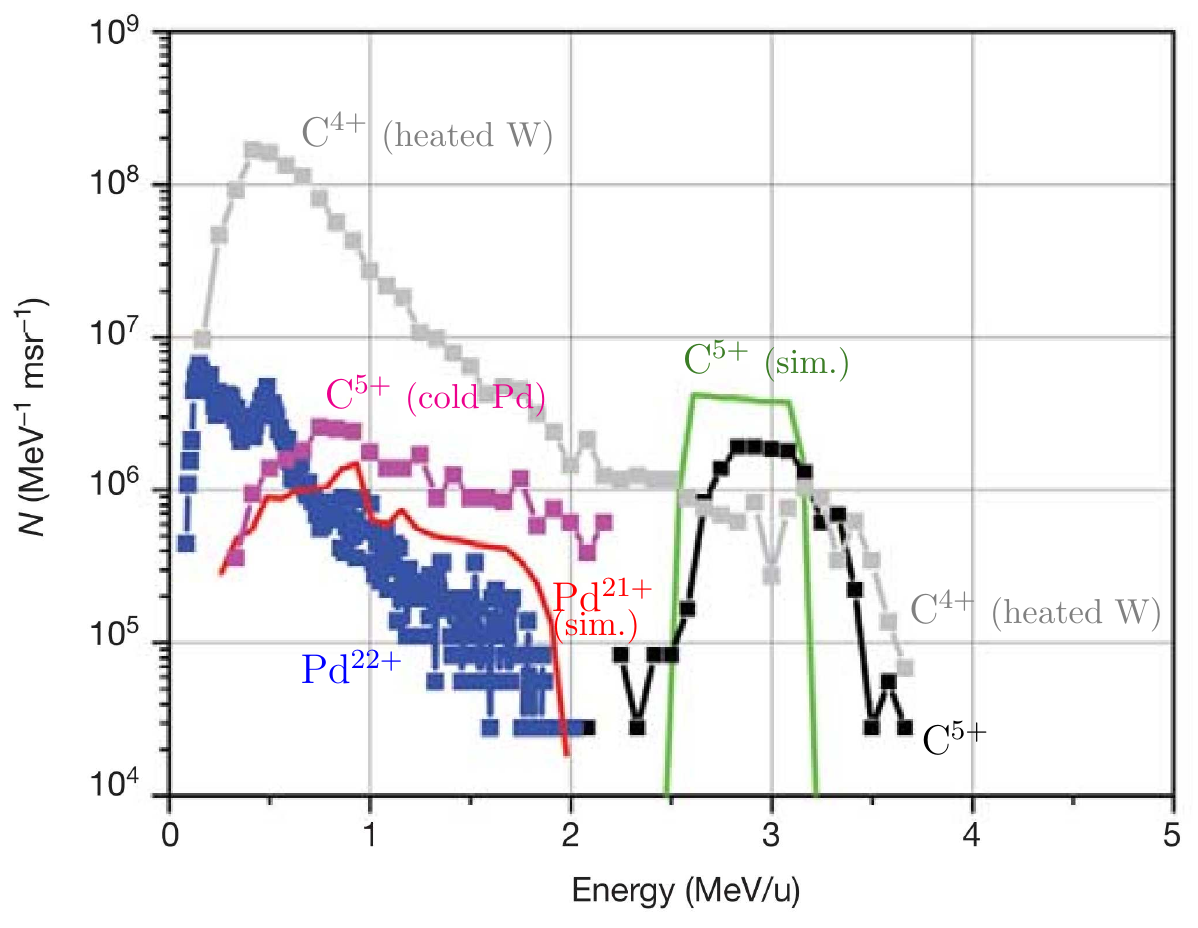}
\caption{(Color online)
Ion spectra from pre-heated Pd substrate targets
from which hydrogen contaminants have been removed \cite{hegelichN06}.
Black curve: spectrum of C$^{5+}$ ions. Blue curve: spectrum of the dominant
substrate charge state Pd$^{22+}$. Green and red curves: simulated C$^{5+}$ and
 Pd$^{21+}$ spectra.
Grey curve: spectrum of dominant C$^{4+}$ ions from a heated W
target. Magenta curve: C$^{5+}$ signal from a cold Pd target.
In the cases of black and blue curves, an ultrathin layer of Graphite is
present on the target surface, and a quasi-monoenergetic spectrum appears.
In the last two cases (grey and magenta curves) the targets have a thick layer
of carbon contaminants and do not form a monolayer source, resulting in
exponential-like spectra.
Reprinted figure by permission from Macmillan Publishers Ltd:
\cite{hegelichN06}, Nature \textbf{439}, 441. Copyright 2006.}
\label{fig_N_hegelich_06}
\end{center}
\end{figure}

{Another} experiment
also relied on the (partial) removal of {Hydrogen} contaminants from the
surface of  a high–$Z$ {Palladium} target \citep{hegelichN06}
{so that protons} did not appear in the spectrum. Instead,
monoenergetic features appeared in the C spectrum (specifically C$^{5+}$),
suggesting that an ultrathin layer of Graphite is formed as a result of phase
changes of the Carbon compounds in the contaminant
{and that}  all C$^{5+}$ ions {from the layer} experience approximately
the same accelerating field,
{as theoretically predicted \cite{esirkepovPRL02,albrightPRL06}}.
Fig.\ref{fig_N_hegelich_06} shows spectra for
{targets with and without contaminant removal},
together with hybrid simulation predictions.

\subsubsection{Beam post-acceleration and control}
\label{sec:TNSA_optimization_control}

Staged acceleration
employing two laser pulses on two separate targets has also been investigated as a possible route to spectral manipulation of laser-driven ion beams.
This idea relies on accelerating a TNSA beam from a first target, and direct it through a second foil, which is irradiated by a second laser pulse at the time that a particular group of TNSA protons crosses the foil.
These protons should thus experience an accelerating field as they transit through the rear surface of the second foil and gain additional energy.
An experiment by \textcite{pfotenhauerNJP10},
also carried out on the JETI laser, has tested this idea.
Peaks and dips in the spectrum were observed  at energies of $\sim 1~\mbox{MeV}$ which correlated well with the time of flight of protons reaching the second target as it is irradiated, showing that the field on the second target slightly boosts protons in a given energy range resulting in the spectral modification.
\textcite{burzaNJP11} reported
a two stages approach employing  spherical shell targets, irradiated by a single laser pulse, in which protons accelerated by TNSA at the front of the shell experience a second accelerating field while they transit through the opposite side of the shell, which modifies the high energy end of the spectrum.
The field is due to a hot electron charge wave spreading along the target surface from the interaction point, as reported in several experiments
\citep{quinnPRL09,toncianS06,mckennaPRL07}.

A different type of two stage approach was tested by \textcite{markeyPRL10}
on the VULCAN laser.
Two pulses of sub-ps duration
were sequentially focused, with controllable delay, on the same target in order to modify the temporal history of the hot electron source driving the TNSA, as suggested originally by \textcite{robinsonPPCF07}.
An optimal delay was seen to result in an increase of energy and conversion efficiency and, additionally, a modification of the slope of the spectral profile.
In this case, besides an optimization of hot electron production by the main pulse in a front-surface plasma gradient, similar to \textcite{kaluzaPRL04,mckennaLPB08},
the authors suggest that an additional modification of the proton spectrum arises from the fact  that proton acceleration by the main pulse takes place in  an already expanding multispecies, plasma sheath at the target rear surface.
Under these conditions, the electrostatic field peaks at the front separating protons from heavier species, and re-accelerates mainly the lower energy part of the spectrum.
Similarly, in a recent experiment, \textcite{dollarPRL11} have obtained spectral modifications, resulting in the appearance of narrow band spectral peaks at $\sim 2-3~\mbox{MeV}$ energies, by focusing a prepulse
($10^{-5}$ of the $10^{21}~\mbox{W cm}^{-2}$ peak intensity) on ultrathin foils a few tens of ps before the peak of the main pulse.

\begin{figure*}
\includegraphics[width=0.98\textwidth]{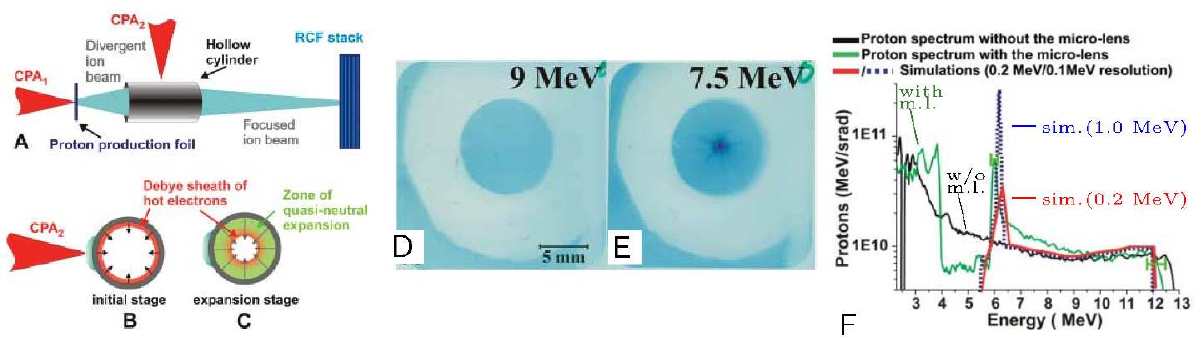}
\caption{(Color online)
A-C) Schematic of laser-driven electrostatic lens. D-E) RCF stack beam profiles for protons of 9 and 7.5 MeV, respectively, showing that the 7.5 MeV protons are focused by the fields inside the cylinder and form a black spot on the RCF. F): Proton spectra. Green line: spectrum obtained under same triggering conditions as in E). Black line: typical exponential spectrum obtained when cylinder is not triggered.
Reprinted figure from \textcite{toncianS06}, Science \textbf{312}, 410. 
Reprinted with permission from AAAS, 2013.
}
\label{fig_N_toncian_06}\label{fig_toncianS06}
\end{figure*}

A staged technique which acts on the protons post-acceleration,
but employing all-optical means has been demonstrated  by
\textcite{toncianS06,toncianAIPADV11}.
A transient electric field is excited at the inner surface of a metal cylinder
(having $\sim\mbox{mm}$ diameter and length) irradiated on the outer surface
by a high-intensity laser pulse while a laser-driven proton beam transits
through it, see Fig.\ref{fig_N_toncian_06}~A-C).
The field acts on the protons by modifying their divergence leading to a
narrow, collimated beamlet.
As the field is transient, typically lasting for $\sim 10~\mbox{ps}$,
it affects only the component transiting through the cylinder within
this time window, affecting only protons within a narrow energy band
{and leading} to a spike in the energy spectrum, as clearly visible in
Fig.\ref{fig_N_toncian_06}~F),
showing a $0.2~\mbox{MeV}$ band at $\sim 6~\mbox{MeV}$.
Further experiments have shown that the position of the spectral peak can
be controlled by varying the delay between the two laser pulses
\citep{toncianAIPADV11}
and confirmed that the focusing is {chromatic}, i.e.
the focal position varies with proton energy.

A similar approach, but employing a single pulse, was developed by \textcite{karPRL08a} for reducing the proton beam divergence.
Also conceptually similar to the above described approach
by \textcite{burzaNJP11},
the scheme employs specially designed targets in which a thin foil is inserted in a thicker frame, so that the charge wave expanding outwards from the acceleration region at the rear of the foil generates on the frame’s surface an electric field transverse to the expanding beam, which partially constrains its natural divergence.

Other proposed methods of optical control of proton beam properties include beam steering triggered by shock waves deforming locally the target surface
\citep{lindauPRL05,lundhPRE07},
an effect also reported by \textcite{zeilNJP10},
and control of the beam homogeneity and cross section profile by focusing an annular beam around the high-intensity interaction region, which modifies the properties of hot electrons refluxing through the target
\citep{carrollPRE07}.

The high degree of beam laminarity, and the fact that ion emission is substantially normal to the target surface, led early on to the suggestion that by appropriately shaping
the surface it should be possible to focus down ballistically the protons to a tight spot \citep{wilksPoP01,ruhlPPR01}, ideally recovering the properties of the virtual source.
The idea is consistent with (and complementary to) observations of TNSA ions from wire targets, where the curvature of the target leads to a highly diverging beam ion with the form of an expanding disk \citep{patelPRL03,begPoP04}.
An indirect experimental demonstration of focusing was obtained via enhanced, localized  heating of a secondary target, as will be discussed in
Sec.\ref{sec:APP_WDM} \citep{patelPRL03,snavelyPoP07}.
Recently, a more direct demonstration of proton beam focusing has been obtained by mesh projection methods in experiments where, employing  thick ($250~\um$) targets with hemicylindrical shape \citep{karPRL11}, beam focusing (down to an estimated $25~\um$ spot) over the whole spectrum (up to 25~MeV) was demonstrated.
The data have highlighted the achromatic nature of the focusing at the different energies, consistent with the energy dependent variations in divergence from a planar foil.

Several groups have implemented conventional accelerator techniques for energy selection or transport of laser-accelerated protons, in view of possible downstream applications of the proton beam (see Sec.\ref{sec:applications}).
Besides simple energy selection with bending magnets, the range of options explored includes the use of pairs of quadrupole magnets for refocusing
protons at distances in the 5--60~cm range and in $\sim 100~\um$ spots
\citep{schollmeierPRL08,nishiuchiAPL09}
or to collimate \citep{ter-avetisyanLPB08}
protons within a given  spectral band, up to 14~MeV as found by \textcite{schollmeierPRL08}.
A crucial parameter in this approach is the acceptance angle of the quadrupole system, which may limit the number of particles that can be focused.
Large acceptance pulsed solenoids ($\sim 9~\mbox{T}$) were also used
\citep{rothPPCF09b,harresPoP10}
for collimation and transport of a large number of
$\sim 10^{12}$ particles.

The use of synchronous RF fields for phase rotation resulted in the appearance of multiple peaks across a broadband spectrum \citep{ikegamiPRSTAB09}.
Although demonstrated only at relatively low energy and over low energy bands, this technique is in principle interesting  as, rather than ``slicing'' a portion of the spectrum, which is effectively what is done in several of the methods above, it can concentrate in a narrow spectral band protons originally contained within a larger spectral region.
The phase rotation can be accompanied under the right conditions by a collimation effect.

Some of these techniques have already been implemented sequentially in test beamlines operating at 1~Hz repetition \citep{nishiuchiPRSTAB10}
with a view to future biomedical applications (Sec.\ref{sec:APP_biomedical}).

\section{Other acceleration mechanisms}
\label{sec:othermech}

\subsection{Radiation Pressure Acceleration}
\label{sec:RPA}
\label{sec:RPA_basics}\label{sec:RPA_circpol}

Electromagnetic (EM) waves carry momentum, which may be 
delivered to a non-transparent (either absorbing or reflecting) medium. 
This is the origin of radiation pressure\footnote{The 
electromagnetic theory of radiation pressure 
is due to James Clerk Maxwell \citep{maxwell1873}. 
It is however interesting that the Italian physicist Adolfo Bartoli 
also obtained independently Maxwell's result in 1875 from 
thermodynamic considerations \citep{bartoliNC1884}.}
whose expression for a plane, monochromatic EM wave of intensity $I$ and 
frequency $\omega$ 
normally incident on the plane surface of a medium at rest,
the radiation pressure $P$ is given by
\begin{equation}
P_{\mbox{\tiny rad}}=(1+R-T)\frac{I}{c}=(2R+A)\frac{I}{c}
\label{eq:basics-radiation_pressure}
\end{equation} 
where $R$, $T$ and $A$ are the reflection, transmission and absorption 
coefficients (with $R+T+A=1$) as defined, e.g., in the 
derivation of Fresnel formulas \citep{jackson-fresnelformulas}
as a function of the refractive index and thus of the wave frequency, i.e. 
$R=R(\omega)$ and so on. 
Radiation pressure is related to the total steady ponderomotive force (PF)
on the medium (see Secs.\ref{sec:nutshell_interaction} and \ref{subsec:JXB}).
The PF effectively acts on the electrons being proportional to the inverse of 
the particle mass. At the surface of an overdense
plasma the electrons are pushed inwards by the PF, leaving a charge separation 
layer and creating an electrostatic, back-holding field that in turn acts on
the ions and leads to acceleration.

In the case of normal incidence of a plane wave on a flat surface the PF 
density is the cycle-averaged value of the ${\bf J}\times{\bf B}$ force.
In the following discussion of Radiation Pressure Acceleration (RPA) 
we refer to such case unless otherwise stated and consider 
only the steady action of radiation pressure. 
As discussed in Sec.\ref{subsec:JXB},
the oscillating component of the 
${\bf J}\times{\bf B}$ drives a sweeping oscillation at $2\omega$ 
of the density profile 
and causes strong absorption and hot electron generation, 
except in the case of circular polarization 
for which the oscillating component vanishes.
In the latter case, on the time scale of ion motion it may be assumed that 
the electrons are mostly in a mechanical equilibrium so that the PF and 
electrostatic force locally balance each other.

\subsubsection{Thick targets. Hole Boring regime}
\label{sec:RPA_holeboring}

The intense radiation pressure of the laser pulse pushes the surface of 
an overdense plasma inwards, steepening the density profile. 
For a realistic laser beam of finite
width, the radiation pressure action drives a parabolic--like deformation
of the plasma surface allowing the laser pulse penetrating deeply into 
the target; this process is commonly named ``hole boring'' (HB), 
{even when referring to a planar geometry}, and it is 
associated with ion acceleration at the front side of the target. 
{Notice that in the literature different definitions, such as 
``sweeping acceleration'' \citep{sentokuPP03} or 
``laser piston'' \citep{schlegelPoP09} are also used to refer to essentially 
to the same process}. 

The recession velocity of the plasma 
surface, also named the HB velocity $v_{hb}$, may be simply 
estimated by balancing the EM and mass momentum flows {in a planar
geometry}
\citep{wilksPRL92,denavitPRL92,robinsonPPCF09a,schlegelPoP09}.
In the instantaneous frame where the surface is at rest we observe incoming 
ions with density $n_i\gamma_{hb}$ and velocity 
$-v_{hb}$ bouncing back at the surface. The EM momentum flow, i.e. the radiation
pressure, must then balance a momentum flow difference equal to 
$n_i\gamma_{hb}(2 m_i\gamma_{hb}v_{hb})v_{hb}$ with $\gamma_{hb}=(1-v^2_{hb}/c^2)^{-1/2}$. 
In this reference frame, the radiation pressure is 
$P_{\mbox{\tiny rad}}=(2I/c)(1-v_{hb}/c)/(1+v_{hb}/c)$ as can be demonstrated
by a Lorentz transformation.\footnote{Notice that the ``relativistic'' correction is equivalent to account for the energy depletion of the incident radiation in the adiabatic approximation. This can be easily shown by the heuristic model of radiation pressure as resulting from the reflection of a number $N$ (per unit surface) of photons with energy-momentum $(\hbar\omega,\xv\hbar\omega/c)$ contained in a short bunch of duration $\tau$, corresponding to an intensity $I=N\hbar\omega/\tau$. If the surface is moving at velocity $V=\beta c$, the frequency of the reflected photons is $\omega_r=\omega(1-\beta)/(1+\beta)$ and the reflection time is $\tau_r=\tau/(1-\beta)$. The resulting pressure is $P=|\Delta{\bf p}|/\Delta t=(N\hbar/c)(\omega+\omega_r)/\tau_r=(2I/c)(1-\beta)/(1+\beta)$.}
The global momentum balance 
thus gives\footnote{For simplicity we assume $I$ to be independent of time. Generalization to a time-dependent profile $I(t)$ is discussed by \textcite{robinsonPPCF09a}.}
\bea
\frac{2I}{c}\frac{1-v_{hb}/c}{1+v_{hb}/c} 
= n_i\gamma_{hb}(2m_i\gamma_{hb} v_{hb})v_{hb} . 
\label{eq:holeboringbalance}\label{eq:holeboringbalancerel}
\eea
Solving for $v_{hb}$ this yields
\bea
\frac{v_{hb}}{c}=\frac{\Pi^{1/2}}{1+\Pi^{1/2}}, \qquad \Pi=\frac{I}{m_in_ic^3}
=\frac{Z}{A}\frac{n_c}{n_e}\frac{m_e}{m_p}a_0^2.
\label{eq:vhbnotrel}\label{eq:vhbrel}
\label{eq:vhbnotrel2}\label{eq:holeboringvelocity}
\eea
The fastest ions are those bouncing back from the surface in the moving frame, 
resulting in a maximum energy per nucleon in the lab frame
\bea
{\cal E}_{\mbox{\tiny max}}=2m_p c^2\frac{\Pi}{1+2\Pi^{1/2}} .
\label{eq:Ehbnotrel}
\label{eq:Ehbrel}
\label{eq:Ehbnotrel2}
\eea
In the non-relativistic regime where $\Pi\ll 1$ and $v_{hb}\ll c$,
we obtain $v_{hb}/c \simeq \Pi^{1/2}$ and 
${\cal E}_{\mbox{\tiny max}} \simeq 2m_p c^2{\Pi}$.

Essentially the same results are obtained by a dynamical model of ion
acceleration in the charge separation region at the surface \cite{macchiPRL05}.
Such model and related PIC simulations show that the ions pile up at the end
of the skin layer producing a sharp density spike and causing 
hydrodynamical breaking and collapse of the electron equilibrium.
This process leads to the production of a narrow bunch of fast ions at the 
velocity $2v_{hb}$ that penetrates into the plasma bulk. Eventually the 
quasi-equilibrium condition is established again and the process repeats 
itself until the laser pulse is on. HB acceleration is thus of pulsed nature, 
although on the average it may be described by a steady model 
\citep{schlegelPoP09}.\footnote{For theoretical or simulation studies 
of HB by circularly polarized laser pulses see also, e.g., 
\textcite{liseikinaAPL07,liseikinaPPCF08,naumovaPRL09,chenPP08,yinPP08}
for single ion species case, and 
\textcite{zhangPRSTAB09,robinsonPPCF09b} for two ion species plasmas. 
} 

\begin{figure}[t]
\includegraphics[width=0.48\textwidth]{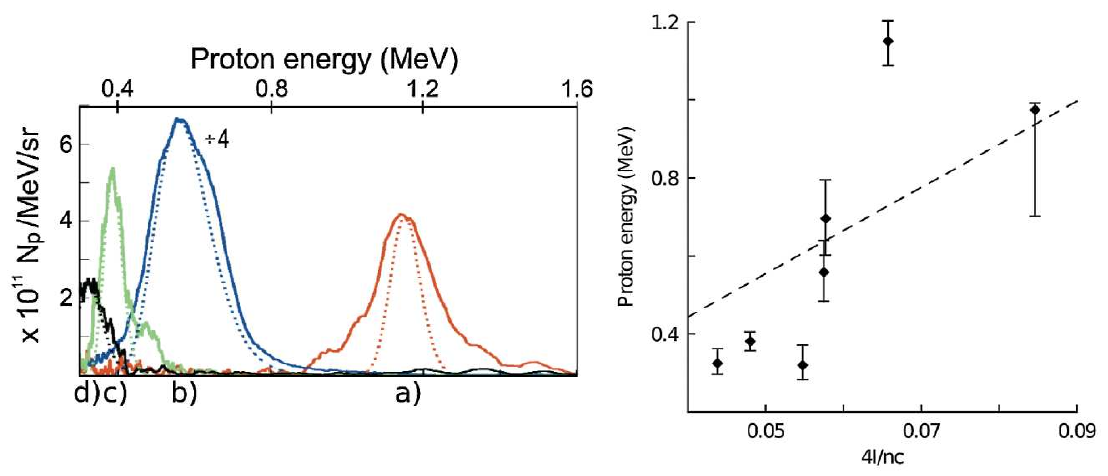}
\caption{(Color online) 
Hole boring acceleration by a CO$_2$ laser pulse in a gas jet. 
Left frame shows ion spectra for various values of the intensity $I_{15}$ 
(in units of $10^{15}~\mbox{W cm}^{-2}$) and the 
electron density $n=n_e/n_c$: 
a) $I_{15}=6.4$, $n=6.1$; b) $I_{15}=5.5$, $n=6.1$;
c) $I_{15}=5.9$, $n=7.6$; d) $I_{15}=5.7$, $n=8.0$.
Right frame shows the observed scaling of ion
energies with the ratio $4I/nc$.
Reprinted figure with permission from 
\textcite{palmerPRL11}, Phys. Rev. Lett. \textbf{106}, 014801.
Copyright 2013 by the American Physical Society.}
\label{fig:palmerPRL11}
\end{figure}

Eq.(\ref{eq:Ehbnotrel2}) indicates that with present-day intensities
high energies may be obtained via HB acceleration if the 
density can be reduced to be slightly above $n_c$, which is possible 
if a gas jet target and a long wavelength laser, i.e. CO$_2$, is used. This 
scheme would be interesting for applications since it allows control of 
the background density, use of a pure proton target and high repetition rate 
since the gas is flowing. In a recent experiment \textcite{palmerPRL11} 
employing $10~\um$ wavelength, $\sim 6 \times 10^{15}~\Wcm2$ circularly 
polarized pulses ($a_0 \simeq 0.5$) and a hydrogen gas jet with density of a 
few times $n_c$,  protons of energy up to $1.2~\mbox{MeV}$ and a narrow energy 
spread were observed (Fig.\ref{fig:palmerPRL11}). The 
observed ion energies were fairly consistent with a linear scaling 
with $I/n_e$ as predicted by the HB model. The energies were actually higher
than expected taking the vacuum laser intensity, suggesting that self-focusing
in the underdense region could have increased the intensity in the plasma.

We notice that \textcite{palmerPRL11} reported on 
``protons accelerated by a radiation pressure driven shock'', 
similarly to several authors who refer 
to HB or ``piston'' acceleration in thick targets as acceleration 
in the electrostatic shock sustained by the laser pressure at the front
surface \citep{zhangPoP07c,zhangPRSTAB09,schlegelPoP09}. 
In the context of ion acceleration by laser, we prefer to reserve the term 
``shock'' for the regime described in Sec.\ref{sec:SHOCK} which implies the 
generation of a ``true'' electrostatic shock wave, able to propagate into the 
plasma bulk and drive a ion acceleration there.  
From the point of view of fluid theory, a 
shock wave launched with some velocity $v_{\mbox{\tiny sho}}$ requires the 
sound speed, and thus the electron temperature, to be hot enough to prevent 
the Mach number $M=v_{\mbox{\tiny sho}}/c_s$ from 
exceeding the critical value $M_{\mbox{\tiny cr}}\simeq 6.5$ above
which one does not have a shock but a ``pure piston'' \citep{forslundPRL71}.
Thus, formation of a ``true'', high speed shock wave may be inhibited for
circular polarization because of the reduced electron heating. 

Experimental evidence of HB acceleration in solid targets is less clear at
present. 
{\textcite{badziakAPL04} reported a series of observations of 
high-density, $\sim\mbox{keV}$ energy ion pulses (plasma ``blocks'') for 
sub-relativistic irradiation ($<10^{18}~\Wcm2$) of solid targets 
(but in the presence of significant preplasma). 
These results were interpreted using a
model of ``ponderomotive skin-layer acceleration'' at the critical
surface, a concept that sounds rather similar to HB-RPA.}
\textcite{akliPRL08} reported on heating of solid density matter due to 
laser-driven density profile sweeping and shock formation at intensities 
up to $5\times 10^{20}~\Wcm2$, and \textcite{henigPRL09c} reported
on ion acceleration by a converging shock in spherical targets irradiated at 
$1 \times 10^{20}~\mbox{W cm}^{-2}$. For both these 
latter experiments, the analysis of data 
and supporting PIC seems also compatible with HB-RPA occurring at the 
front surface, although the electron heating due to the use of linear 
polarization complicates the picture. 
Indications of strong radiation pressure effects were also obtained from the
modeling of collimated, high-density plasma jets at the rear side of targets 
with a few microns thickness, at intensities up to 
$3\times 10^{19}~\Wcm2$ \citep{karPRL08b}. 
It may be noticed that although the scaling of Eq.(\ref{eq:Ehbnotrel2})
leads to relatively modest energies in solid-density targets, the foreseeable
values are of interest for applications requiring large number of ions
at energies of only a few MeV (see Sec.\ref{sec:applications}).

\subsubsection{Thin targets. Light Sail regime} 
\label{sec:RPA_thin}
\label{sec:RPA_lightsail}

Hole Boring RPA
applies to a ``thick'' target, i.e. much thicker than the skin layer in which
ion acceleration by the space-charge field occurs.
The laser pulse penetrates deeper as far as adjacent surface layers
are pushed into the target by a repeated cycle of ion bunch acceleration.
The situation changes when a target is thin enough that all the ions are
accelerated before the end of the laser pulse, i.e. a complete hole boring
occurs. In such a case, the laser pulse is able to further
accelerate ions to higher energies since the ions are not screened
by a background plasma anymore.

The thin target regime of RPA has been named ``Light Sail'' (LS) as the
term is appropriate to refer to a \emph{thin} object
of finite inertia, having large surface and low mass, so that it can receive a
significant boost from radiation pressure.
The invention of the laser soon stimulated possible applications
of the LS concept, including visionary ones such as
laser-driven spacecraft propulsion \citep{forwardJS84}.
To support this idea \textcite{marxN66} used calculations based on
the simple model of a flat, perfect mirror boosted by a plane wave.
The analytical solution and scaling
laws provided by such basic model \citep{simmonsAJP92}
are very useful to illustrate the most
appealing features of LS-RPA,
such as high conversion efficiency in the relativistic limit and the possibility
to reach very high energies with foreseeable laser and target technology.

The equation of motion for a moving target (``sail'')
in the laboratory frame can be obtained with the help
of a Lorentz transformation, similarly to Eq.(\ref{eq:holeboringbalancerel}).
Neglecting absorption for simplicity ($A=0$) we obtain
\bea
\frac{d}{dt}(\beta\gamma)
                         =\frac{2I(t_{\mbox{\tiny ret}})}{\sigma c^2}R(\omega')
                            \frac{1-\beta}{1+\beta},
\label{eq:lightsail-LS}
\qquad
\frac{dX}{dt}=\beta c \label{eq:lightsail-motion},
\eea
where $X$ is the position of the sail, $\beta= V/c$ is its velocity
in units of $c$, $\gamma=(1-\beta^2)^{-1/2}$,
$\sigma=m_in_i\ell$ is the mass density per unit surface, and
$\omega'=\omega[{({1-\beta})/({1+\beta})}]^{1/2}$ is the EM wave (laser)
frequency in the rest (sail) frame.
Notice that the intensity $I$ is in general
a function of the retarded time $t_{\mbox{\tiny ret}}=t-X/c$.

\begin{figure}
\begin{center}
\includegraphics[width=0.48\textwidth]{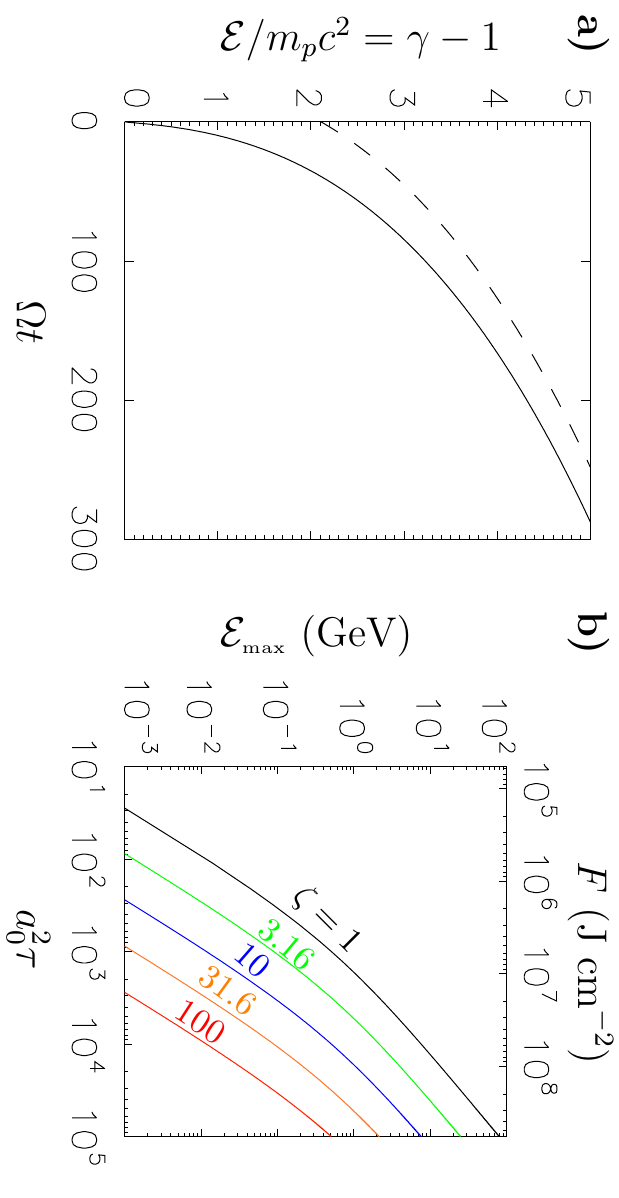}
\end{center}
\caption{a) Energy per nucleon vs. time from the analytical solution
Eq.(\ref{eq:LSenergy_t}) of the LS model with $R=1$. The dashed line gives the
asymptotic $\sim t^{1/3}$ behavior.
b) Scaling of the energy per nucleon
as a function of the dimensionless pulse fluence
$a_0^2\tau$ (where $\tau$ is the pulse duration in units of the laser period)
and of the surface density $\zeta$ [Eq.(\ref{eq:SITthinfoil})] for
$\zeta=1$ (black line), $3.16$ (green), $10$ (blue), $31.6$ (orange) and
$100$ (red). The values on the upper horizontal axis give the fluence in
$\mbox{J cm}^{-2}$ corresponding to $a_0^2\tau$ for
$\lambda=0.8~\mu\mbox{m}$.
}
\label{fig:RPA_lightsail_scaling}
\end{figure}

Analytical solutions to Eqs.(\ref{eq:lightsail-LS}) exist depending on
suitable expressions for $R(\omega)$, the simplest case being that of a
perfectly reflecting mirror ($R=1$) and
a pulse of constant intensity $I$ \cite{simmonsAJP92}.\footnote{For a constant intensity $I$, Eqs.(\ref{eq:lightsail-LS}) are identical to those for a charge accelerating during the Thomson scattering from a plane wave: see \textcite{landau-mirror} who leave the solution as an exercise for the reader.}
The $\gamma$ factor as a function of time is given by
\beq
\gamma(t)=\sinh(u)+\frac{1}{4\sinh(u)} , \quad
u\equiv \frac{1}{3}\mbox{asinh}(3\Omega t+2) ,
\label{eq:LSenergy_t}
\eeq
where $\Omega\equiv ({Zm_ea_0^2}/{Am_p\zeta})\omega$ and
$\zeta$ has been defined in Eq.(\ref{eq:SITthinfoil}).
Asymptotically, $\gamma(t) \simeq (3\Omega t)^{1/3}$
(Fig.\ref{fig:RPA_lightsail_scaling}~a).

The most {significant} quantities can be obtained for an arbitrary pulse
shape $I(t)$ as a function of the dimensionless pulse fluence ${\cal F}$
(the pulse energy per unit surface):
\bea
{\cal F}(t_{\mbox{\tiny ret}})=\frac{2}{\sigma c^2}
                             \int_0^{t_{\mbox{\tiny ret}}}I(t')dt'.
\label{eq:lightsail-Ecal}
\eea
The sail velocity $\beta$, the corresponding energy per nucleon
${\cal E}=m_pc^2(\gamma-1)$ and the instantaneous efficiency $\eta$
(i.e. the ratio between the mechanical energy delivered to the sail and the
incident pulse energy)\footnote{The expression for $\eta$ also follows from ``photon number'' conservation and frequency downshift (see Section~\ref{sec:RPA_holeboring}). In the reflection of $N$ photons from the mirror, the
energy transferred to the mirror is
$N\hbar(\omega-\omega_r)=[2\beta/(1+\beta)]N\hbar\omega \equiv \eta(N\hbar\omega)$.}
are given by
\bea
\beta(t_{\mbox{\tiny ret}})&=&\frac{[1+{\cal F}(t_{\mbox{\tiny ret}})]^2-1}
                               {[1+{\cal F}(t_{\mbox{\tiny ret}})]^2+1}
\label{eq:lightsail-beta0}, \\
{\cal E}(t_{\mbox{\tiny ret}})&=&m_p c^2 \frac{{\cal F}^2(t_{\mbox{\tiny ret}})}
              {2[{\cal F}(t_{\mbox{\tiny ret}})+1]} ,
\label{eq:lightsail-Enuc} \\
\eta(t_{\mbox{\tiny ret}})
&=&\frac{2\beta(t_{\mbox{\tiny ret}})}{1+\beta(t_{\mbox{\tiny ret}})}
=1-\frac{1}{[{\cal F}(t_{\mbox{\tiny ret}})+1]^2}.
\label{eq:lightsail-efficiency}
\eea
Thus, $\eta\rightarrow 1$ when $\beta(t_{\mbox{\tiny ret}})\rightarrow 1$.
The final energy per nucleon ${\cal E}_{\mbox{\tiny max}}$
is obtained from the total fluence
${\cal F}_{\infty}={\cal F}(t_{\mbox{\tiny ret}}=\infty)$.
For a constant intensity ${\cal F}_{\infty}=\Omega\tau_p$
where $\tau_p$ is the duration of the laser pulse.
In practical units
${\cal F}_{\infty}=2.2 F_{1e8}\rho_1^{-1}\ell_{10}^{-1}$
where $F_{1e8}$ is the fluence in units of $10^8~\mbox{J cm}^{-2}$,
$\rho_1=m_in_i/1~\mbox{g cm}^{-3}$ and $\ell_{10}={\ell}/{10~\mbox{nm}}$.
The scalings for ${\cal E}_{\mbox{\tiny max}}$ are summarized in
Fig.\ref{fig:RPA_lightsail_scaling}~b).
With present-day or near-term laser technology, fluence values of
$10^8~\mbox{J cm}^{-3}$ seem affordable, while target manufacturing can
produce films of few nm thickness, e.g. Diamond-Like Carbon (DLC) foils. These
values yield ${\cal F}_{\infty}>1$
allowing to approach a regime of high efficiency,
relativistic ions, and favorable scaling with the pulse energy.

The above estimates have been obtained assuming a perfectly reflecting sail
($R=1$) that, for a given surface density parameter $\zeta$ limits the
laser amplitude to $a_0<\zeta$ due to the onset of relativistic
transparency [Eq.(\ref{eq:SITthinfoil})] that reduces the
boost on the foil. This effect suggests $a_0=\zeta$ as an ``optimal'' condition
for LS acceleration \citep{macchiPRL09,tripathiPPCF09}\footnote{Some authors give a similar condition for the optimal thickness but with slightly different numerical factors \citep{yanPRL08,jiPRLcomm09,yanPRLcomm09}} that might be
however relaxed by the effect of frequency decrease in the moving foil frame,
increasing $R(\omega')$ [see Eq.(\ref{eq:lightsail-LS})].
For $a_0>\zeta$, all electrons are pushed away from the foil.
In this regime the ions in the foil undergo a
Coulomb explosion producing a broad ion spectrum.
In a composite target the
ion field after electron expulsion might be used for monoenergetic
acceleration of a proton layer \citep{bulanovjrPRE08,grechNJP09}.

\label{sec:RPA_dominance}

\begin{figure}[t]
\includegraphics[width=0.48\textwidth]{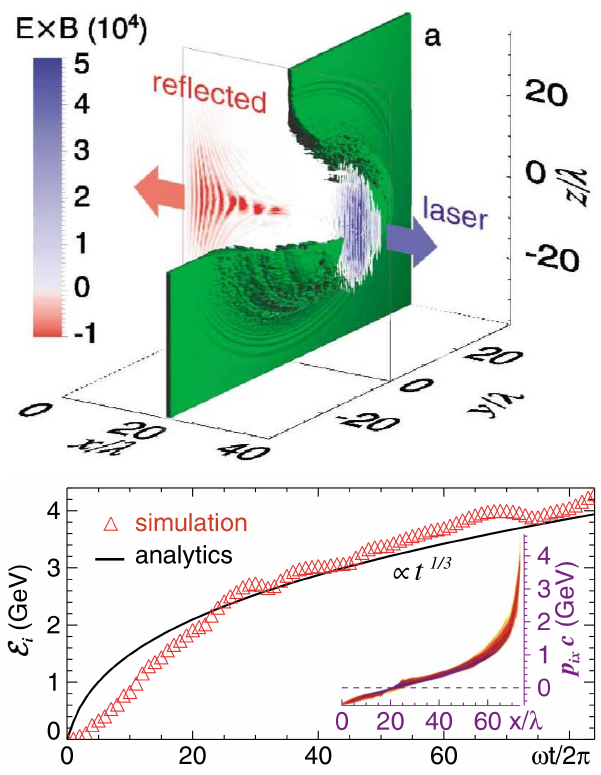}
\caption{Three-dimensional simulations of thin foil acceleration in the
Radiation Pressure Dominant regime
(see text for parameters).
Top frame: snapshots at $t=40T$ of ion density isosurface and
Poynting vector in the $y=0$ plane.
Bottom frame: The maximum ion kinetic energy versus time
and the ion phase space projection ($x,p_x$) at
$t=80T$. The solid line corresponds to the analytical calculation according
to the LS model.
Reprinted figure with permission from 
\textcite{esirkepovPRL04}, Phys. Rev. Lett. \textbf{92}, 175003.
Copyright 2013 by the American Physical Society.}
\label{fig:esirkepovPRL04}
\end{figure}

The interest in the LS regime was greatly stimulated by
three-dimensional PIC simulations of thin foil acceleration
by \textcite{esirkepovPRL04} which showed that the temporal dependence and
typical values of the ion energy were well described by the LS model.
The simulations assumed a laser pulse with peak amplitude $a_0=316$
($I\lambda^2=1.4 \times 10^{23}~\mbox{W cm}^{-2}$)
and 8 cycles duration, and a proton slab of density $49n_c$ and
$1\lambda$ thickness. Most of the ions in a thin foil target are
accelerated coherently up to relativistic energies ($\sim 1.5~\mbox{GeV}$)
as shown in Fig.\ref{fig:esirkepovPRL04}. According to
\textcite{esirkepovPRL04}, in order for RPA to become the dominant
acceleration mechanism the ions have to acquire relativistic
energies already within one laser cycle, so that they can promptly
follow electrons which are displaced in the longitudinal direction by the
ponderomotive force.
Later theoretical studies of such so-called Radiation Pressure Dominant
(RPD) regime include Rayleigh-Taylor-like instability of the foil
\citep{pegoraroPRL07} and the effects of radiation friction which play
a significant role at ultrarelativistic intensities
\citep{tamburiniNJP10}. Of particular interest is the possibility of a
self-regulated regime where the transverse expansion of the foil decreases
the density along the axis (while the frequency downshift in the foil frame
compensates the effect of decreasing $\zeta$ on $R(\omega')$), allowing for an
increase of the ion energy at the expense of the total number of accelerated
ions \citep{bulanovPRL10,bulanovPoP10}.
For a 3D expansion, theory predicts an asymptotic scaling with time of
kinetic energy $K(t)/mc^2 \simeq (3\Omega t)^{3/5}$ that is more favorable than
for plane acceleration.
This effect has been recently confirmed by 3D simulations \citep{tamburiniPRE12}
showing a \emph{higher} peak energy than found in lower dimensionality
simulations.

\label{sec:RPA_lightsail_CP}

The ultra-high intensities needed for RPD acceleration are still above
present-day laser technology.
However, after the proposal of \textcite{esirkepovPRL04}
it was realized that exploring the concept using pulses with circular
polarization (CP)
at normal laser incidence would enable an investigation of a RPD
regime at lower intensities as theoretically discussed
by \textcite{macchiPRL05} in thick targets.
Three papers \citep{zhangPP07,robinsonNJP08,klimoPRSTAB08} independently
showed that the use of CP allowed an optimal
coupling with an ultrathin foil target as well as rather monoenergetic spectra.
Much theoretical work has been then devoted to LS-RPA with CP pulses,
unfolding a dynamics that is much richer than what
is included in the simple ``accelerating mirror'' model.
In particular, formation of a monoenergetic ion distribution is not
straightforward
\citep{macchiPRL09,eliassonNJP09,macchiNJP10}
and may require to control or engineer both the
pulse and target properties \citep{qiaoPRL09,qiaoPRL10,yuPRL10,grechNJP12}.

Several multi-dimensional simulation studies suggested to use flat-top
transverse profiles to keep a quasi-1D geometry
\citep{robinsonNJP08,klimoPRSTAB08,liseikinaPPCF08,qiaoPRL09} in order to
avoid target bending that would favor electron heating, to prevent early
pulse breakthrough due to transverse expansion and to keep a monoenergetic
spectrum against the inhomogeneous distribution of the laser intensity; for this
last issue, a target with modulated surface density has been also proposed
\citep{chenPRL09}. In contrast to these studies
\textcite{yanPRL09} use a Gaussian intensity profile and find the formation of
a narrow, high energy ion bunch via a self-organization mechanism somewhat
similar to that inferred by \textcite{bulanovPRL10}.
Another open issue is the stability of the foil against transverse
perturbations\footnote{See e.g. \textcite{pegoraroPRL07,klimoPRSTAB08,tikhonchukNIMA10,chenPop11,yuPRL10,adusumilliPoP12}}.
Very recent simulation studies characterized regimes of efficient LS-RPA
for linearly polarized pulses at irradiances
$\sim 10^{21}~\mbox{W cm}^2{\um}^2$ \citep{qiaoPRL12,doverHEDP12}.

\begin{figure}
\includegraphics[width=0.48\textwidth]{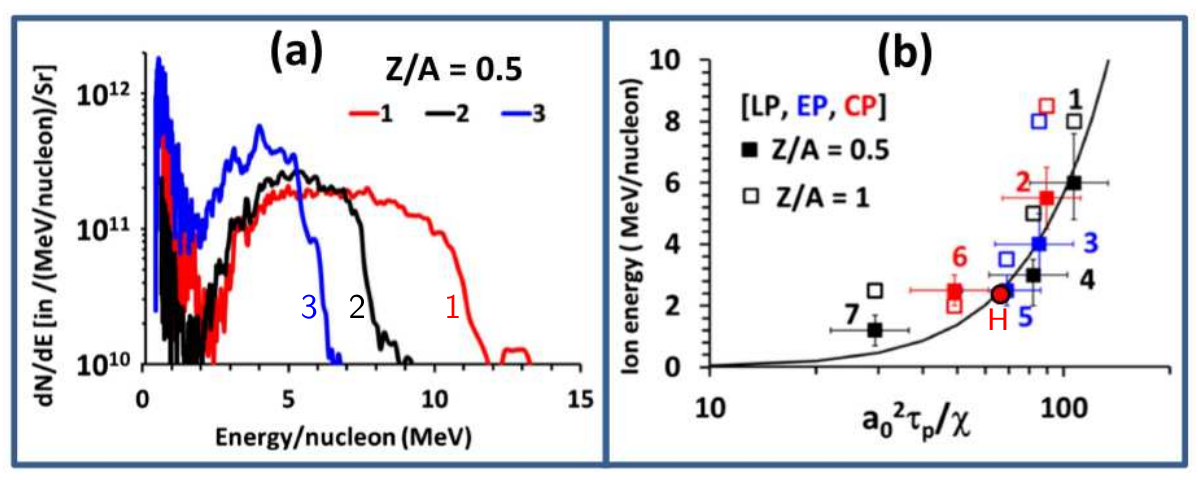}
\caption{a) Three typical narrow band spectra for $Z/A=1/2$ impurity ions
observed from thin ($0.1~\um$) metallic targets (for parameters of shots 1-3 see explanation in part (b)).
b) Peak energies for both ions with $Z/A=1/2$ (filled squares) and protons with
$Z/A=1$ (empty squares)
for seven shots with different polarization parameters
[see e.g. Eq.(\ref{eq:ellipticpol})]:
$\epsilon=0$ (LP), $0.47$ (EP) and $0.88$ (CP).
The peak energy is plotted
as a function of the parameter $a_0^2\tau_p/\chi$, which
corresponds to $(m_p/2m_e)(\Omega\tau_p)$ in our notations.
The parameter set [$a_0$, target material, thickness ($\um$), polarization]
for the data points 1–7 is
[15.5, Cu, 0.1, LP],
[10, Cu, 0.05, CP],
[13.8, Cu, 0.1, EP],
[7.5, Al, 0.1, LP],
[6.9, Al, 0.1, EP],
[13.6, Al, 0.5, CP], and
[14.1, Al, 0.8, LP], respectively.
The red circle marked with ``{\sf H}''
is the data point from \textcite{henigPRL09b}.
The solid line is the LS scaling (\ref{eq:lightsail-Enuc}).
Reprinted figure with permission from 
\textcite{karPRL12}, Phys. Rev. Lett. \textbf{109}, 185006.
Copyright 2013 by the American Physical Society.
}
\label{fig:karXXX12}
\end{figure}

{Possible} indications of the onset of LS regime
have been recently provided in an experiment
performing using $800~\mbox{fs}$, $3 \times 10^{20}~\mbox{W cm}^{-2}$ high
contrast ($10^{9}$) pulses
from the VULCAN laser and very thin ($\sim 0.1~\um$) metallic targets
\cite{karPRL12}.
Narrow-band spectra with  peak energies up to
$\simeq 10~\mbox{MeV}$ per nucleon
were observed both for proton and heavier ions
$Z/A=1/2$ ions present as surface impurities [Fig.\ref{fig:karXXX12}~a)],
while heavier bulk ions had a broad spectrum at lower energies.
The peak energies scaled with
the fluence parameter as $\sim {\cal F}^2_{\infty}$ {$\sim a_0^4t_p^2$}
[Fig.\ref{fig:karXXX12}~b)],
in agreement with Eq.(\ref{eq:lightsail-Enuc}) for non-relativistic
ions{, and differently from scalings as $a_0$, $a_0^2$ or $a_0^2t_p$
which have been inferred for TNSA or for other mechanisms effective for
ultrathin targets (see Sec.\ref{sec:OTHER_transparency})}.
The $Z/A=1$ peaks are at slightly higher energy than the $Z/A=1/2$ ones,
suggesting that the LS stage is followed from a multispecies expansion
(Sec.\ref{sec:TNSA_multispecies})
in the sheath field where protons gain additional energy and the spectral peak
separation may be further enforced.

The scaling plot in Fig.\ref{fig:karXXX12}~b) also contains data from
\textcite{henigPRL09b} who investigated LS using 45~fs, CP pulses at
ultrahigh contrast ($\sim 10^{11}$) and $\sim 5 \times 10^{19}~\Wcm2$ intensity,
and few-nm DLC foils.
Experimental
spectra of fully ionized C$^{6+}$ ions show a difference between
linearly and circularly polarized pulses, with a broad
peak at $\simeq 30~\mbox{MeV}$ appearing in the latter case,
and reduced electron heating for CP.
More recent experimental data by \textcite{dollarPRL12} using tightly focused
($f/1$) pulses with intensity up to $2 \times 10^{21}~\mbox{W cm}^{-2}$ showed
weak difference between CP and LP, which was attributed to the
early deformation of the thin targets causing excessive electron heating.
Very recently, preliminary indications of a transverse instability resulting
in spatial modulations of the accelerated proton beam have been reported
\citep{palmerPRL12}.

To summarize the experimental evidence, so far there is a fairly clear
confirmation of the expected LS scaling, but also indications of significant
detrimental effects.
The observed ion spectrum is relatively
broad, suggesting that transverse inhomogeneity and heating effects need to be
reduced. In perspective, the relatively slow scaling of the energy gain with
time might pose the challenge to increase the acceleration length against
the effect of, e.g., pulse diffraction and instabilities.

\subsection{Collisionless Shock Acceleration}
\label{sec:shock}\label{sec:SHOCK}

Acceleration of particles by shock waves (briefly, shocks) in plasmas
is a problem of central interest in astrophysics \citep{martinsApJ09}.
The existence of an ion component that is reflected by the shock front
is actually integral to the formation of the collisionless,
electrostatic shock waves in basic fluid theory
where the electrons are assumed to be in a Boltzmann equilibrium
\citep{tidman-krall,forslundPRL70,forslundPRL71}.
In the frame moving at the shock velocity, ions are reflected by the shock
if the height of the electrostatic potential barrier $\Phi_{\mbox{\tiny max}}$
at the front is such that $Ze\Phi_{\mbox{\tiny max}}>m_iv_1^2/2$,
being $v_1$ the velocity of the
ion component in the shock frame. Behind the shock front, the fields have
an oscillatory behavior.
Reflected ions initially at rest acquire a velocity in the lab frame equal to
$2v_{\mbox{\tiny sho}}$ where $v_{\mbox{\tiny sho}}$ is the shock front
velocity.

Collisionless shock acceleration (CSA) was
proposed as an ion acceleration mechanism in superintense laser
interaction with an overdense plasma on the basis of PIC simulations
by \textcite{denavitPRL92} and \textcite{silvaPRL04}.\footnote{In experiments on underdense plasmas created either by using gas jet targets \citep{weiPRL04} or by the effect of long prepulses in solid targets \citep{habaraPRE04}, the observation of ion acceleration along the radial direction has been attributed to radial shock generation in a laser-driven channel.}
In the latter work simulation showed the generation of shocks with high
Mach numbers $M=v_{\mbox{\tiny sho}}/c_s=2-3$, where the sound speed is estimated
using for the ``temperature'' the \emph{hot} electron energy, i.e.
$T_h\simeq {\cal E}_p $ (\ref{eq:ponderomotive}).
The shocks are generated at the front surface with a velocity close to $v_{hb}$
given by Eq.(\ref{eq:holeboringvelocity}), consistently with the assumption
that they are driven by the piston action of radiation pressure.
By estimating $v_{\mbox{\tiny sho}} \simeq v_{hb}$,\footnote{Notice that $v_{\mbox{\tiny sho}} \simeq v_{hb}$ implies that ``reflected'' ions directed into the bulk will have a velocity $\sim 2v_{hb}$, i.e. twice the surface recession velocity, as the fastest ions generated by the piston action in HB acceleration (Sec.\ref{sec:RPA_holeboring}. This similarity may explain why HB and CSA are often confused in the literature.} in the strongly relativistic limit
$a_0\gg 1$ the condition to obtain radiation-pressure driven supersonic shocks
($M>1$) can be written as $\sqrt{2}a_0>n_e/n_c$.
The reflected ions
may get further acceleration by the transient sheath field
at the rear surface as in TNSA, eventually producing a plateau in the ion
spectrum. A similar signature was observed experimentally by
\textcite{zepfPRL03} and thus intepreted as evidence of front side
contribution to ion acceleration, in contrast to pure TNSA at the rear
side of the target.
In particular conditions, the staged CSA-TNSA acceleration might produce the
highest energy component in the ion spectrum as observed in simulation studies
\cite{dhumieresPoP05,chenPoP07}
which however also suggest lower
efficiency and brilliance with respect to pure TNSA.

Very recently, CSA has been indicated as the mechanism responsible for
monoenergetic acceleration of protons up to $22~\mbox{MeV}$ in the
interaction of CO$_2$ laser pulses with Hydrogen gas jets at intensities up to
$6.5 \times 10^{16}~\mbox{W cm}^{-2}$ corresponding to $a_0=2.5$
\citep{haberbergerNP12}. The particular
temporal structure of the laser pulse, i.e. a 100~ps train of 3~ps pulses,
was found to be essential for the acceleration mechanism, since no spectral
peaks were observed for a smooth, not modulated pulse. Comparison with PIC
simulations suggested that the multiple pulses lead to efficient generation
of suprathermal electrons, and that the latter (rather than radiation
pressure) drive the shocks which eventually accelerate protons. Simulations
also suggest that the process could scale in order to produce 200~MeV protons
at $10^{18}~\mbox{W cm}^{-2}$, that may foreseeable with future CO$_2$ laser
development. Such scheme based on gas lasers and gas jet target would have
the remarkable advantage of high-repetition rate operation, but the
efficiency per shot might be low with respect to other approaches:
in the experiment of
\textcite{haberbergerNP12} the number of ions ($\sim 2.5 \times 10^5$ )
in the narrow spectral peak at $\simeq 22~\mbox{MeV}$ for a
60~J pulse energy implies a conversion efficiency of $\sim 10^{-8}$.

\begin{figure}
\includegraphics[width=0.48\textwidth]{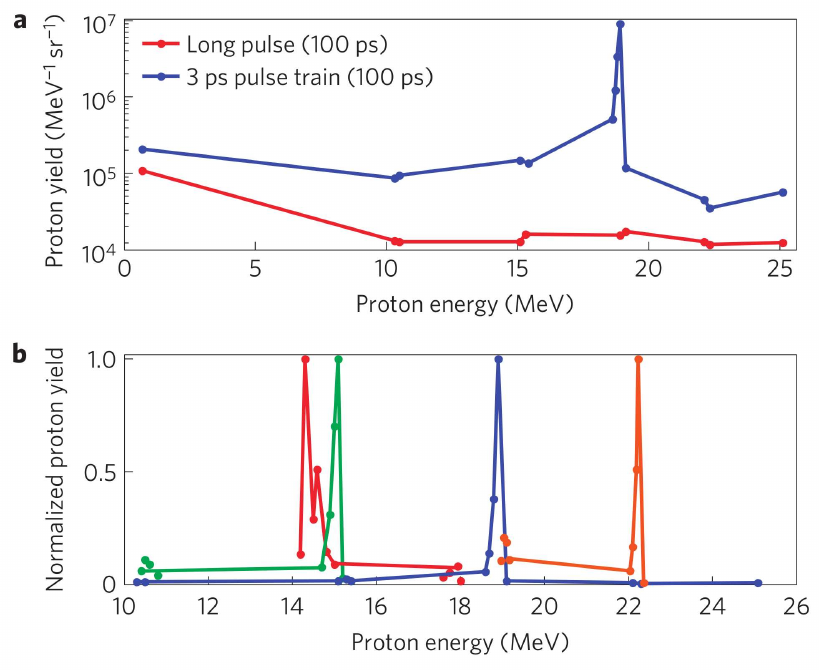}
\caption{Proton spectra from CO$_2$ laser interaction with a Hydrogen gas
jet \cite{haberbergerNP12}.
Frame~a) shows different spectra for a smooth (long) pulse (lower line) and a
pulse train of 3~ps spikes (upper line); in the latter case, a peak appears in the spectrum.
Frame~b) shows narrow spectra obtained in
different shots. See text for parameters.
Reprinted by permission from Macmillan Publishers Ltd:
\cite{haberbergerNP12}, Nature Physics \textbf{8}, 95. Copyright 2012.}
\end{figure}

In addition to collisionless shocks, the standard fluid theory also predict
solitons \citep{tidman-krall}
propagating at the velocity $v_{\mbox{\tiny sol}}$ with
$1 < v_{\mbox{\tiny sol}}/c_s \lesssim 1.6$. These solitons are characterized by
$Ze\Phi_{\mbox{\tiny max}}<m_iv_{\mbox{\tiny sol}}^2/2$ and are thus
transparent to background ions ``by construction''.
However, generation of electrostatic solitons
may lead to ion acceleration in some circumstances, e.g.
when the soliton breaks in the expanding rear sheath due to the effect of the
plasma flow \citep{zhidkovPRL02b}.
Additional simulation studies of shock and solitary wave acceleration
are reported by \textcite{hePRE07,liuLPB09,macchiPRE12}.

\subsection{Transparency regime. Break-out afterburner}
\label{sec:OTHER_transparency}

If ultrathin foils are used as targets (which requires ultrahigh-contrast, 
prepulse-free conditions), the expansion of the foil may lead to the onset
of transparency during the short pulse interaction, when the electron density
$n_e$ is further decreased down to the cut-off value (of the order of 
$\gamma n_c$ due to relativistic effects, see 
Sec.\ref{sec:nutshell_interaction}).
While this effect limits the energy attainable via RPA 
(Sec.\ref{sec:RPA_lightsail_CP}), it can lead to enhanced ion acceleration
via different mechanisms. 

\begin{figure}[t!]
\includegraphics[width=0.48\textwidth]{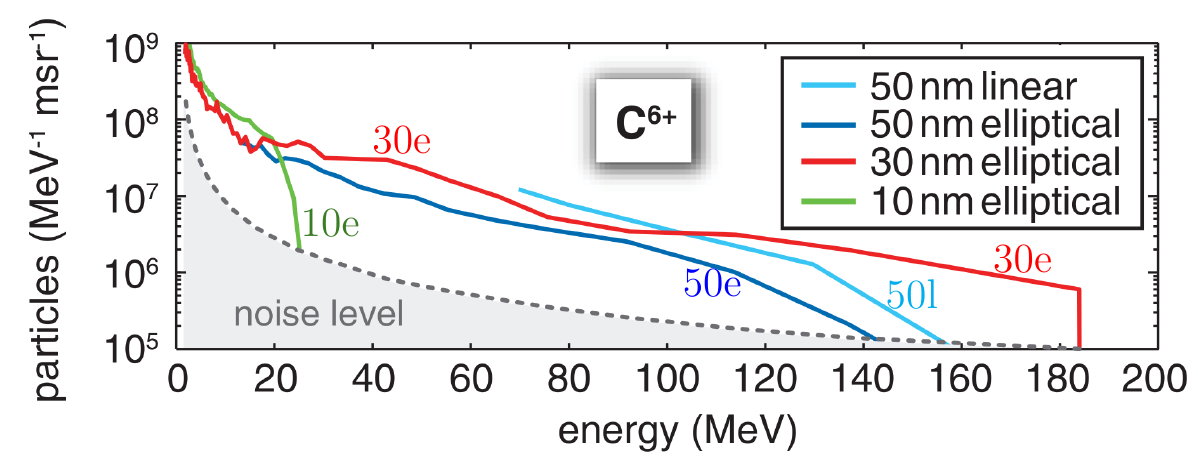}
\caption{(Color online)
Spectra of $\mbox{C}^{6+}$ ions from laser interaction with ultrathin
targets in the regime of relativistic transparency, as a function of target
thickness and laser polarization. 
Reprinted figure with permission from 
\textcite{henigPRL09a}, Phys. Rev. Lett. \textbf{102}, 095002.
Copyright 2013 by the American Physical Society.}
\label{Fig2_PRL_henig_09a}
\end{figure}

Several related experiments were performed at the 
TRIDENT laser facility at Los Alamos National Laboratory (LANL), using
pulse durations in the 500-700~fs range. $\mbox{C}^{6+}$ ions with 
energies up to $15~\mbox{MeV}$ per nucleon were observed by irradiating 
DLC foils with $\sim 40~J$, $\sim 7 \times 10^{19}~\mbox{W cm}^{-2}$ pulses, 
for an optimal thickness of 30~nm that is determined by the condition 
that relativistic transparency occurs at the pulse peak
\citep{henigPRL09a}. 
Fig.\ref{Fig2_PRL_henig_09a} shows spectra for 
different polarizations. 
For more energetic and intense pulses ($\sim 80~J$, $10^{21}~\mbox{W cm}^{-2}$)
and thicker targets (140~nm), broad $\mbox{C}^{6+}$ spectra  with higher 
cut-off energies beyond $40~\mbox{MeV}$ per nucleon were observed, and
the inferred conversion efficiency was $\sim 10\%$ \citep{hegelichNF11}.
Narrower $\mbox{C}^{6+}$ spectra ($\Delta{\cal E}_i/{\cal E}_i\simeq 15-20\%$)
at lower energies ($\sim 3-10~\mbox{MeV}$) were observed using either loose 
focusing or circular polarization \citep{hegelichNF11,jungPRL11}.
Very recently, energies up 80~MeV per nucleon for Carbon and 
120~MeV per protons have been communicated \citep{hegelichBAPS11}.
{The onset of relativistic transparency in these conditions has
been recently investigated in detail with ultrafast temporal resolution 
\cite{palaniyappanNP12}.}

Simulation studies of this regime
show that the increase of the cut-off energy is related to enhanced and
volumetric heating of electrons as the target becomes transparent, leading 
to a stronger accelerating field for ions; the name ``Break-Out Afterburner''
(BOA) has been proposed for such regime by the Los Alamos group 
\citep{yinPoP07}. 
3D PIC simulations of BOA have been reported by \textcite{yinPRL11}.
Modeling of BOA is not simple as the process appears to 
involve different stages. Analytical descriptions of BOA have been reported by 
\textcite{yanAPB10,albrightJPCS10} and a scaling of the maximum 
ion energy ${\cal E}_{\mbox{\tiny max}} \simeq (1+2\alpha)ZT_e$, 
with $T_e$ the electron temperature
and  $\alpha$ a phenomenological parameter (estimated to be $\sim 3$ from 
simulations), has been proposed.
A fastly growing relativistic Buneman instability, excited due to the relative 
drift between electron and ions, has been invoked as a mechanism enhancing 
the coupling with ions \cite{albrightPoP07}.
Theoretical explanations for narrow $\mbox{C}^{6+}$ spectra, based on 
an electromagnetic ``ion-soliton'' model (fundamentally different from 
electrostatic solitons described in Sec.\ref{sec:shock}) 
are discussed by \textcite{yinPoP11}.

\subsection{Acceleration in near-critical and underdense plasmas}
\label{sec:OTHER_underdense}

A number of studies has been 
devoted to ion acceleration in ``near-critical'' plasmas 
with electron density close to the cut-off value ($n_e \simeq n_c$),
in order to allow a more efficient generation of hot electrons to drive TNSA. 
Production of a low-density plasma by a laser prepulse 
has been investigated for laser and target parameters such that
at the time of interaction with the main short pulse the preplasma was either 
underdense \citep{matsukadoPRL03} or slightly overdense \citep{yogoPRE08}; in
the latter experiment, protons up to $3.8~\mbox{MeV}$ are observed at 
$10^{19}~\mbox{W cm}^{-2}$ intensity, and directed slightly off the normal
to the target rear side. 
The analysis of these experiments gave indication of a regime where the 
pressure due to self-generated magnetic field at 
the rear surface strongly contributes to charge separation. 

An alternative strategy to reduce the electron density is to use special 
target materials such as foams, which may be manufactured in order to have an 
average value of $n_e$ slightly larger, or even lower than $n_e$ (the average
is meant over a length larger than the typical sub-micrometric scale of
inhomogeneity). 
Experimentally, proton acceleration in low-density foams
($n_e=0.9-30n_c$)
has been investigated by \textcite{willingalePRL09,willingalePoP11} 
at intensities up to $10^{21}~\mbox{W cm}^{-2}$, showing that the proton 
energy is close to that obtained for solid foils and the same laser pulse
for the lowest density value ($n_e \simeq 0.9n_c$). In this experiment,  
proton acceleration has been mostly investigated 
as an indication of the onset of relativistic transparency, leading to 
enhanced laser penetration and collimation of hot electrons and ions by 
self-generated magnetic fields. 
Recent simulation studies of ion acceleration in solid target covered with 
foam layers 
have been also reported \cite{nakamuraPoP10,sgattoniPRE12}.

Experimental investigations of ion acceleration using gas jet targets, with 
typical densities below $10^{20}~\mbox{cm}^{-3}$, have been also performed.
These experiments include the already described investigations of 
hole boring RPA (Sec.\ref{sec:RPA_holeboring}) and shock acceleration
(Sec.\ref{sec:shock}) using CO$_2$ lasers for which gas jets are near-critical
targets. Using optical or near-infrared lasers, several experiments of 
high-intensity laser interaction with underdense gas jets have reported 
observations of energetic ions accelerated
in the \emph{radial} direction with respect to the laser pulse propagation axis
\citep{sarkisovPRE99,krushelnickPRL99,weiPRL04}
by the electric field created by the electron displacement
in the channel drilled by the ponderomotive force. The ion spectrum may provide
information on the self-focusing and channeling dynamics of the laser pulse
and the acceleration mechanism shows indeed some similarity with those
active in the interaction with solid targets \citep{macchiPPCF09}. However,
radial acceleration of ions is of modest interest for applications since 
the ions are not collimated at all. 

\begin{figure}[t]
\begin{center}
\includegraphics[width=0.48\textwidth]{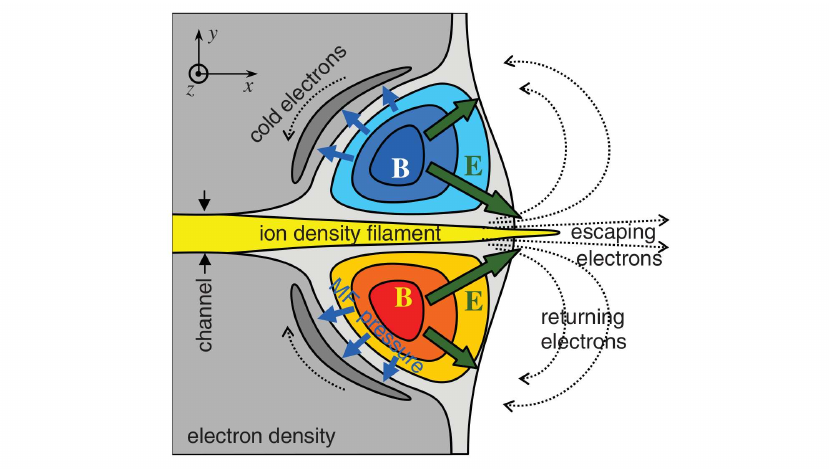}
\caption{(Color online)
Sketch of magnetic-field sustained acceleration of ions, showing the
topology of the magnetic and electric fields and the flows of escaping and 
returning electron currents. 
Reprinted figure with permission from 
\textcite{bulanovPRL07}, Phys. Rev. Lett. \textbf{98}, 049503.
Copyright 2013 by the American Physical Society.}
\label{Fig1_PRL_bulanov-esirkepov_07}
\end{center}
\end{figure}

A collimated emission in the forward direction from an
underdense He gas jet was reported by \textcite{willingalePRL06}.
Using a laser pulse of 1~ps duration, energy up to 340~J and intensity up to 
$6 \times 10^{20}~\mbox{W cm}^{-2}$, He ions up to 40~MeV were observed 
collimated in a beam with $<10^{\circ}$ aperture. The data were interpreted 
by assuming that a large electric field was generated at the rear side of the
gas jet by escaping hot electrons. Differently from TNSA in solid targets,
the mechanism was considered to be effective despite the relatively long 
density scalelength at the rear surface because a larger fraction of 
hot electrons was generated by electron acceleration in the underdense plasma.
Further analysis of simulations of the experiment \citep{willingalePRL07}
showed also a significant contribution due to the generation of a 
quasi-static magnetic field at the rear surface, which in turns enhances the
accelerating electric field via magnetic pressure and induction effects 
according to the model by \textcite{bulanovPRL07} that was also used to 
explain the above mentioned 
experimental results by \textcite{yogoPRE08} in a near-critical
plasma.
Fig.\ref{Fig1_PRL_bulanov-esirkepov_07} shows a sketch of such mechanism.

Magnetic-field sustained ion acceleration was also indicated as the dominant 
mechanism in an experiment by \textcite{fukudaPRL09}, where ions in 
a 10-20~MeV range and collimated in a $3.4^{\circ}$ aperture cone were 
observed in the 
interaction of a $7 \times 10^{17}~\mbox{W cm}^{-2}$, 40~fs laser pulse
with an underdense gas jet where $\mbox{CO}_2$ clusters were formed. 
The role of the clusters was apparently that of enhancing the self-channeling 
and focusing of the laser pulse, leading to an increase of the intensity
in the plasma, rather than contributing to ion acceleration via cluster 
explosions. 

Generation of collimated ions from underdense plasmas 
at ultrahigh intensities ($>10^{21}~\mbox{W cm}^{-2}$)
was investigated 
theoretically and with numerical simulations already more than a 
decade ago \citep{esirkepovJETPL99,sentokuPRE00,bulanovJETPL00}.
In particular in these papers it was predicted that, for 
$a_0>(m_i/m_e)^{1/2}\simeq 43A^{1/2}$, the effective inertia of the 
highly relativistic electrons in the laser field becomes comparable to those
of ions. As a consequence the ions closely  follow 
the electron displacement
due to the ponderomotive action and the acceleration process may become
similar to what is observed in an overdense plasma. 
A few more recent simulation studies investigated a regime where a small 
ion target is placed in an underdense plasma
\citep{shenPRSTAB09,yuNJP10}. 
The superintense laser pulse
accelerates and overruns the ion target and then generates a wakefield in 
the underdense plasma, where ions may be trapped and accelerated 
in a way similar to the well-known scheme for
laser acceleration of electrons \citep{esareyRMP09}.
In those simulations GeV energies were reached,
but the required laser pulses should have multi-petawatt power
and multi-kJ energy, that is
still far beyond present-day laser technology.

\subsection{Resistively enhanced acceleration}
\label{sec:OTHER_resistive}

Already during the ``front vs rear side acceleration'' debate related to 
experiments reported in 2000 (Sec.\ref{sec:intro}), it was suggested that 
protons may also be accelerated in the target bulk through a mechanism 
depending on the target resistivity $\eta$ \cite{daviesLPB02}. 
The electric field generated in the target bulk 
to provide the return current ${\bf E}={\bf j}_r/\eta$
(see Sec.\ref{sec:nutshell_fastelectrons_transport}) 
increases for low $\eta$ reducing the penetration of hot electrons through 
the target and at the same time favoring acceleration in the front and bulk 
regions versus TNSA. The mechanism has been theoretically investigated by 
\textcite{gibbonPRE05} using a collisional tree-code approach. 

Indications of dominant front side acceleration due to resistivity effects 
have been reported in solid plastic targets \citep{leePRE08,leePoP11}
and also in low-density foams \citep{liPRE05} where an anomalously high 
resistivity might be due to spatially localized fields in the locally 
inhomogenous material.

\section{Current and Future Applications}
\label{sec:applications}

\subsection{Proton radiography}
\label{sec:applications_radiography}
\label{sec:APP_radiography}

The use of ion beams, and particularly proton beams, for radiographic 
applications was first proposed in the 1960s
\citep{koehlerS68}.
Quasi-monochromatic beams of ions from conventional accelerators have been 
used for detecting aereal density variations in samples
{via modifications of the proton beam density cross section, 
caused by differential stopping of the ions, or by scattering.}
Radiography with very high energy protons ($\sim 1-10~\mbox{GeV}$) 
is being developed as a tool for weapon testing \citep{kingNIMA99}.
Ion beams from accelerators have also been employed in some occasions for 
electric field measurements in plasmas, via the detection of the proton 
deflection, e.g. in \textcite{mendelPRL75}.
In practice, the difficulties and high cost involved in coupling externally 
produced particle beams of sufficiently high energy to laser-plasma experiments 
(or indeed magnetic confinement experiments) and the relatively long duration 
of ion pulses produced from conventional accelerators have limited the 
application of such diagnostic techniques.

The unique properties of protons from high intensity laser-matter interactions, 
particularly in terms of spatial quality and temporal duration, have opened up 
a totally new area of application of  proton probing or radiography. 
As seen in Sec.\ref{sec:TNSA_characterization}, the protons emitted from a 
laser-irradiated foil by TNSA can be described as emitted from a virtual, 
point-like source located in front of the target \citep{borghesiPRL04}.
A point-projection imaging scheme is therefore automatically achieved with 
magnification $M$ set by the geometrical distances at play. 
Backlighting with laser-driven protons has intrinsically high spatial 
resolution, which, for negligible scattering in the investigated sample, is 
determined by the size $d$ of the virtual proton source 
and the width $\delta s$ of the point spread function of the detector 
(mainly due to scattering near the end of the proton range), offering the 
possibility of resolving details with spatial dimensions of a few $\um$. 
As discussed in Sec.\ref{sec:nutshell_diagnostics} 
multilayer detector arrangements employing RCFs or CR39 layers offer the 
possibility of energy-resolved measurements despite the broad 
spectrum. Energy dispersion provides the technique with an intrinsic 
multi-frame capability. 
In fact, since the sample to be probed is situated at a finite distance from 
the source, protons with different energies reach it at different times. 
As the detector performs spectral selection, each RCF layer contains, in a 
first approximation, information pertaining to a particular time,
so that a movie of the interaction made up of discrete frames can be taken 
in a single shot. 
Depending on the experimental conditions, 2D proton deflection map frames
spanning up to 100~ps can be obtained. 
The ultimate limit of the temporal resolution is given by the duration of the 
proton burst at the source, which is of the order of the laser pulse duration. 

Several radiographic applications of laser-produced protons have been reported 
to date {and radiographs of objects for various size and thickness (down
to a few $\sim\um$) have been obtained \citep{rothPRSTAB02,cobbleJAP02,borghesiPRL04,mackinnonPRL06}.}
Density diagnosis via proton radiography 
has potential application in Inertial Confinement Fusion (ICF)
A preliminary test studying the compression of empty CH shells under 
multi-beam isotropic irradiation at the moderate irradiance of 
$10^{13}~\mbox{W cm}^{-2}$ has been studied in an experiment carried out at 
the Rutherford Appleton Laboratory \citep{mackinnonPRL06}.
Radiographs of the target at various stages of compression were obtained. 
Modelling of proton propagation through target and detector carried out using 
Monte Carlo codes permits the retrieval of density and core size at maximum 
compression (3 g/cc, 80 $\um$) in good agreement with hydrodynamic simulations.
{Radiographic analysis of cylindrically compressed matter \citep{volpePPCF11} 
and of shock wave propagation \citep{ravasioPRE10} has been also carried out, 
although the available detail with low energy protons was limited.}   

The most successful applications to date of proton probing are related to the 
detection of electric and magnetic fields in plasmas 
\citep{borghesiPoP02,mackinnonRSI04}.
Jointly with a parallel technique using monoenergetic protons from fusion reactions driven from laser-driven compressions
\citep{liPRL06},
proton probing with laser-accelerated protons has provided in this way novel and unique information on a broad range of plasma phenomena. 
The high temporal resolution is here fundamental in allowing the detection of highly transient fields following short pulse interaction. 

Two main arrangements have been explored. 
In \emph{proton imaging}, i.e. simple backlighting projection of the sample
\citep{borghesiPoP02,borghesiPPCF01},
the deflections cause local modulations in the proton density $n_p$ across the 
proton beam cross section, which, under the approximation of small deflections, 
can be written as 
\bea
\frac{\delta n_p}{n_p}
\simeq -\frac{eL}{2\varepsilon_pM}\int_{-b/2}^{+b/2}
        \nablav_{\perp}\left({\bf E}+\frac{{\bf v}_p}{c}\times{\bf B}\right)dx
\label{eq:protimag}
\eea  
where $v_p$ and $\varepsilon_p$ are the proton velocity and  energy, $M$ the 
projection magnification, $L$ the distance between the plasma and the detector, 
and the integral is along the trajectory of the protons, crossing a region 
$|x|<b/2$ where the fields ${\bf E}$ and ${\bf B}$ to be probed are present. 
Under simplified assumptions the above formula 
can be used to yield line-averaged values of the fields \citep{sarriNJP10}.
In \emph{proton deflectometry}, thin meshes are inserted in the beam between 
the proton source and the object as ``markers'' of the different parts of the 
proton beam cross sections \citep{mackinnonRSI04}.
The meshes impress a modulation pattern in the beam before propagating through 
the electric field configuration to be probed. 
The beam is in this way effectively divided in a series of beamlets, and their 
deflection  can be obtained directly from the pattern deformation. 
A technique employing two grids to generate a set of Moir\'e 
fringes has also been proposed as a way to increase the sensitivity to small 
electric fields \citep{mackinnonAPL03}.

{The proton probing technique has provided uniquely detailed information 
on nonlinear phenomena in high-intensity laser-plasma interactions, such as 
ion acoustic solitons and collisionless shock waves \citep{romagnaniPRL08},
phase-space electron holes \citep{sarriPoP10a}, 
the charge-displacement channel formation dynamics following relativistic 
self-focusing of laser pulses \citep{karNJP07,sarriPoP10c,willingalePRL11}
and the evolution of remnants of coherent electromagnetic structures and 
instabilities of various type 
\citep{borghesiPRL02,borghesiPRL05,romagnaniPRL10,sarriPRL10,sarriPoP11}.
Application to ns laser-produced plasmas of ICF interest has also allowed
to investigate laser filamentation in underdense plasmas 
\citep{sarriPRL11b,lanciaPoP11},
plasma expansion inside hohlraums \citep{sarriNJP10} 
and self-generation of magnetic fields 
\citep{nilsonPRL06,cecchettiPoP09,willingalePRL10,sarriPRL11b}.}
As an example of the use of time-resolved proton diagnostic,
Fig.\ref{Fig1_PRL_romagnani_05} reports data from an experiment 
where the protons are used to probe the rear of a foil following ultraintense 
irradiation of the front of the foil \citep{romagnaniPRL05}.
The probe proton pattern is modified by the fields appearing at the target rear 
as a consequence of the interaction, and the technique effectively allows 
spatially and temporally resolved mapping of the electrostatic fields 
associated to TNSA acceleration from the foil (see Sec.\ref{sec:TNSA_scenario}).
Fig.\ref{Fig1_PRL_romagnani_05}~a) shows the set-up for both imaging and 
deflectometry measurements. 
Frames b)-g) of Fig.\ref{Fig1_PRL_romagnani_05} correspond to proton images
at different times taken in a single shot, resolving the expansion of the 
plasma sheath and highlighting the multiframe capability of this diagnostic. 

\begin{figure}[t!]
\begin{center}
\includegraphics[width=0.48\textwidth]{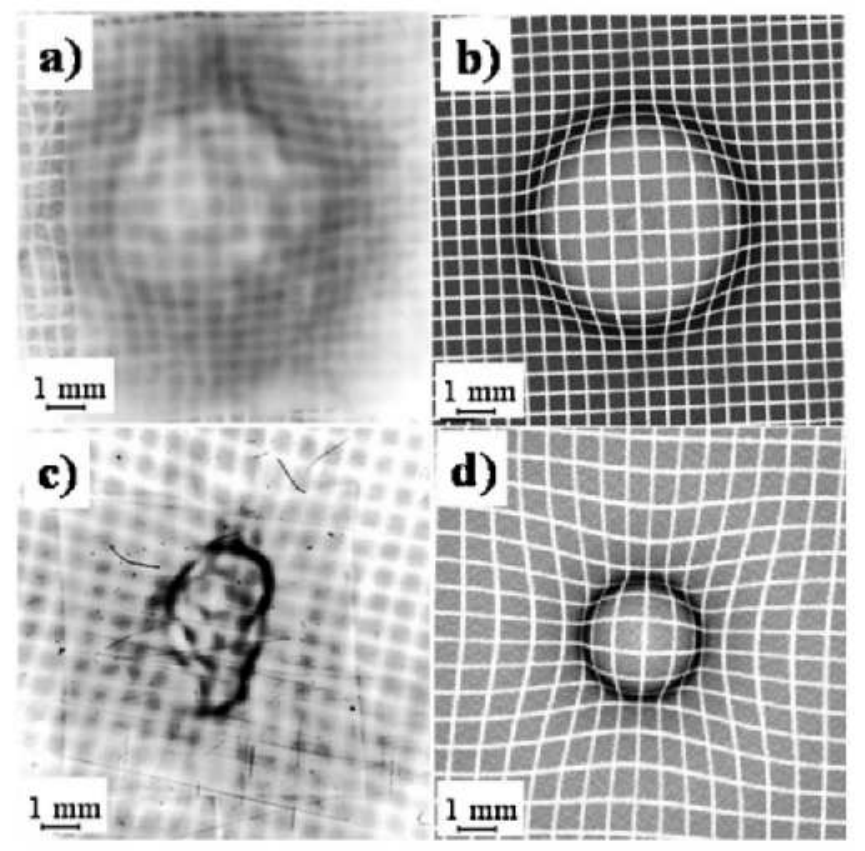}
\end{center}
\caption{Proton probing of magnetic fields \cite{cecchettiPoP09}.
\textbf{a)}, \textbf{c)}: probing deflectogram of a laser-irradiated foil (ns pulse, $10^{15}~\mbox{W cm}^{-2}$  on a $6~\um$ Al foil) obtained with a 5.5~MeV proton beam facing the foil, entering from the non-irradiated side in \textbf{a)} and from the opposite, laser-irradiated side in \textbf{c)}. 
The inversion of the deflection pattern reveals the effect of a toroidal ${\bf B}$-field (the asymmetrical pattern in c) is due to a nonideal intensity distribution in the focus). 
\textbf{b)}, \textbf{d)}: particle tracing simulations for the conditions of \textbf{a)}, \textbf{c)} assuming a suitably parametrized ${\bf B}$-field.
Reprinted with permission from Physics of Plasmas: 
\textcite{cecchettiPoP09}, Phys. Plasmas \textbf{16}, 043102.
Copyright 2009, American Institute of Physics.}
\label{Fig3ab_PoP_cecchetti_09}
\end{figure}

It could be noticed than on the basis of Eq.(\ref{eq:protimag}) it may not be 
possible in principle to attribute unambigously the measured deflections to the 
sole action of either electric or magnetic fields. 
Confidence in the interpretation of observed patterns can be increased by supporting the analysis method of both imaging and deflectometry data with particle tracing codes. Such codes simulate the propagation of the protons through a given space- and time-dependent field configuration, which can be modified iteratively until the computational proton profile reproduces the experimental ones. 
State-of-the-art tracers allow realistic simulations including experimental proton spectrum and emission geometry, as well as detector response.
Moreover, in some specific experiments it was possible to give evidence of magnetic fields, discriminating their effect on probe protons from that due to electric fields, thanks to the possibility of different probing directions \citep{cecchettiPoP09} or even exploiting the divergence of the probe beam \citep{romagnaniPRL10}. 
An example is given in Fig.\ref{Fig3ab_PoP_cecchetti_09} where the presence of an azimuthal ${\bf B}$-field has been revealed with mesh deflectometry by either a compression or outwards dilation of the mesh lines, depending on whether the  ${\bf B}$-field has clockwise or counter clockwise direction compared to the propagation direction of the probe beam. 

\begin{figure}[b!]
\begin{center}
\includegraphics[width=0.48\textwidth]{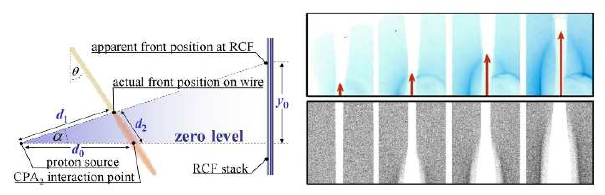}
\end{center}
\caption{(Color online)
Left frame shows the proton imaging set-up with increased dynamical 
range in the time domain \citep{quinnPRL09,quinnRSI09}. By placing a wire 
target at an angle with respect to the probe beam it is possible to resolve 
the propagation at a velocity close to $c$ of a field front along the 
laser-irradiated wire, as shown in the right frames 
(top: experimental images, bottom: particle tracing simulations). 
Reprinted figure with permission from 
\textcite{quinnPRL09}, Phys. Rev. Lett. \textbf{102}, 194801.
Copyright 2013 by the American Physical Society.}
\label{Fig_PRL_quinn_09}
\end{figure}

The divergence of the probe beam also implies that the effective probing time
is a function of the position on the image plane because of the different time
of flight for protons at different angles. 
This effect has to be taken into account for measurement of field structures propagating at relativistic speeds \citep{karNJP07,quinnPRL09} and actually may improve the capability to characterize such structures, as it was obtained by a slightly modified arrangement \citep{quinnRSI09}. 
This allowed to characterize the ultrafast, transient field front associated to the early stage of TNSA where electromagnetic effects come into play 
(see Fig.\ref{Fig_PRL_quinn_09}). 
A \emph{proton streak deflectometry} technique for obtaining continuous temporal mapping (but only one spatial dimension is resolved) has also been proposed, in which the energy resolution is done by means of magnet dispersion \citep{sokollikAPL08}.

\subsection{Production of Warm Dense Matter}
\label{sec:applications_WDM}
\label{sec:APP_WDM}

Laser-driven ions have found application in a number of experiments aimed to heat up solid density matter via isochoric heating, and create so-called Warm Dense Matter (WDM) states, i.e. matter at 1-10 times solid density and  temperatures up to 100 eV \citep{koenigPPCF05}
of broad relevance to material, geophysical and planetary studies
\citep{ichimaruRMP82,leeJOSAB03}.
The high-energy flux and short temporal duration of laser-generated proton 
beams are crucial parameters for this class of applications. 
WDM states can be achieved by several other means, 
e.g. X-ray heating \citep{tallentsHEDP09}
and shock compression \citep{kritcherS08}. 
However, when studying fundamental properties of WDM, such as equation 
of state or opacity, it is desirable to generate large volumes of uniformly 
heated material; ion beams, which can heat the material in depth, are in 
principle better suited to this purpose than the methods described above.

Heating of solid density material with ions can be achieved with 
accelerator-based or electrical-pulsed ion sources, see e.g.
\textcite{baileyLPB90,hoffmannNIMB00,tahirHEDP06}.
However the relatively long durations of ion pulses from these sources (1-10 ns) means that the materials undergo significant hydrodynamic expansion already during the heating period. 
On the contrary, laser-generated proton beams,  emitted in ps bursts,  provide a means of very rapid heating, on a timescale shorter than the hydrodynamic timescale.
By minimizing the distance between the ion source and the sample to be heated, it is possible to limit the heating time to tens of ps. 
The target then stays at near-solid density before significant expansion occur, 
and the WDM properties can be investigated within this temporal window. 

\begin{figure}
\begin{center}
\includegraphics[width=0.48\textwidth]{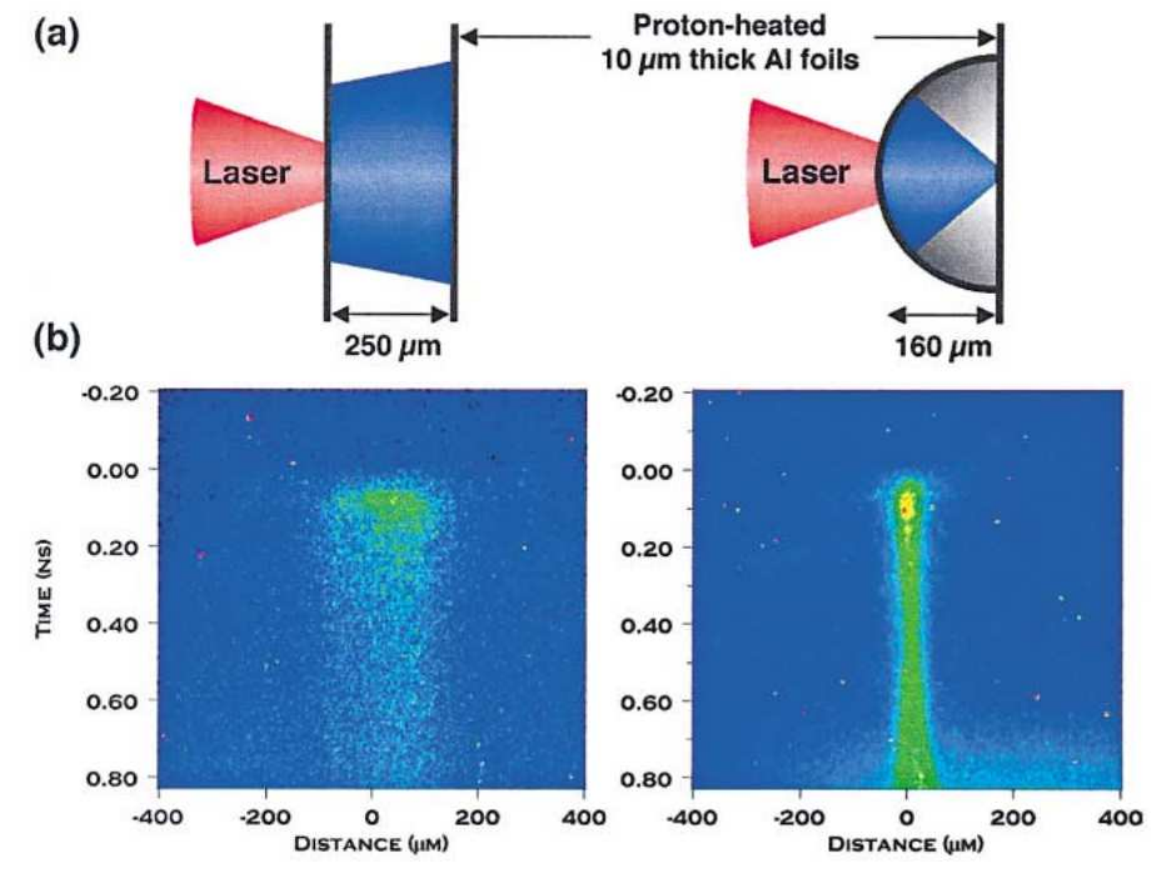}
\end{center}
\caption{(Color online) Heating of solid targets by protons. 
(a) Experimental setup for flat and focusing target geometries.
Each target consists of a flat or hemispherical
$10~\um$ thick Al foil irradiated
by the laser, and a flat $10~\um$ thick
Al foil to be heated by the protons.
(b) Corresponding streak camera images
showing space- and time-resolved
thermal emission at 570 nm from the
rear side of the proton-heated foil. Proton focusing by the 
hemispherical foil leads to a stronger,
more localized heating. 
Reprinted figure with permission from 
\textcite{patelPRL03}, Phys. Rev. Lett. \textbf{91}, 125004.
Copyright 2013 by the American Physical Society.}
\label{Fig1_PRL_patel_03}
\end{figure}

The first demonstration of laser-generated proton heating was obtained by 
\textcite{patelPRL03}.
In this experiment a 10 J pulse from the 100 fs JanUSP laser at LLNL was focused onto an Al foil producing a 100-200 mJ proton beam {used to heat a 
second Al foil}
Target heating was monitored via time-resolved rear surface emission, as shown 
in Fig.\ref{Fig1_PRL_patel_03}.
A focused proton beam, produced from a spherically-shaped target 
(Sec.\ref{sec:TNSA_optimization_control} ), was seen to heat a smaller region 
to a significantly higher temperature,
{($\sim$ 23 eV vs 4 eV) with respect to a focused beam}.  
With a similar ion focusing arrangement on a higher energy laser system, Gekko at ILE Osaka, \textcite{snavelyPoP07}
demonstrated secondary target heating up to 80 eV by imaging both visible and extreme-ultraviolet Planckian emission from the target’s rear surface.

Subsequent experiments 
{have investigated}
the properties of the WDM produced in this way 
with a number of diagnostics, either passive or in pump-probe configurations, 
combined  to self-consistent modeling of sample heating and 
expansion.
Warm solid Al at temperatures up to 15-20 eV \citep{dyerPRL08,mancicHEDP10}
and Carbon up to $\sim 2~\mbox{eV}$ \citep{rothPPCF09b} 
have been produced in this manner. 
\textcite{dyerPRL08} reconstructed the Equation of State (EOS) of the heated 
material by measuring the temperature and expansion rate of the heated target, 
via streaked thermal emission and chirped pulse interferometry.  

\begin{figure}
\begin{center}
\includegraphics[width=0.48\textwidth]{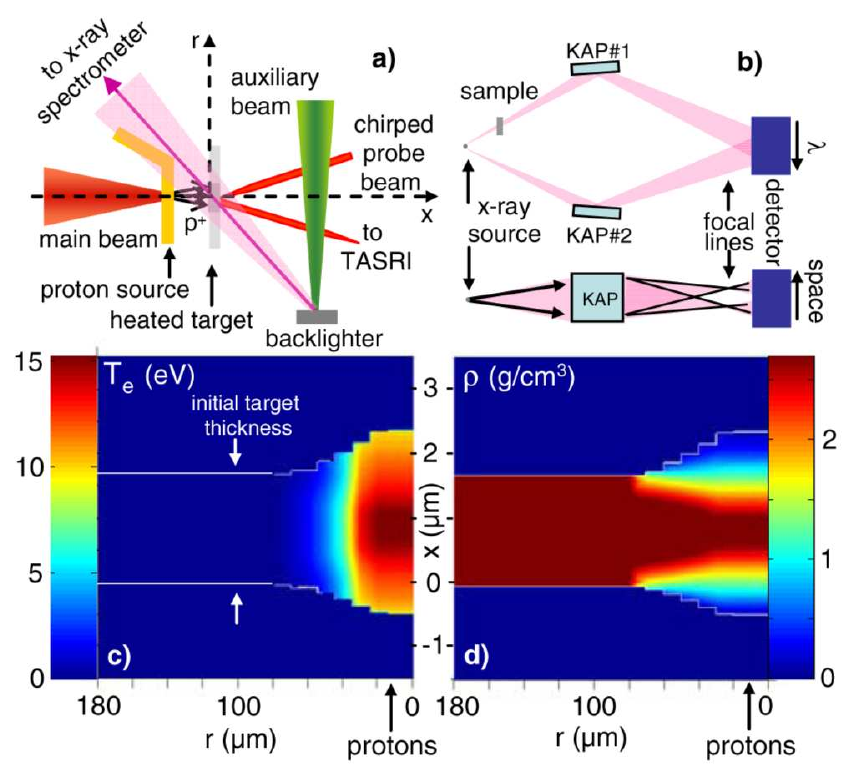}
\end{center}
\caption{(Color online) 
Set-up for X-ray probing of a solid Al target heated by protons: 
(a) top view of the experiment, 
(b) schematic of the x-ray spectrometer;
(c, d) snapshots of temperature  and density profiles 
of the heated $1.6~\um$ thick Al foil 
as given by self-consistent simulations, demonstrating isochoric heating up to
15~eV temperatures. 
Reprinted figure with permission from 
\textcite{mancicPRL10}, Phys. Rev. Lett. \textbf{104}, 035002.
Copyright 2013 by the American Physical Society.}
\label{Fig1_PRL_mancic_10}
\end{figure}

Pump-probe arrangements have been used in recent, more sophisticated experiments, which have provided totally novel information on the transition phase between cold solid and plasmas in isochorically heated Al and C targets.
\textcite{mancicPRL10}
have investigated the short range disordering of warm Al at solid density by applying time-resolved X-ray absorption near-edge spectroscopy
(see Fig.\ref{Fig1_PRL_mancic_10}).
Progressive smoothing of spectral features near the K-edge allowed to place an upper bound on the onset of ion lattice disorder within the heated solid-density medium of 10 ps. 
\textcite{pelkaPRL10} have recenty diagnosed ultrafast melting of  carbon samples, by X-ray scattering techniques, which allowed determination of the fraction of melted carbon in the heated sample.
Comparison to predictions based on different theoretical descriptions of the EOS of Carbon, indicates a departure from existing models, with implications for planetary core studies.

In all the experiments mentioned above the isochoric heating by the protons is 
volumetric, but not uniform \citep{brambrinkPRE07}, 
see e.g. Fig.\ref{Fig1_PRL_mancic_10}~c). 
Uniform heating would require some degree of proton energy selection, and 
choosing the sample thickness so that the Bragg peak of the selected protons 
does not fall within the sample, as suggested for example by 
\textcite{schollmeierPRL08}.

\subsection{Fast Ignition of fusion targets}
\label{sec:APP_fastignition}

The traditional route to ICF 
\citep{atzeni-book}
relies on the driven implosion of a pellet of thermonuclear fuel (a DT mixture).
Ignition occurs 
in a central ``hot spot'' following pulse compression. This approach
requires an extremely high symmetry and is prone to hydrodynamics 
instabilities, making ICF a historically difficult goal. 

In the Fast Ignition (FI) concept (see \textcite{keyPoP07} for a compact review)
ignition is driven by an external trigger, 
creating the hot spot in a time much shorter than
the typical fuel disassembly time.  Hence, ignition is separated from pulse 
compression. 
The FI approach might relax symmetry and stability requirements,
reduce the energy need for ignition
and allow fuel burn in a isochoric regime with high fusion gain.

In the original FI proposal by \textcite{tabakPoP94}, the foreseen ignitor beam
was composed by multi--MeV electrons accelerated by a petawatt
laser pulse
via the mechanisms described in Sec.\ref{sec:nutshell_fastelectrons}. 
Subsequent research showed that, besides generating an electron beam with
enough power to ignite, most problematic were the issues of energy transport
and deposition in the core. Concerning the latter issue, the energy deposition
profile of electrons is a smooth function, making it difficult to produce
a localized hot spot. 

The observation of efficient generation of multi--MeV proton beams in 
Petawatt experiments \citep{snavelyPRL00,hatchettPoP00} soon stimulated the
proposal of the use of such protons as an ignitor beam \citep{rothPRL01}. 
The most promising features of proton beam ignition as claimed in the paper 
were the highly localized energy
deposition profile (see Fig.\ref{fig:braggpeak}), the low emittance of the beam
and its focusability, for instance by parabolically shaping the rear side of 
the proton-producing target as suggested by numerical simulations 
\citep{ruhlPPR01,wilksPoP01}. 

\begin{figure}
\begin{center}
\includegraphics[width=0.48\textwidth]{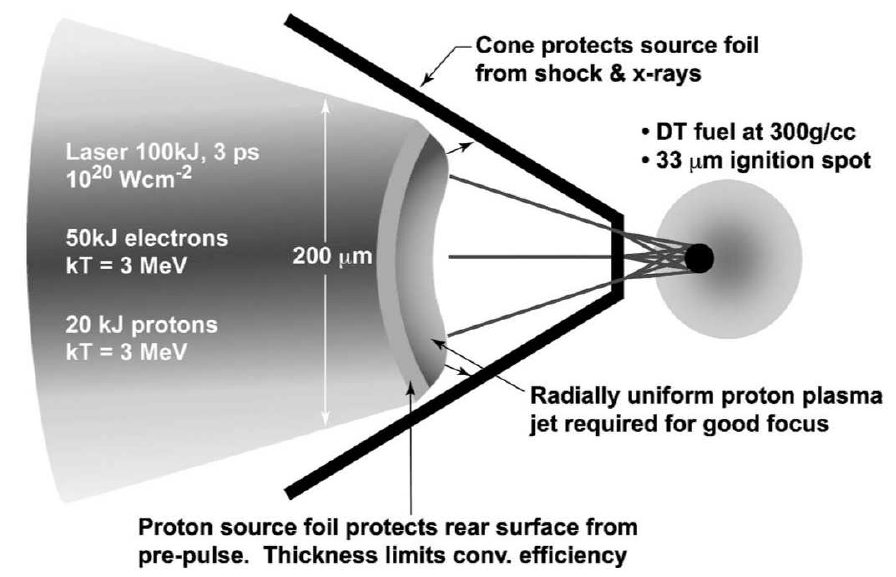}
\end{center}
\caption{Concept of proton-driven Fast Ignition in the TNSA-based, cone guided
scheme \cite{keyPoP07}. 
Typical parameters required for the ion beam and optimization issues
are also indicated. 
Reprinted with permission from Physics of Plasmas: 
\textcite{keyPoP07}, Phys. Plasmas \textbf{14}, 055502.
Copyright 2007, American Institute of Physics.}
\label{Fig16_PoP_key_07}
\end{figure}

Detailed calculations by \textcite{atzeniNF02,temporalPoP02}
addressed in particular the effects of the quasi-thermal energy distribution
typical of TNSA protons, and of the related temporal dispersion.
The latter could be beneficial for energy deposition since the 
proton stopping range increases with plasma temperature. Hence, heating 
due to the more energetic protons favors energy deposition by the
less energetic ones which arrive later in the dense fuel region. 
A fit of simulations for a proton temperature of 
$5~\mbox{MeV}$ provided the following estimate of the ignition 
energy\footnote{${\cal E}_{\mbox{\tiny ig}}$ includes only the
energy of the proton beam. The total energy of the laser driver is 
 ${\cal E}_{\mbox{\tiny ig}}/\eta_p$ with $\eta_p<1$ the conversion efficiency
into protons.
} 
${\cal E}_{\mbox{\tiny ig}}$ as 
a function of fuel density $\rho$ and distance  $d$ 
between proton source and fuel core:
\bea
{\cal E}_{\mbox{\tiny ig}} 
\simeq 90(d/\mbox{mm})^{0.7}(\rho/100~\mbox{g cm}^{-3})^{-1.3}~\mbox{kJ}.
\eea
Integration of the foil inside the cone of conically-guided ICF targets 
already designed for electron FI was then proposed
{in order to reduce $d$ and thus ${\cal E}_{\mbox{\tiny ig}}$.}
This raised the issue
of shielding the foil from pre-heating caused, e.g., by external radiation, 
which may jeopardize efficient TNSA 
(see Sec.\ref{sec:TNSA_optimization_energy});
a preliminary analysis is mentioned by \textcite{geisselNIMA05}. 
Fig.\ref{Fig16_PoP_key_07} sketches the target and foil assembly and summarizes
suitable parameters for protons FI with cone targets. 
\textcite{temporalPoP06,temporalPoP08} investigated a similar scheme but using 
two proton beams with suitably shaped radial profiles, obtaining a 40\% 
reduction of the ignition energy. 

Independently from the paper by 
\textcite{rothPRL01}, fast ignition by 
laser accelerated light ion beams was proposed by \textcite{bychenkovPPR01}.
{The} use of deuterons and Beryllium ions was investigated, 
and it was suggested that using ions with $Z>1$ 
could be advantageous because of their higher stopping power with respect to 
protons. Calculations by \textcite{honrubiaPoP09} showed that, 
differently from
the proton-based scheme, ions with a narrow energy spread 
{$\delta{\cal E}/{\cal E}$}
would allow to lower the ignition threshold: 
{for $\delta{\cal E}/{\cal E}=0.1$, 
${\cal E}_{\mbox{\tiny ig}}<10~\mbox{kJ}$ might be obtained}. 
This feature might also relieve the need to place the ion 
producing foil in a reentrant
cone, yielding a simpler target design \citep{fernandezNF09}.
The estimated parameters for a C ignitor beam are $400-500~\mbox{MeV}$ energy
per ion, $\gtrsim 10\%$ efficiency, and $\delta{\cal E}/{\cal E}<0.2$.
{To achieve such figures, mechanisms such as 
RPA (Sec.\ref{sec:RPA}) or BOA (Sec.\ref{sec:OTHER_transparency}) 
might be more suitable than TNSA}. 
Progress in related research 
has been recently reported by \textcite{hegelichNF11}
where separated experiments approaching the three above mentioned requirements 
are described. A fourth and so far unexplored issue might be the need to focus
the ion beam.

Another approach \citep{naumovaPRL09}
to ion-driven FI is based on RPA in the hole boring regime 
(Sec.\ref{sec:RPA_holeboring}). In such scheme, differently from the above 
described ones, ion acceleration occurs \emph{in situ} by direct interaction
of an ultraintense, {circularly polarized} 
laser with the corona of the fusion plasma. 
\textcite{tikhonchukNF10} report calculations on this 
scheme,
assuming direct acceleration of Deuterons and
characterizing possible high gain regimes with 
${\cal E}_{\mbox{\tiny ig}} \simeq 12-17~\mbox{kJ}$.
This corresponds to 
an overall ignition energy $>100~\mbox{kJ}$ and 
a required laser intensity exceeding $10^{22}~\mbox{W cm}^{-2}$. 
FI by laser-accelerated ions has been
investigated theoretically in several other works.\footnote{See e.g. 
\textcite{barrigaPRE04,ramisNF04,hosseiniJFE08,shmatovFST03,shmatovJPCS08,shmatovLPB11,badziakPoP11}}
Integrated {FI} studies, on the route to ignition-class 
experiments, could be perfomed in either the electron or the ion
approach in facilities to be developed, equipped with Petawatt-class laser 
{systems.}

\subsection{Biomedical applications}
\label{sec:APP_biomedical}

Hadrontherapy is the radiotherapy technique that uses protons, neutrons or carbon ions to irradiate cancer tumors. 
The use of ion beams in cancer radiotherapy\footnote{The use of energetic protons in radiology was first proposed by \textcite{wilsonR46} and demonstrated by \textcite{lawrenceC57}. Recent reviews on the state of the art in ion beam therapy are given, e.g., by \textcite{amaldiRPP05,smithMP09,schardtRMP10}. Focus on research and possible improvements in therapy with heavy ions is given e.g. by \textcite{kraftNJP09}.}
exploits the advantageous energy deposition properties of ions as compared to 
more commonly used X-rays (see Fig.\ref{fig:braggpeak}): 
the range for a proton or ion is fixed by its energy, which avoids irradiation 
of healthy tissues at the rear side of the tumor, while the well localized 
Bragg peak leads to a substantial increase of the irradiation dose in the 
vicinity of the stopping point.
The proton energy window of therapeutical interest ranges between 60 and 250~MeV, depending on the location of the tumor (the required Carbon ion range extends up to {400}~MeV/nucleon).
{The typical dose of a treatment session is in the 1-5 Gray range, 
and typical currents are 10 nA for protons and 1.2 nA for singly charged 
Carbon ions.}

Ion beam therapy has proven to be effective and advantageous in a number of tumours and several clinical facilities, employing mainly protons from synchrotron, cyclotron or linac accelerators are operational and routinely treating a significant number of patients. 
While protons are the most widespread form of ion treatment, facilities using carbon ions also exist, as their higher biological effectiveness makes them  very effective in treating radioresistant and hypoxic tumours
\citep{schardtNPA07}.
In treatment centers, magnetic steering systems (Gantries) are employed for multi-directional irradiation of a laying patient. 
Gantries are costly, large and massive, with a weight exceeding 100 tons for proton systems, and 500 tons for Carbon systems \citep{enghardtSPIE11}.

The use of laser based accelerators has been proposed by several authors as an alternative to RF accelerators in proton and ion therapy systems
\citep{bulanovPLA02,bulanovPPR02,fourkalPMB03,malkaMP04},
with potential advantages in terms of compactness and costs. 
Proposed options range from using laser-driven protons as high quality injectors in a RF accelerator \citep{anticiPoP11}
to all-optical systems, in which the ion beam acceleration takes place in the treatment room itself and ion beam transport and delivery issues are thus minimized \citep{bulanovPLA02}. 
It is recognized that there are significant challenges ahead before laser-driven ion beams meet therapeutic specifications, both in terms of maximum energy, energy spectrum, repetition rate and general reliability, to the levels required by the medical and therapeutic standards, as critically reviewed by \textcite{linzPRSTAB07} where specific issues are mentioned and a comparison with existing accelerator technologies is made. 
At present, the ion beam parameters are still far from the requirements
and it is clear that an extensive, long term activity will be needed to ascertain if and how laser-driven ion beams may become a competitive option. 
Several projects are currently active worldwide to explore the potential of laser-driven proton and  ion sources for biomedical applications, see e.g. \textcite{boltonNIMA10,borghesiSPIE11,enghardtSPIE11}. 
In view of future applications, several authors have started to design possible delivery systems, including target chamber and shielding \citep{maMP06}, particle energy selection and beam collimation systems to enable operation with the broadband and diverging laser-driven beams
\citep{fourkalMP03,nishiuchiJPCS10,hofmannPRSTAB11}. 

While currently a relative energy spread $\Delta{\cal E}/{\cal E} \simeq 10^{-2}$ is required for optimal dose delivery over the tumour region, authors have also modelled approaches in which the native broad spectrum of laser-accelerated ions is used to obtain directly the Spread Out Bragg peak distributions which are normally used to cover the tumour region \citep{fourkalMP07,luoPMB08} and more in general advanced methods exploiting the properties of laser-accelerated beams \citep{schellMP10}.

\begin{figure}
\begin{center}
\includegraphics[width=0.48\textwidth]{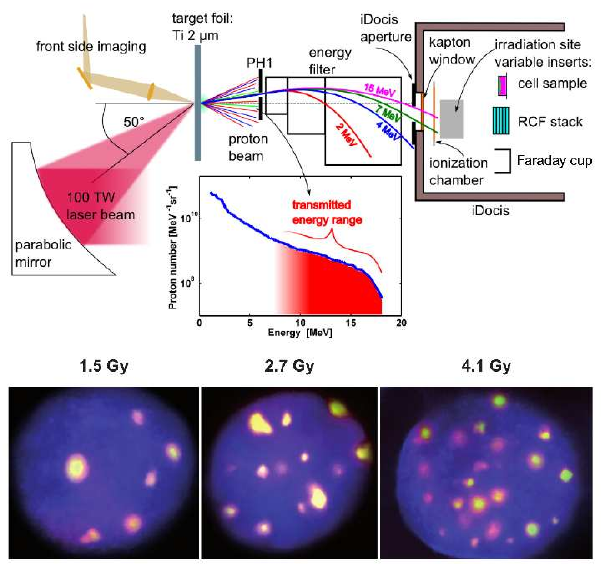}
\end{center}
\caption{(Color online) 
Top: overview of an experimental setup for integrated dosimetry and cell irradiation system by laser accelerated protons.
Bottom: fluorescence microscopy view of SKX tumour cell nuclei irradiated with such system, showing that the number of DNA double-strand breaks (bright spots, yellow-pink in the color version) increases with the delivered dose.
Reprinted figure from \textcite{kraftNJP10}, New J. Phys. \textbf{12}, 085003.
By permission from Institute of Physics Publishing (2013).
}
\label{Fig_NJP_kraft_10}
\end{figure}

An important step in view of  their future medical use
is to assess the biological effect of laser-driven ions, 
and highlight any anomaly associated to their pulsed, ultrashort  temporal profile.
\textcite{yogoAPL09} have first demonstrated the feasibility of cell irradiation studies using laser-driven protons, employing a suitable beam transport set-up
and then applied a refined technique to infer, via a clonogenic assay, the Relative Biological Effectiveness (RBE) of $\sim 2~\mbox{MeV}$ laser-accelerated protons, as compared to irradiation with a standard X-ray source \citep{yogoAPL11}, 
in human cancer cells.
The RBE observed ($1.2 \pm 0.1$) is comparable with literature results employing RF-accelerated protons of comparable Linear Energy Transfer (LET)
\citep{folkardIJRB96}.
The dose required to cause significant cell damage (typically 1 to several Gray) was obtained in several irradiations taking place at 1~Hz repetition rate.
\textcite{kraftNJP10} also carried out proton irradiations of cells, 
highlighting 
dose-dependent incidence  of double strand DNA break in the cells 
(Fig.\ref{Fig_NJP_kraft_10}).
 
The peculiar characteristics of laser-driven protons 
have required the development of innovative dosimetric approaches, 
as described for example
by \textcite{richterPMB11,fioriniPMB11}.
In all these experiments, the dose (1-10 Gy) is delivered to the cells in short bursts of $\sim\mbox{ns}$ duration. 
In experiments by \textcite{yogoAPL11,kraftNJP10} 
the dose is fractionated and the average dose rate is comparable to the one used in irradiations  with conventional accelerator sources 
($\sim 0.1~\mbox{Gy s}^{-1}$). 
In a recent experiment \citep{fioriniPMB11}
employing a high energy ps laser system, it has been possible to reach up to 5~Gy in a single exposure, reaching dose rates as high as $10^{9}~\mbox{Gy s}^{-1}$. 
This allows access to a virtually unexplored regime of radiobiology, where, in principle, nonlinear collective effects
\citep{fourkalPMB11}
on the cell due to the high proton density in the bunch may become relevant.

Besides cancer therapy, application of laser-driven ion beams in medical diagnosis has also been proposed.
Multi-MeV proton beams can induce nuclear reactions in low-$Z$ materials 
(Sec.\ref{sec:APP_nuclear})
in order to produce neutrons, of possible interest for Boron Neutron Capture Therapy for cancer, or short-lived positron emitting 
isotopes which may be employed in Positron Emission tomography (PET). 
PET has proven to be extremely useful in medical imaging of blood flow and amino acid transport and in the detection of tumors. 
Usually, reactions for PET are carried out by using up to 20~MeV protons or similar energy deuterons from cyclotrons with the concomitant problems of large size and cost and extensive radiation shielding. 
Production of  short-lived  isotopes  via laser-driven proton beams may be feasible in the near future with the possibility of employing moderate energy, ultrashort, high-repetition  table-top lasers. 
Extrapolations based on present results point to the possibility of reaching the GBq activities required for PET if laser systems capable of delivering 1~J, 30~fs pulses focused at $10^{20}~\mbox{W cm}^{-2}$ with kHz repetition will become available \citep{fritzlerAPL03,lefebvreJAP06} and economically competitive with existing technologies.

\subsection{Nuclear and particle physics}
\label{sec:APP_nuclear}

The interaction of laser-driven high-energy ions with secondary targets can initiate nuclear reactions of various  type,  which, as mentioned  
before (Sec.\ref{sec:nutshell_diagnostics})
can been used as a tool to diagnose the beam properties. 
This also presents the opportunity of carrying out nuclear physics experiments in laser laboratories  rather than in accelerator or reactor facilities, and to apply the products of the reaction processes  in several areas.  
Reactions initiated by laser-accelerated high-$Z$ ions have been studied in a number of experiments. 
\textcite{mckennaAPL03,mckennaPRL03,mckennaPRE04} have shown that fusion reactions between fast heavy ions from a laser-produced plasma and stationary atoms in an adjacent ``activation'' sample create compound nuclei in excited states, which de-excite through the evaporation of protons, neutrons, and $\alpha$-particles. 
A similar experiment with protons driving nuclear reactions and
excitations in a Cu target has been reported by \textcite{hannachiPPCF07}.
Nuclear reactions of interest for spallation physics have also been investigated by employing the multi-MeV proton beams \citep{mckennaPRL05}.
The broad energy distribution of the beams is in this case advantageous for the determination of residual nuclide generation arising from specific spallation processes such as evaporation.
In addition, MeV proton interaction with low-$Z$ materials can produce short-lived isotopes\footnote{Typical short lived positron emitters include  $^{11}\mbox{B}$, $^{11}$C, $^{13}$N, $^{15}$O, ${^{18}}$O and $^{18}$F. Related experiments have been reported e.g. by \textcite{nemotoAPL01,fritzlerAPL03,ledinghamS03,clarkeAPL06,fujimotoAPL08,fujimotoRSI09,oguraAPE09}.} 
 of medical interest, e.g. for PET diagnostic (see Sec.\ref{sec:APP_biomedical}).

Recently, a scheme of a ``fission-fusion'' process driven by RPA 
(Sec.\ref{sec:RPA}) has also been proposed to 
produce neutron-rich nuclei in
the range of the $r$-process \citep{habsAPB11}; 
such studies, of relevance for astrophysical 
nucleosynthesis, would require intensities above 
$3 \times 10^{22}~\mbox{W cm}^{-2}$ that may be available with next-generation
short-pulse laser facilities.

Neutrons are an important product of the nuclear reactions produced above, with potential applications in cancer therapy, neutron radiography, radiation damage of materials and transmutation of nuclear waste. 
The potential for laser-driven neutron sources is considerable and offers advantages over accelerator- and reactor-driven sources in term of cost, compactness, brightness and short duration for applications such as fast neutron radiography \citep{lancasterPoP04}
and studies of impulsive damage of matter \citep{perkinsNF00}.
This has motivated several experiments on the production of neutrons initiated by laser-driven proton beams on secondary targets. 
Experiments carried out at the VULCAN laser facility have revealed
neutron yields  up to 
$4 \times 10^{9}~\mbox{sr}^{-1}$ per pulse at a laser intensity of 
$3 \times 10^{20}~\mbox{W cm}^{-2}$ \citep{yangJAP04},
produced via the $^{11}\mbox{B}(p,n){^{11}}\mbox{C}$ and
$^{7}\mbox{Li}(p,n){^{7}}\mbox{Be}$ reactions. The latter was also investigated
by \textcite{youssefPoP06} as a diagnostic of proton acceleration.
Neutron production has also been observed in interaction with solid targets containing deuterium (typically deuterated plastic), which can be either directly irradiated by high intensity laser pulses
\citep{norreysPPCF98,disdierPRL99,habaraPoP03,habaraPRE04,willingalePoP11}
or irradiated by ions accelerated on a separate target
\citep{fritzlerPRL02,karschPRL03}. 
In both cases the neutrons are produced in the course of fusion reactions of the type $\mbox{D}(d,n){^{3}}\mbox{He}$ involving laser-accelerated  deuterium ions as also observed in gaseous targets
\citep{ditmireN97,grillonPRL02}.
Numerical modeling and theoretical investigations
of laser-driven neutron production have been carried out by several authors,
see e.g. 
\textcite{toupinPoP01,macchiAPB06,davisPoP08,davisPoP11,ellisonPoP10}.

\label{sec:APP_particle}

The application of laser-accelerated ions in particle physics requires
``by definition'' the ions to be ``relativistic'', i.e. their total energy must
exceed the rest energy whose value per nucleon is 
$\sim m_pc^2 \simeq 0.94~\mbox{GeV}$. Presently, observed cut-off energies 
are more than an order of magnitude below this threshold. Nevertheless, 
the scalings inferred from either experiments or theoretical models and the 
foreseen availability of higher laser powers in a few years suggests that
GeV ions may eventually be produced and applied in selected particle physics
experiments. 
Moreover, it may be noticed that the very low emittance that can be obtained 
for laser accelerated ions make them suitable for post-acceleration, e.g. as
an injection source for heavy ion accelerators \citep{krushelnickIEEE00}.
Specific advantages might be the high number of ions produced per shot combined
with the short duration.

\textcite{bychenkovJETPL01} estimated the threshold for production of pions 
by protons accelerated in a solid target, obtaining that at intensities above 
$10^{21}~\mbox{W cm}^{-2}$ the flux of pions may be much higher than obtained
with conventional accelerator techniques. It may be noticed that the 
prompt laser-driven, high field-gradient acceleration of pions is of high 
interest because of the finite lifetime of such particles; a related discussion
is reported by \textcite{mourouRMP06}.

\textcite{pakhomovJPG02} proposed the use of laser-accelerated protons at 
intensities of $\simeq 10^{23}~\mbox{W cm}^{-2}$ to drive, via pion generation
and decay, pulses of 20~MeV muon-neutrinos of interest for, e.g., studies 
of neutrino oscillations. 
\textcite{bulanovNIMA05} further explored this concept suggesting the
radiation pressure dominated acceleration regime (Sec.\ref{sec:RPA_dominance})
as suitable for this class of experiments. A more general discussion of the
required laser developments is reported by \textcite{terranovaNIMA06}.

\section{Conclusions and outlook}
\label{sec:conclusions}
We have reviewed about twelve years of research on ion acceleration driven by superintense laser pulses. The most investigated acceleration mechanism, namely the so-called Target Normal Sheath Acceleration, has been extensively discussed, surveying both the main experimental achievements and the underlying theoretical interpretation. In addition, we have provided an outlook to other proposed acceleration schemes, like Radiation Pressure Acceleration, Shock acceleration, Break-out afterburner, acceleration in near-critical and underdense plasmas and resistively enhanced acceleration; for these mechanisms, the fundamental theoretical ideas have been presented, together with the most promising experimental results.  A brief presentation of the most promising possible applications of the laser-generated ion beams has been finally given. 
While we were completing our work, another extended review on this topic has been published \cite{daidoRPP12}, which we are delighted to recommend as a 
complementary reading.

This field of research has attracted an enormous interest and has shown unique potential for both innovative investigations and for applicative purposes. 
The impressive development in laser technology and the increasing use of advanced methods of material science for target manufacturing has resulted in a high level of sophistication of current experiments, with new physical issues continuously emerging from the experimental investigations. 
At the same time, theory and simulation naturally have found a fertile field, which poses original problems and suggest unexplored paths for reaching their solution. 
The vitality of this research area is well demonstrated by the need for frequent updates during the preparation of this Review, as new significant results continuously appeared in the literature.

As discussed in the Introduction, future developments and achievements are 
naturally linked to foreseen developments in laser technology, providing for 
the first time extreme laser intensities close
to, or even beyond the limit at which the ions become relativistic. Emerging 
laser projects will enable to verify
the scaling of ion acceleration physics in the ultra-high intensity regimes 
and its suitability for proposed applications,
as well as to test theoretical ideas and provide to fundamental physics an 
example of ``relativity in action''
in a macroscopic, many-body system. The expected progress is not exclusively 
related to further developments
of the solid state laser technology which has been so far the preferred route 
to producing high intensity
laser pulses, as for example, recent experiments have attracted novel interest 
also in “old” technology such as
CO2 laser systems. Smart and advanced target engineering, e.g. development of 
multilayer, low-density, micro- and nanostructured
targets, will also play an important role in establishing future directions of ion 
acceleration.

The successful developments in this first period leads to a series of open questions, which will inform research in this field over  the coming years. 
Will it be possible to reach and break the GeV/nucleon threshold? 
Will researchers achieve an active and satisfactory experimental control on the physics of laser-ion acceleration working on the detailed properties of laser pulses and of target material and structure? 
Is there hope to pursue front-edge nuclear and particle physics research in small scale laboratories thanks to the use of laser-driven ion beams? 
Will the research on ion acceleration  result in practical, technological applications of direct societal benefit? 
We leave to the future experimental and theoretical research the answer to these and many other questions, some of which we probably do not even imagine today.

\begin{acknowledgments}
We acknowledge support of  EPSRC, grant EP/E035728/1 (LIBRA consortium)
and from the Italian Ministry of University and Research via
the FIRB project ``Superintense Laser-Driven Ion Sources''.
\end{acknowledgments}

\hyphenation{Post-Script Sprin-ger}

\end{document}